\documentclass{aa} % for a referee version
  \def\swift{{\it Gehrels Swift~}} 
   
  \def\nustar{{\rm NuSTAR~}}

  \def\snxv{${\sigma^2_{NXV}}$}
  \def\mbh{${\rm M}_{\rm BH}$}
  \usepackage{xcolor}
  \usepackage{natbib}
  \usepackage{graphicx}
  \usepackage{txfonts}
\begin{document}

     \title{Black hole mass estimation using X-ray variability measurements in Seyfert galaxies}

%     \subtitle{}
      \titlerunning{Black hole mass estimation of Seyfert galaxies}
      \authorrunning{A. Akylas et al.}
     
     \author{A. Akylas
            \inst{1}
            \and
            I. Papadakis
            \inst{2,3}
            \and
            A. Georgakakis
            \inst{1}
                        }

     \institute{Institute for Astronomy Astrophysics Space Applications and Remote Sensing (IAASARS), National Observatory of Athens, I. Metaxa \& V. Pavlou, Penteli, 15236, Greece \\ \email{aakylas@noa.gr}
     \and
     Department of Physics and Institute of Theoretical and Computational Physics, University of Crete, 71003 Heraklion, Greece
     \and
     Institute of Astrophysics - FORTH, N. Plastira 100, 70013 Vassilika Vouton, Greece}

 %   \date{}

 \abstract
  % context heading (optional)
  % {} leave it empty if necessary  
   {}
  % aims heading (mandatory)
   {Our objective is to critically assess the X-ray flux variability as a tool for measuring the black hole (BH) mass in active galactic nuclei (AGN). We aim to establish a prescription for estimating BH masses based on measurements of the normalised  excess variance from X-ray data. We discuss the minimum requirements in terms of the light-curve duration and X-ray signal-to-noise ratio (S/N)  to enable a reliable determination that is comparable to what can be derived from the continuum and emission line reverberation studies.}
  % methods heading (mandatory)
   {We used the light curves of local Seyfert from the  Nuclear Spectroscopic Telescope Array hard X-ray mission ($\rm NuSTAR$), to compute the normalised excess variance (\snxv) in the 3-10 and 10-20 keV bands, thus extending the analysis to an energy band higher than 10 keV. The excess variance measurements were then combined with independent BH mass estimates from the literature to establish the \mbh\ versus \snxv\ relation for different samples and weigh its accuracy in terms of the light-curve duration and X-ray S/N.}
  % results heading (mandatory)
   {We find that it is possible to accurately measure the BH mass in AGN using excess variance measurements in the 3-10 and the 10-20 keV bands, however, strong quality requirements should be applied. The minimum necessary S/N and duration of the light  curves used to compute the excess variance ought to be $\sim$3 and $\sim 80 - 100$ ks, respectively. We provide a linear relationship between the normalised excess variance and the black hole mass that can be used to estimate the latter, with an average uncertainty of the order of  $0.4 - 0.25$ dex (depending on the adopted light-curve segment duration). In general, BH mass estimates from 3-10 keV and 10-20 keV band light curves  are expected to be similar.  The 10–20 keV band is preferred for sources that are heavily absorbed and the 3–10 keV band is preferred for sources that may be dominated by the X–ray reflection component at energies above 10 keV. }
  % conclusions heading (optional), leave it empty if necessary
   {}


    \authorrunning{Akylas et al.}

    \maketitle
  %
  %________________________________________________________________

\section{Introduction}

Super-massive black holes (SMBHs) reside in the centre of most (if not all) galaxies and 
are responsible for their most energetic face, namely, of active galactic nuclei (AGN). 
According to the current paradigm, the incredibly high luminosity of these objects 
is powered by the accretion of matter in the vicinity of the SMBHs. 
As matter  spirals inward, copious amounts of energy are  released, over a wide 
range of wavelengths (from radio to gamma rays) due to 
the conversion of gravitational potential energy into radiation.
The emitted power in AGN depends on the black hole (BH) mass and the accretion rate. Their luminosity can reach $10^{15}$ 
times that of the Sun, while the accretion of matter may significantly 
contribute to the growth of the BH mass.

Studies have made it clear that SMBHs  play a major role in regulating star 
formation in galaxies and, thus, affect their surroundings. One of the  strongest pieces of empirical 
evidence of the mutual interaction between AGN and its host galaxy 
is demonstrated by the strong correlation between the mass of SMBHs and the bulge stellar velocity dispersion, $\rm \sigma_{\ast}$, \citep[e.g.][]{ferrarese2000, gebhardt2000}. 
This relation can be established through the 
interaction between the energy and radiation generated by accretion and the gas in the 
host galaxy,  known as AGN  feedback \citep[e.g.][]{fabian2012}. 

In order to understand and investigate the role of SMBHs in galaxy formation and evolution 
processes, we need to monitor the  growth of the SMBHs across the cosmic timescale. Therefore, it 
is not surprising that a lot of effort has been focussed on finding ways of measuring the mass of SMBHs in 
galaxies -- overall, and, in particular, in the case of AGN. 

Direct mass measurements of SMBHs are possible with stellar kinematics and gas dynamics, 
although these methods require good spatial resolution and are presently only feasible for a small number 
of nearby galaxies. These methods can be used for weak AGN, such as low-luminosity AGNs, LINERs, 
and Seyfert-2s; this is because for more luminous AGNs, the strong nuclear emission weakens the stellar 
features in the spectrum. 

Based on the assumption that the motion of the gas in the broad line region (BLR) of AGNs is dominated 
by the gravitational influence of the SMBH, we can use the virial equation $\rm M_{BH}$ = ($\rm f R_{BLR} \Delta V^2$ )/G 
to estimate SMBHs. Here,  $\rm R_{BLR}$ is the average radius of the emitting gas in the broad line region, 
usually determined with reverberation mapping \citep[e.g.][]{peterson2004}, while the velocity dispersion 
of the gas ($\rm \Delta V$) is measured from the width of the broad emission lines. The dimensionless factor, f, 
is referred to as a virial coefficient and encapsulates the unknown geometry and dynamics of the broad line region gas. 
Its  true value for each AGN is unknown and, in most cases, an average value is used, based on the assumption that 
AGNs follow the \mbh\ -- $\rm \sigma_{\ast}$ relation observed in  quiescent galaxies \citep[e.g.][]{grier2013}.

Black  hole masses can also be inferred indirectly from observable quantities that are correlated with the mass of the SMBHs,
such as the velocity dispersion of bulge stars, that is, the value of the \mbh\ -- $\rm \sigma_{\ast}$ scaling relationship (mentioned above) or the bulge luminosity \citep[e.g.][]{gultekin2009}.

In this work, we focus on the flux variability of AGN as a means of measuring SMBHs. Stochastic variations of the radiated energy 
is one of the main observational characteristics of the accretion flows onto compact objects  and SMBHs in particular. 
The origin of these variations is still not well understood and could be related to accretion flow instabilities, 
a flaring corona, or hot-spots orbiting the central compact object 
\citep[e.g.][]{GravityCollaboration2018, GravityCollaboration2020}. 

Although the AGN flux variability is observed across the entire electromagnetic spectrum, analysing light curves at X-ray wavelengths 
offers several advantages. Stellar processes emit (very) little X-ray radiation and, therefore, observations at these energies arguably provide 
the cleanest diagnostic of active SMBHs at the centres of galaxies over a broad redshift range and accretion luminosity baselines \citep[e.g.][]{brandt2015}. 
The X-ray photons can also penetrate relatively dense gas clouds and be virtually unaffected, thereby providing a representative sampling 
of the obscured AGN in the Universe.  Therefore, X-ray observations enable flux variation measurements across a broad range of AGN (including low-luminosity and obscured objects) 
that are challenging or even impossible to perform at other wavebands, where the host galaxy emission is dominant. The proximity of the X-ray 
emitting region to the active black hole implies the possibility of a direct connection between the X--ray flux variations and the 
physical properties of the system (e.g. black-hole mass or accretion rate). 

Past observations have indicated that such a relation exists in AGN. The X-ray power spectral density (PSD) of AGN has been modelled
using  a bending  power-law with a slope of -1 at low frequencies and then a slope of -2 \citep[e.g.][]{mchardy2004}. Various studies 
have indicated that the PSD bend frequency may be aptly correlated with BH mass and (potentially) with the bolometric luminosity 
\citep[e.g.][]{czerny2001, mcHardy2006, kording2007, Gonzalez2012}. 
However, it is difficult to estimate the PSD in AGN (and, even more so, detect the bending frequency) as this requires long, high 
signal-to-noise ratios (S/N) and uninterrupted light curves. Therefore, it is not practical to use the bending-frequency versus BH mass relation 
to measure the BH mass in AGN. 

Using the normalised excess variance, \snxv, \citep[e.g.][]{nandra1997} as an estimator of the intrinsic
'band-variance' of the source, a tight anti-correlation between \snxv\ and \mbh\ 
has also been found in the local  universe \citep[e.g.][]{papadakis2004, oneill2005, zhou2010, ponti2012}
as well as the high-redshift universe \citep[e.g.][]{lanzuisi2014}. In particular, the value of \snxv\ can be estimated much more easily with the available 
data sets for many types of AGN, both  at low and high redshifts. It is for this reason that most of the SMBH mass measurements 
in AGN are based on the use of the \snxv\ vs \mbh\ relation \citep[e.g.][]{ponti2012}.

Our objective is to critically assess the X-ray flux variability as a tool for measuring AGN SMBHs. 
We used the light curves of local Seyfert from the $\rm NuSTAR$ observatory to compute \snxv\ in the energy bands 
3-10 and 10-20 keV. The use of $\rm NuSTAR$ data has allowed us to extend our analysis to an energy band higher than 10 keV, 
which may be important for obscured AGN. The excess variance measurements are then combined with independent BH mass estimates 
from the literature to establish the \mbh\ vs \snxv\ relation in AGN. All studies in the past have 
focused on the reverse relation (i.e. \snxv\ vs \mbh) as the main goal was the study of the X--ray variability, 
while our focus here is the estimation of the BH mass in AGN from the excess variance measurements. 

 We present a prescription to measure SMBH masses using the excess variance measurements and we discuss, 
 for the first time,  the minimum requirements in terms of light-curve duration and X-ray S/N values 
 that enable a reliable determination, similar to those derived from the continuum and emission line reverberation studies 
 or the  \mbh\ -- $\rm \sigma_{\ast}$ relation. 

%-----------------------------------------------------------------
%-----------------------------------------------------------------
%%%%% SECTION 2

\section{The sample}\label{thesample}
    
Our sample consists of the Seyfert galaxies detected in the 105-month survey of the Burst Alert Telescope, \citep[BAT;][]{Barthelmy2004} 
on board the \swift Gamma-Ray Burst observatory \citep{gehrels2004}, along with archival observations from  the \nustar mission \citep{harrison2013}.

The BAT 105 month catalogue \citep{oh2018} contains 379 Seyfert-1s and  448 Seyfert-2s galaxies. There are 161 Seyfert-1s and 253 Seyfert-2s 
(414 in total) that have been observed by \nustar until September 2020, with a duration greater than 10 ks. We found 664 independent, 
archival \nustar observations for these sources. 
In these \nustar observations, a total of 114 are flagged as having been affected by increased solar activity and these have been removed from  further analysis. 
Moreover, we  restricted our analysis to the low-redshift regime. This is meant to minimise the inconsistencies on the excess variance estimation 
when using intervals with a fixed length in the observers frame. Therefore, sources  with redshift greater than 0.07 (i.e.\, the maximum 
redshift for the sources in the reverberation sample; see  Section \ref{revsample} below for details) are not included in the sample.  Our final sample 
consists of 473 independent observation of 300 local AGN -- of which 108 are Seyfert-1s and 192 are Seyfert-2s galaxies. 

In order to provide a recipe for the \mbh\ estimation, we need to study sources with known BH mass. Thus, we kept sources with BH measurements, 
which are based either on the so-called reverberation mapping technique or on the use of good-quality, host galaxy velocity dispersion measurements, 
as we explain below. 

%------------------------
%%%%%%%%%%%%%%%%% S 2.1
\subsection{The reverberation sample}\label{revsample}

Reverberation mapping \citep[e.g.][]{blandford1982} is one of the most direct ways to obtain BH masses in AGNs. It provides an estimate of the size of the 
BLR from the lag between the variability in the photo-ionising continuum and the broad emission lines. On this basis, it is possible to infer a virial mass for the SMBH in the central region of the AGN. 

We cross-correlated the original sample  with the sources included in the AGN BH mass database \citep{massdb} to obtain a sample with BH mass measurements 
based on reverberation mapping technique (hereafter, the 'rev sample'). 
We adopt the weighted average of the individual BH mass estimates determined from all the emission lines, assuming that <{\rm f}>=4.3, as suggested by \citet{grier2013}. 
The rev sample consists of 26  AGN  with 86 \nustar observations. 
The average duration per \nustar observation in the rev sample is $\sim80$ ks.
The rev sample log is presented in Table \ref{rev_sample}.

%---------------------------
%%%%%%%%%%%% S 2.2
\subsection{The velocity dispersion sample}\label{vdsample}

Reverberation mapping data are currently available only for a small number of AGN, but the empirical scaling relation between black hole mass and host-galaxy properties, namely, the \mbh\ -- $\rm \sigma_{\ast}$ relation \citep[e.g.][]{grier2013} offers an alternative  way to obtain BH mass estimates. To this end, we used the central  velocity dispersion measurements ($\sigma_{\ast}$) presented  in the first catalogue of the Swift-BAT AGN Spectroscopic Survey \citep[BASS,][]{koss2017}. These authors provided measurements of $\sigma_{\ast}$ for many AGN in the 70-month Swift BAT all-sky catalogue. We considered only secure velocity dispersion measurements by selecting  the data with spectral fitting quality flag value of 1 or 2. Dual AGN systems, as reported in  \citet{koss2012} are excluded from the analysis. There are 84 sources fulfilling these criteria, with 111 \nustar observations. 

\begin{figure}
    \begin{center}
    \includegraphics[height=0.62\columnwidth]{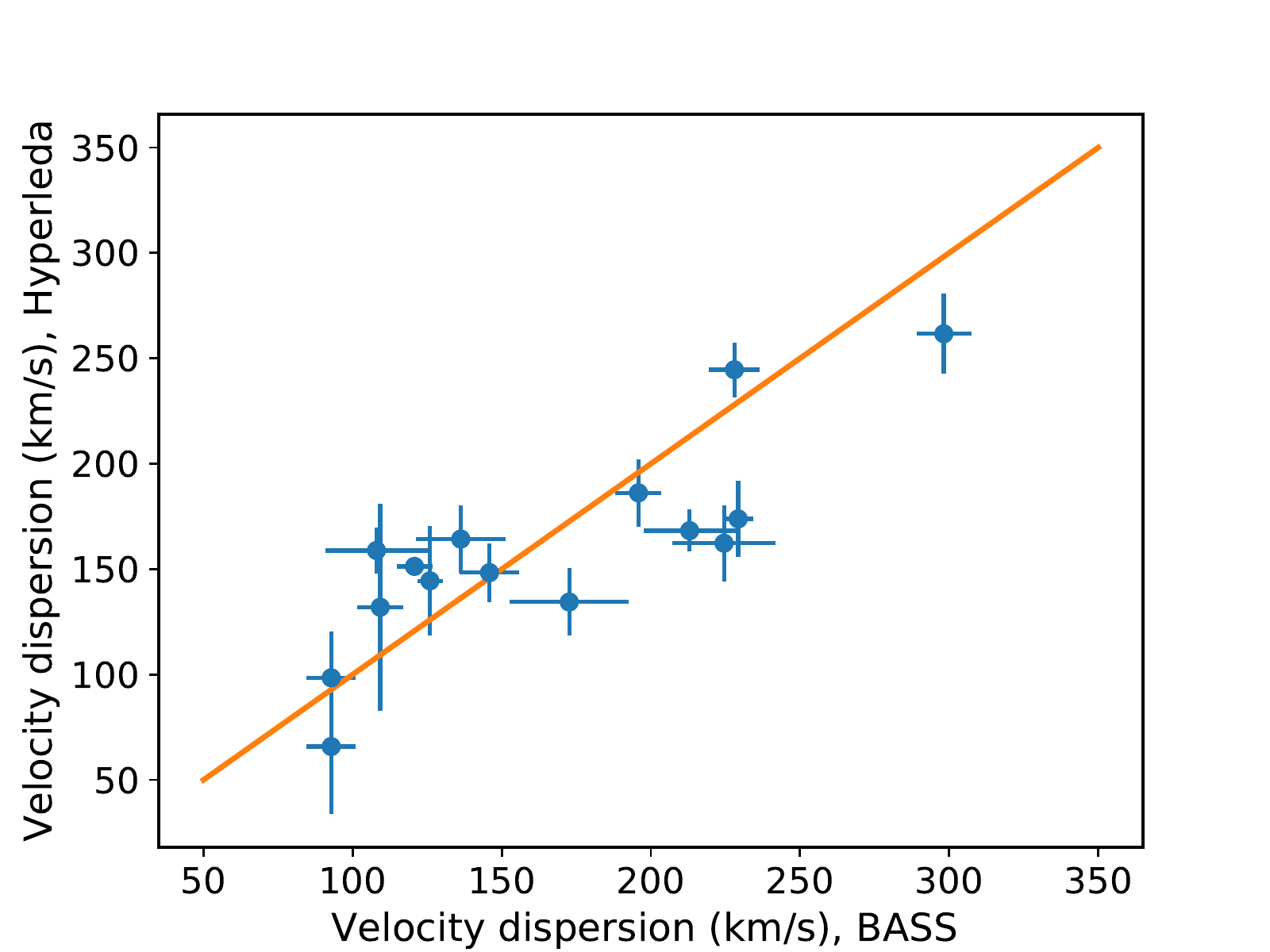}
    \end{center}
    \caption{Comparison of the velocity dispersion measurements of the common sources from the Hyperleda database and the BASS catalogue. The error-bars correspond to 1$\sigma$ confidence interval. The solid line show the one-to-one ($y=x$) relation.}
    \label{hypervskoss}
\end{figure}

We also searched the Hyperleda database (\citet{paturel2003}) to obtain additional central velocity dispersion measurements and increase our sample. 
We find $\rm \sigma_{\ast}$ measurements for 24 new sources (i.e without $\rm \sigma_{\ast}$ measurements in the BASS survey sample) with 49 \nustar observations. Before merging the targets from both the BASS project and the Hyperleda database, we checked whether there are any systematic differences in the velocity dispersion estimates between the two samples.  We then identified the common sources, that is, sources with $\rm \sigma_{\ast}$ measurements in both the Hyperleda and the BASS databases and compare the corresponding $\rm \sigma_{\ast}$ values. There are 15 common sources in these samples. Figure\, \ref{hypervskoss} shows a plot of $\rm \sigma_{\ast, {\rm Hyperleda}}$ vs $\rm \sigma_{\ast, BASS}$. The plot clearly shows that the BASS and the Hyperleda $\rm \sigma_{\ast}$ measurements are very well correlated. There are no large amplitude or systematic deviations from the one-to-one relation and, therefore, we chose to merge the two samples.

%%%%%%% FIGURE 1

In the last step, we cross-correlated the data with \cite{ricci2017} in order to obtain information on the X-ray column density ($\rm N_H$) of each source. We excluded from the sample the Compton thick (CT) sources, namely, the sources with $\rm N_H$>$10^{24}$ $\rm cm^{-2}$. The final velocity dispersion sample (hereafter, the VD sample) is comprised of 84 AGN with reliable measurements of $\rm \sigma_{\ast}$ and 107 \nustar observations. The VD sample log is presented in Table \ref{vd_sample}.

For each source in the VD sample, we can estimate the BH mass using the \mbh\ -- $\rm \sigma_{\ast}$ relation of \citet{woo2013}:

\begin{equation}
{\rm
logM_{BH}=8.37+5.31 \cdot log \left(\frac{\sigma_\ast}{200~km~s^{-1}} \right)}
.\end{equation}

\noindent As shown by \citet{grier2013}, BH mass estimates using Eq.\,1 are fully compatible  with the masses obtained from the reverberation technique, when assuming a mean virial factor <{\rm f}> of 4.3. The mass estimates for the VD sample are also listed in Table \ref{vd_sample}.

%--------------------------------------
%%%%%%%%%% S 2.3
\subsection{The Compton-thick sample}\label{ctsample}
The sources that have been excluded from the VD sample, due to being CT in nature, make up our CT sample. There are in total 24 CT sources with 53 \nustar observations available. These sources are treated separately since their continuum emission, particularly in the 3-10 keV band is severely suppressed, due to heavy absorption, and their spectrum may be dominated by reflected emission \citep[e.g.][]{balok2014}. The log of the CT sample is presented in Table \ref{ct_sample}. 

%-----------------------------------------------------------------
%-----------------------------------------------------------------
%%%%%%%%%%%%%%%%%%% SECTION 3
\section{X-ray data reduction}\label{data_reduction}

The observations were processed with the \nustar data analysis pipeline
({\sc nupipeline}) in order to produce cleaned, calibrated event list files. 
For this standard pipeline processing, we used \nustar data analysis software 
{\sc NuSTARDAS} v2.0.0 and CALDB version 20170222\footnote{$\rm https://heasarc.gsfc.nasa.gov/docs/nustar/analysis/nustar\_swguide.pdf$}. 
From the cleaned event files, 
we extracted source and background light curves for each of the two \nustar focal 
plane modules (FPMA \& FPMB) using the {\sc nuproducts} script. 
We adopted a source radius of 60 arcsec for the source light curve extraction, for both FPMA \& FPMB observations and applied the live-time,  point spread function (PSF), and vignetting corrections. The background light curves are extracted from a four times larger (120 arcsec radius) source-free region of the image at an off-axis angle similar to the source position. We extracted light curves in the 3–10 and 10–20 keV bands,  using a bin size of 250 s. 

The light curves obtained for the FMPA and FMPB modules were first background subtracted and then summed using the {\sc LCMATH} tool within {\sc FTOOLS}. For the subtraction of the background from the source light curves, the proper scaling factors, accounting for the differences in extraction radius for the source and background light curves,  have been taken into account. The combined error in the count rate during subtraction or addition of the light curves is always calculated  as:  $err=\sqrt{err1^2  + err2^2,}$ where err1 and err2 are the  uncertainties in the count rate in each light curve bin obtained during the \nustar data reduction process.

%-----------------------------------------------------------------
%-----------------------------------------------------------------
%%%%%%%%%%%%%%%%% SECTION 4
\section{The normalised excess variance}\label{calc_nxv}

To measure the variability power of the sources in our sample we compute the  normalised excess variance \cite[e.g.][]{nandra1997} using:

\begin{equation}
{\rm
\sigma_{NXV}^{2}=\frac{1}{N\mu^2} \sum_{i=1}^{N} \left( X_i - \mu )^2 -\sigma_i^2 \right),}
\label{snvx}
\end{equation}

\noindent where $N$ is the number of bins in the light curve, $\rm X_i$ and $\rm \sigma_i$ are the count rates and their uncertainties, respectively, and $\mu$ is the unweighted  mean count rate.  

The $\sigma_{NXV}^{2}$ has been measured using light curve segments with a duration of $\rm \Delta t=10$, 20, 40, and 80 ks. When more than one valid segments are available for a source (for a given timescale, $\rm \Delta t$), we compute the mean of the individual excess variances. This approach should reduce the uncertainty due to the stochastic nature of the X-ray variations \citep[see e.g.][]{allevato2013}. Following \citet{ponti2012}, all time bins in the light curve segments with fractional exposure lower than 0.35 have been excluded. We note that due to the restriction of our analysis in the local universe (the median redshift of the sample equals 0.02 and the maximum redshift equals 0.07), the impact of the source redshift in the estimation of $\sigma^2_{NXV}$ in fixed length intervals is negligible and it is also very similar for all the  samples.   

%--------------------------------------------------
%%%%% TABLE 1

\begin{table*}
\caption{Best-fit results for the rev and VD $M_{BH}$ vs $\sigma_{NXV}^2$ relations} % title of Table
\label{fitting1}      % is used to refer this table in the text
\centering                          % used for centering table
\begin{tabular}{c c c c c c c c c }        % centered columns (4 columns)
\hline                % inserts double horizontal l
 & \multicolumn{4}{c}{3-10 keV} & \multicolumn{4}{c}{10-20 keV} \\
 & \multicolumn{2}{c}{Rev} & \multicolumn{2}{c}{VD} & \multicolumn{2}{c}{Rev}  & \multicolumn{2}{c}{VD} \\
\hline
segment & slope & intercept & slope & intercept & slope & intercept & slope & intercept\\    % table heading 
\hline \hline                       % inserts single horizontal line
10 ks &-0.76$\pm$0.12 & 6.78$\pm$0.07 & 0.09 $\pm$0.18  & 7.95 $\pm$0.13 & -0.91 $\pm$0.13 & 6.71$\pm$0.09 & 0.33$\pm$0.16 & 7.96$\pm$0.10 \\
20 ks &-0.91$\pm$0.09 & 7.01$\pm$0.06 &-0.05 $\pm$0.13  & 7.80 $\pm$0.11 & -0.96 $\pm$0.14 & 6.94$\pm$0.11 & 0.23$\pm$0.16 & 7.84$\pm$0.09 \\
40 ks &-0.78$\pm$0.15 & 7.19$\pm$0.08 &-0.40 $\pm$0.15  & 7.59 $\pm$0.10 & -1.16 $\pm$0.15 & 7.25$\pm$0.09 &-0.10$\pm$0.21 & 7.56$\pm$0.11 \\
80 ks &-0.90$\pm$0.09 & 7.36$\pm$0.07 &-1.00 $\pm$0.05  & 7.42 $\pm$0.07 & -1.13 $\pm$0.13 & 7.36$\pm$0.08 &-0.96$\pm$0.21 & 7.39$\pm$0.08 \\
\hline                                   %inserts single line
\end{tabular}
\end{table*}

%-----------------------------------------------------------------
%%%%%%%%%%% FIGURES 2 & 3
    \begin{figure*}
    \begin{center}
    \includegraphics[height=0.62\columnwidth]{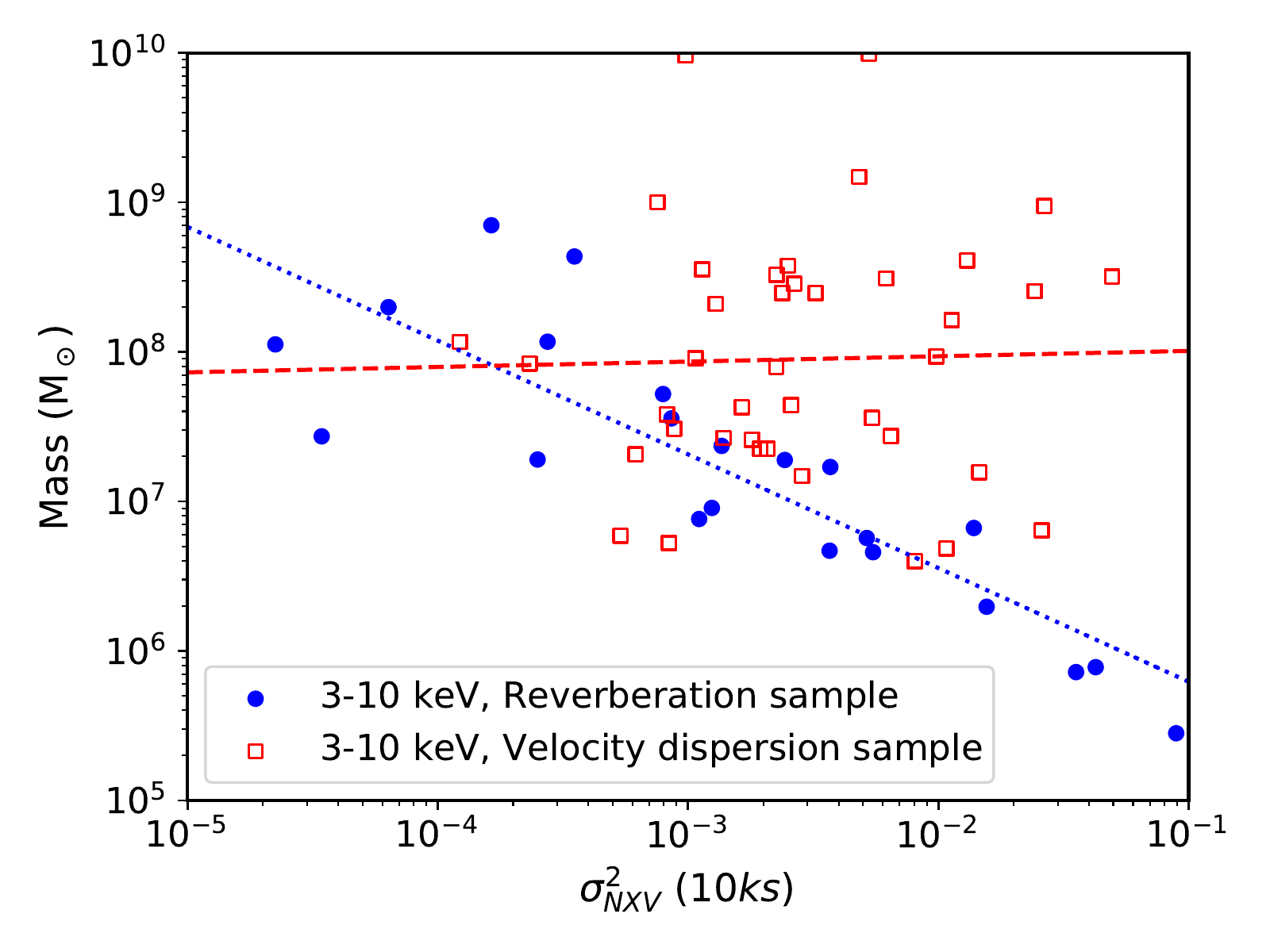}
    \includegraphics[height=0.62\columnwidth]{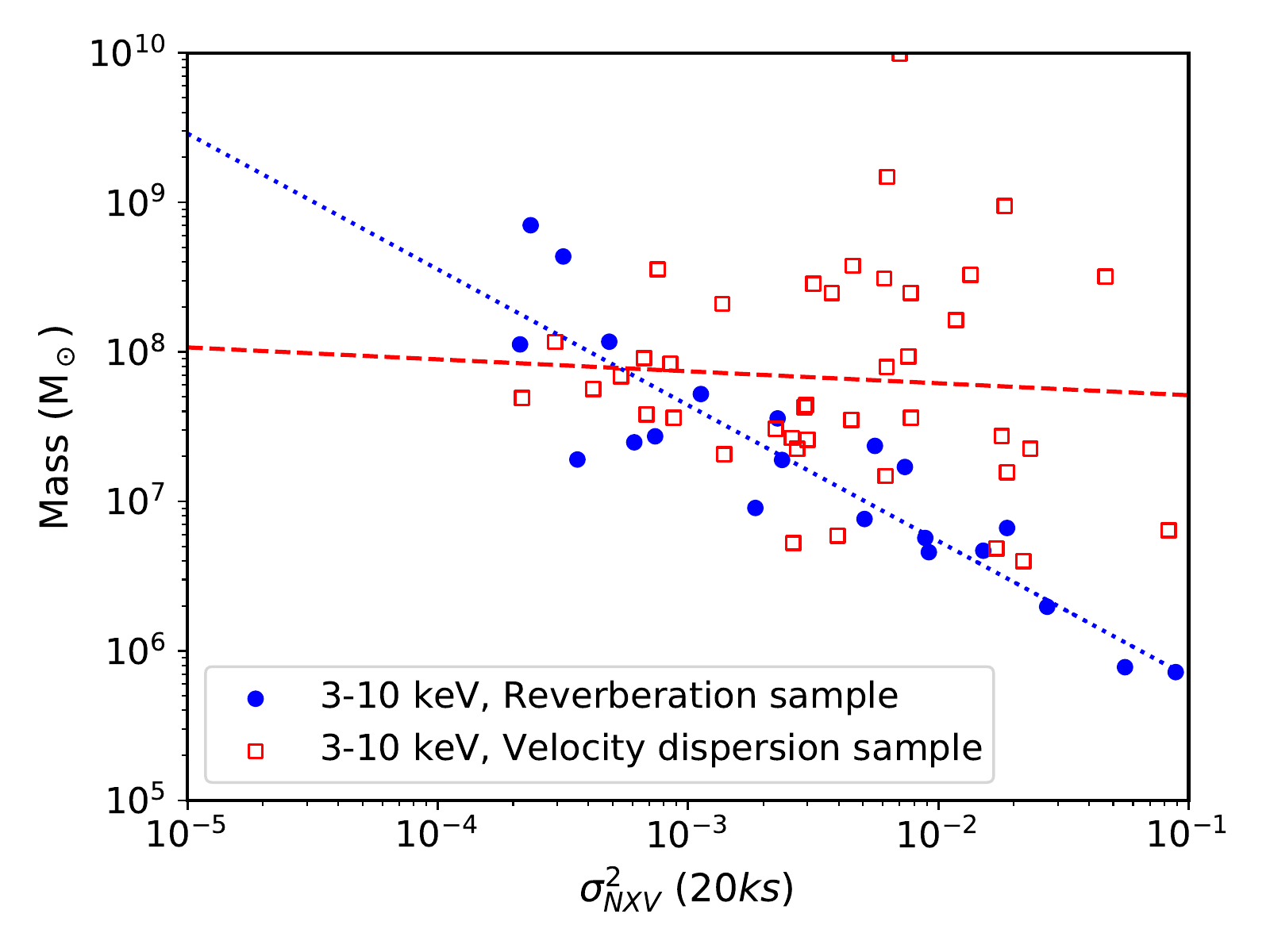}
    \includegraphics[height=0.62\columnwidth]{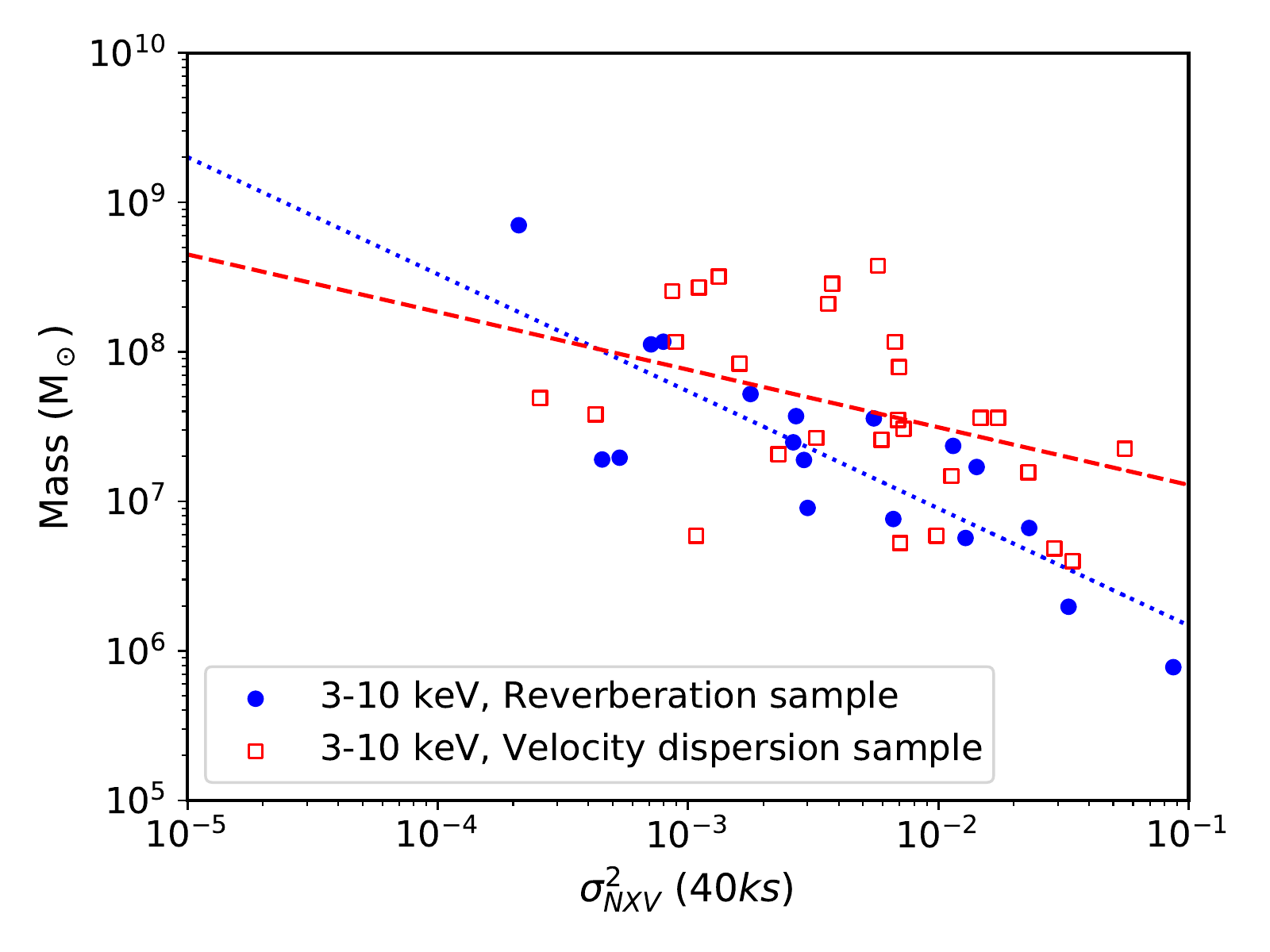}
    \includegraphics[height=0.62\columnwidth]{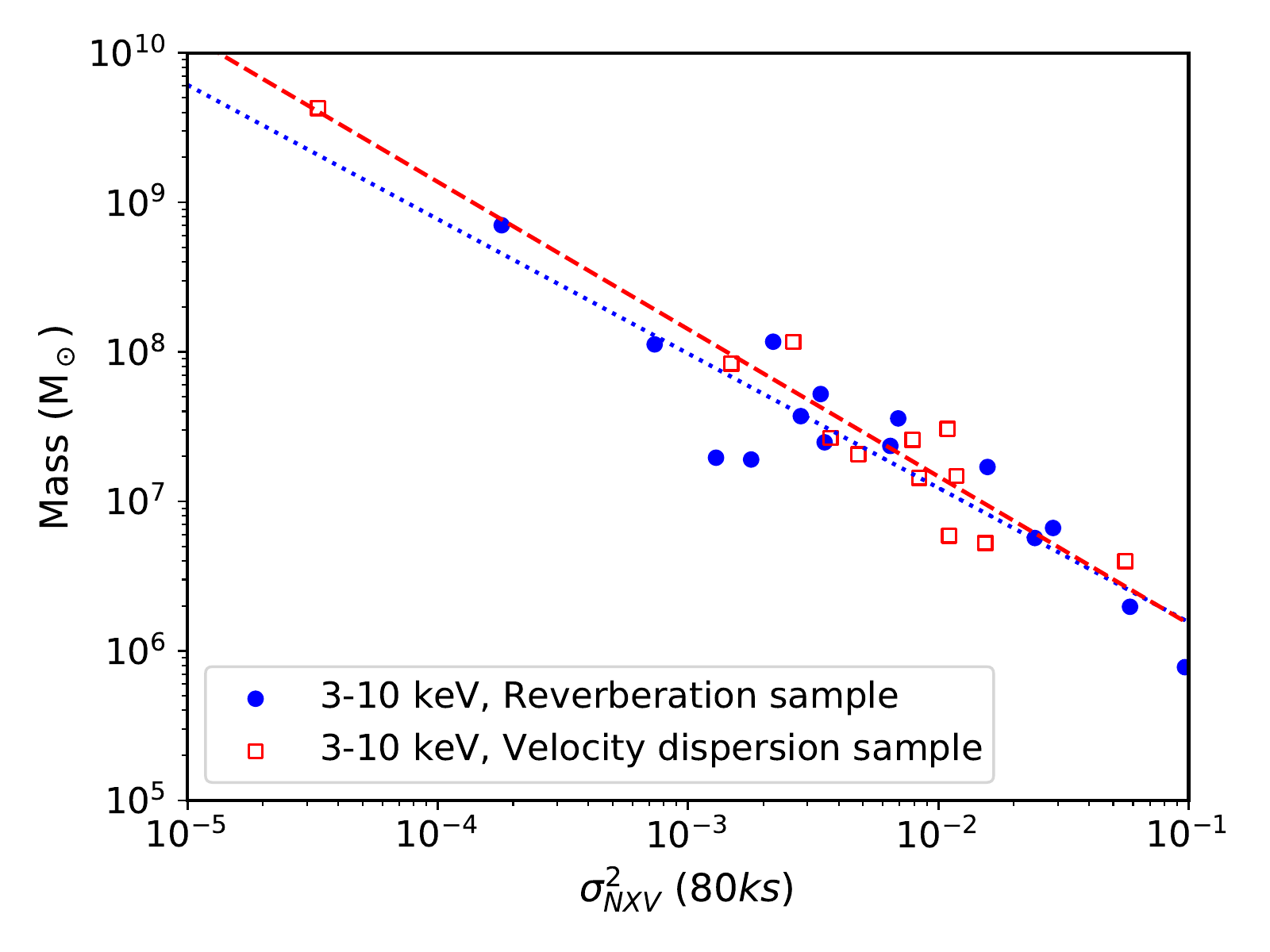}
    \end{center}
    \caption{\mbh\ vs {\snxv} in the 3--10 keV band for the rev sample (filled-blue circles) and the VD sample (open-red squares). The dotted blue line and the red dashed line show the best-fit results for the rev and the VD sample, respectively.}
    \label{comparison1}
    \end{figure*}

    \begin{figure*}
    \begin{center}
    \includegraphics[height=0.62\columnwidth]{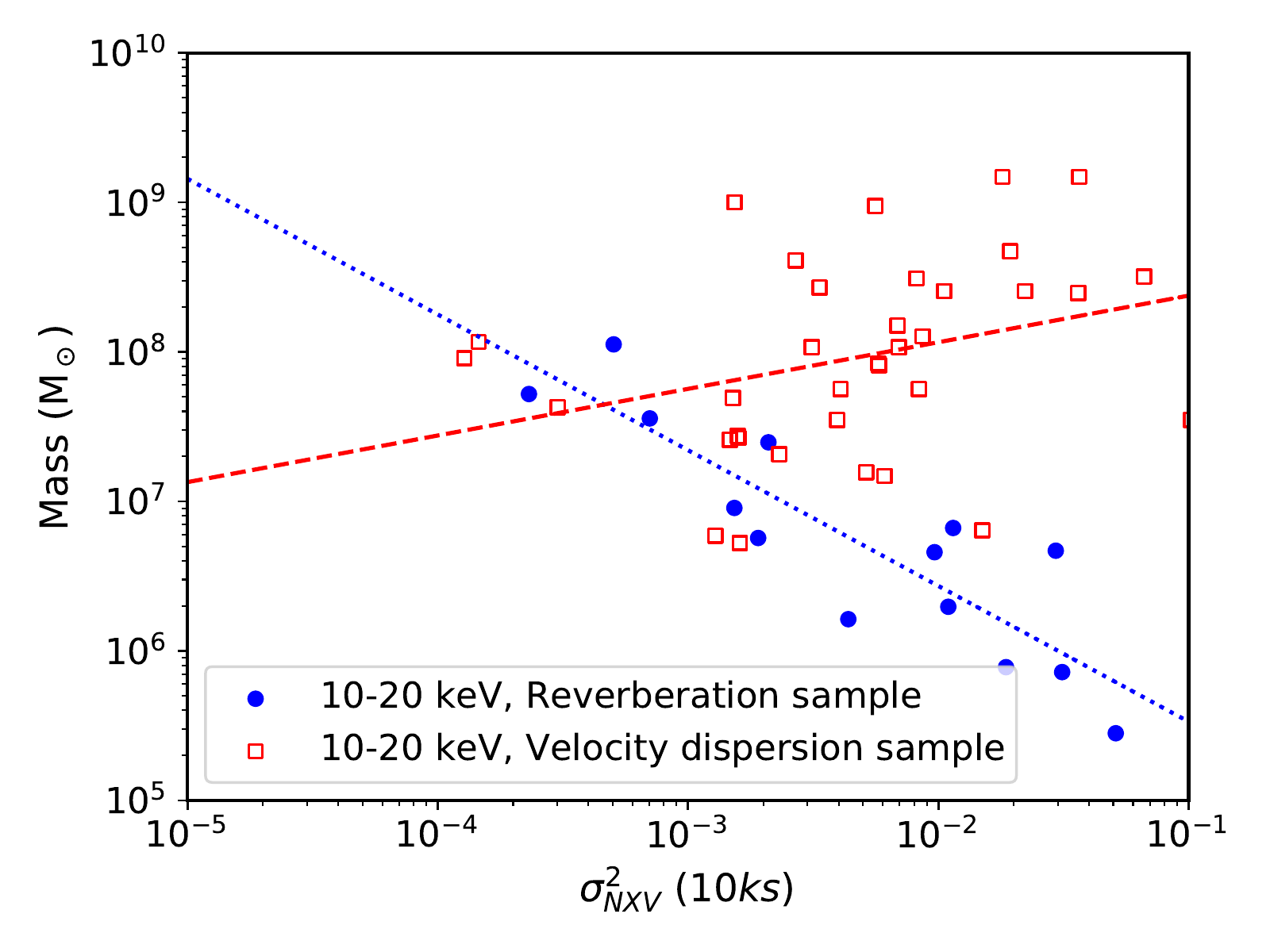}
    \includegraphics[height=0.62\columnwidth]{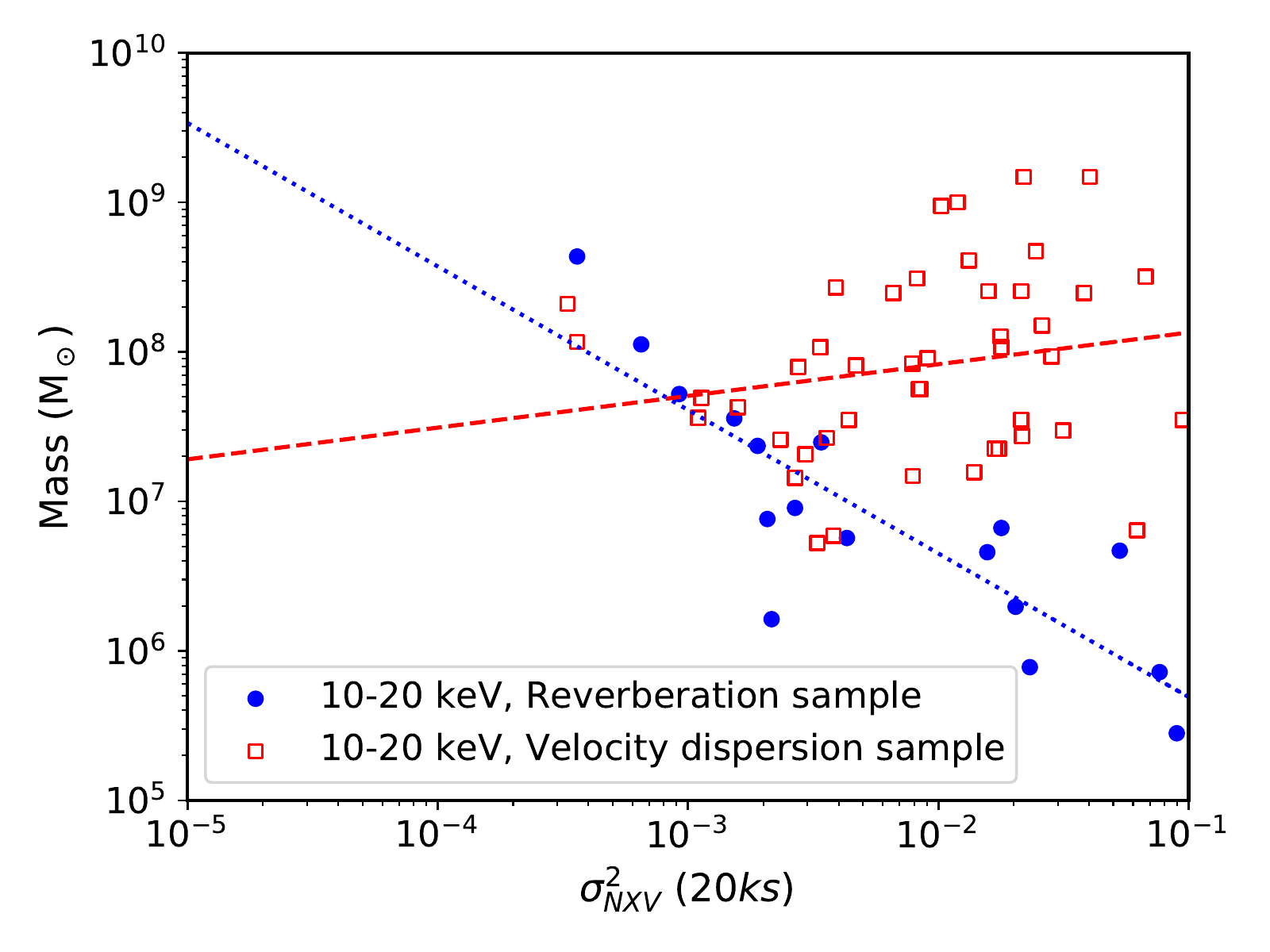}
    \includegraphics[height=0.62\columnwidth]{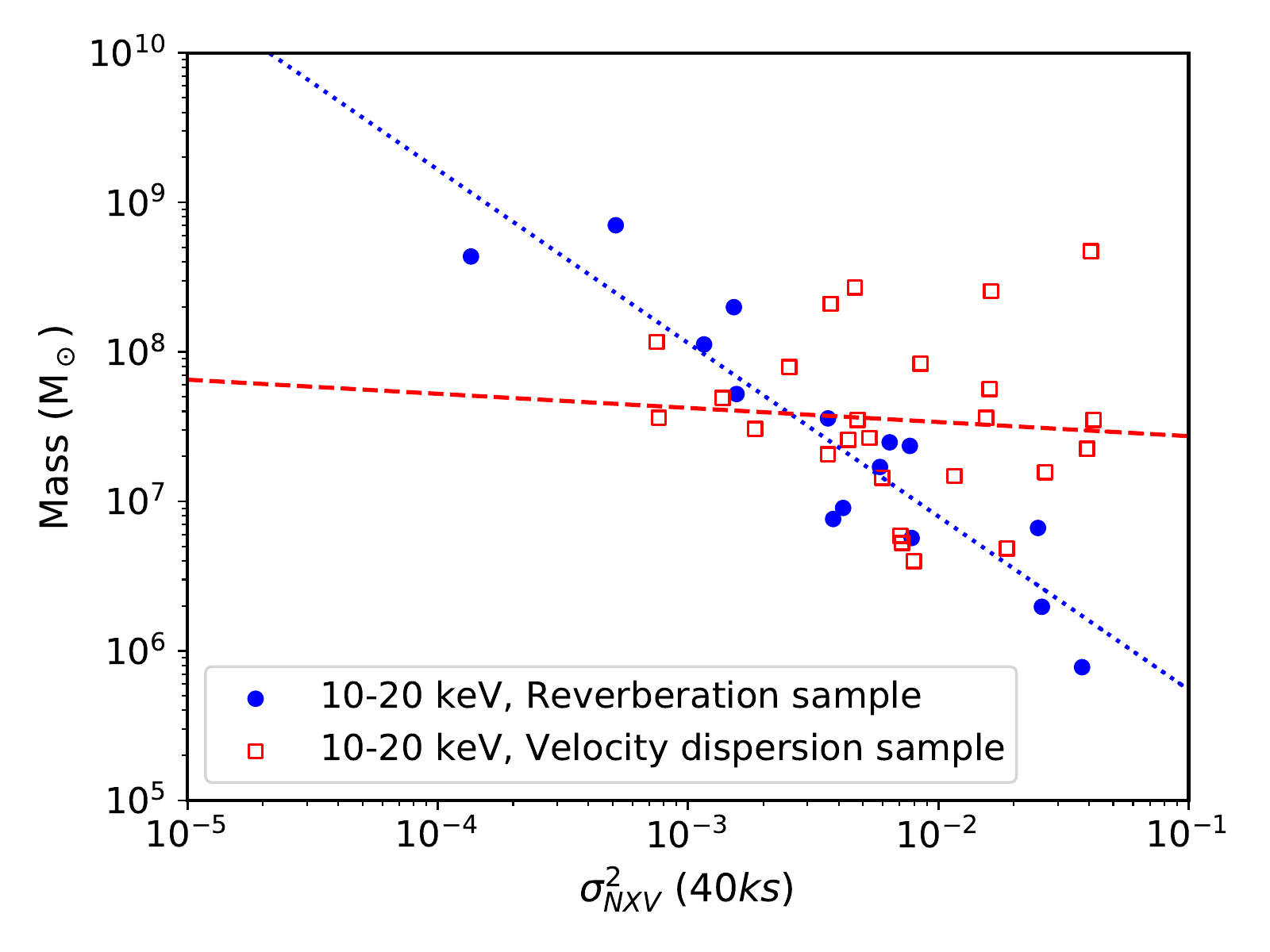}
    \includegraphics[height=0.62\columnwidth]{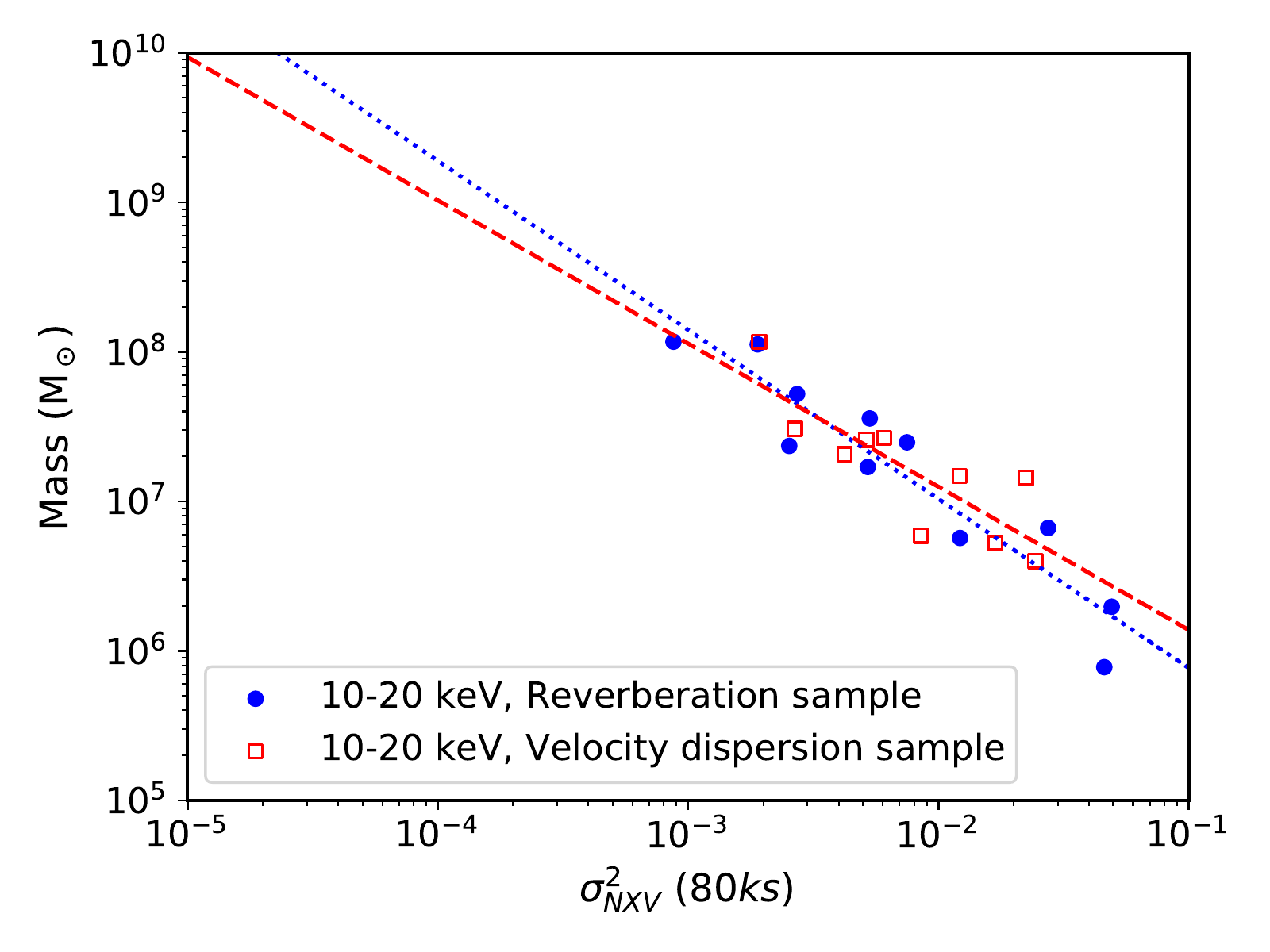}
    \end{center}
    \caption{\mbh\ vs {\snxv} in the 10--20 keV band for the rev sample (filled-blue circles) and the VD sample (open-red squares). The dotted blue line and the red dashed line show the best-fit results for the rev and the VD sample, respectively.}
    \label{comparison2}
    \end{figure*}

%-----------------------------------------------------------------
%-----------------------------------------------------------------
%%%%%%%%%%%%%%% SECTION 5
\section{Measuring \mbh\ using $\sigma^2_{NXV}$}

Previous studies have explored the dependence of the normalised excess variance on the BH mass, ultimately suggesting a linear correlation in the log-log space between the two quantities. 
We aim to establish a prescription to estimate BH masses given the measurements of the normalised  excess variance from X--ray data, therefore, we chose to establish the reverse relation, that is, the relation of {\mbh} versus {\snxv}. Our objective is to investigate what are the requirements for the light curves that will be used to compute {\snxv}, 
so that the resulting BH mass estimates will be as unbiased as possible and of known variance.

Instead of applying, a priory,  a certain cut to the  S/N value, we simply consider only the sources where the AGN variability is well detected, namely, the computed {\snxv} is positive (either the mean or a single value, if there is only one light curve segment available).  In this way, observations where the S/N ratio is low or the intrinsic amplitude is weak were excluded from further analysis. Then we carried out a separate study of the {\mbh} versus {\snxv} relation for the rev and VD samples in four timescales ($\rm \Delta t=10$, 20, 40, and 80 ks) and we investigate how the {\mbh} versus {\snxv} relation changes with respect to the total duration of the light curves used to compute \snxv. In this way (as we show in the following), we are able to establish the minimum requirements for the light curves (i.e. the necessary number of segments and  S/N values) in order to acquire a reliable \mbh\ estimate. 

%----------------------------------------------
%%%%%%% S 5.1
\subsection{\mbh\ -- \snxv\ relation for the rev and VD samples}\label{comparison_section}

The four panels of Fig. \ref{comparison1} show the {\mbh} vs {\snxv} plots for the rev and the VD samples (filled-blue circles and open-red squares respectively) in the 3--10 keV band, when using segments with durations of 10, 20, 40 \& 80 ks. The four panels of Fig. \ref{comparison2} present the same results in the 10--20 keV band.  

A strong anti-correlation between \mbh\ and \snxv\ is observed in all timescales for the rev sample. Smaller BH masses show the largest variability amplitude, as has been observed many times in the past. However, this is not the case for the VD sample, especially at timescales shorter than 40 ks (top panels in both figures). 

We fit a straight line to each plot, in log-log scale, of the following form,

\begin{equation}
{\rm
\log(M_{BH})= \alpha \cdot \log(\sigma_{NXV,0.005}^2) + \beta,}
\label{linefit}
\end{equation}

\noindent
to  model the observed behaviour. The \snxv\ values have been normalised to the value of 0.005 in order to minimise the error
on the line parameters, $\alpha$ and $\beta$. Since our objective is to predict {\mbh} from {\snxv} we used the ordinary least-squares 
regression of Y on X, OLS(Y|X), where Y is the variable to be predicted (i.e.\, $\rm M_{BH}$) and X is the measured variable 
($\rm \sigma_{NXV}^2$), following the prescription of \citet{isobe1990}. 

Table \ref{fitting1} lists the best-fit results and the corresponding 1$\rm \sigma$ uncertainties to the {\mbh}--{\snxv} plots, both for the rev and the VD samples, at all timescales, in both bands. The best-fit slope for the rev sample remains roughly constant, within the errors, in both bands, while the intercept constantly increases as we move to higher timescales. This is fully consistent with the expectations, as the variance provides a measure of the (intrinsic) PSD integral from the shortest to the longest sampled timescale. In the case of red-noise PSDs, this integral should increase with increasing segment duration, hence, the positive correlation of the best-fit line normalisation with the segment duration ($\rm \Delta t$).  

The VD data points shown in Figs.\,\ref{comparison1} and \ref{comparison2} and the best-fit results show that the situation is more complicated in the VD sample. 
There is almost a complete lack of correlation between {\mbh} and {\snxv} on the shortest timescales ($\rm \Delta t=10$ \& 20 ks), in both energy bands. The situation improves progressively as we move to longer timescales where the intrinsic variability is higher. A clear and strong anti-correlation between \mbh\ and \snxv\  is present only in the $\rm \Delta t=80$ ks plots. In fact, the best-fit results to the rev and VD plots are remarkably similar in the case of $\rm \Delta t=80$ ks.

The inconsistency of the {\mbh} versus {\snxv} relations between the rev and the VD samples at $\rm \Delta t\lesssim 40$ ks could not be attributed to any differences in the accuracy of the BH mass measurements between the two cases. After all, the rev BH mass measurements are based on relations that are 'forced' to agree with the $\rm M_{BH}-\sigma_{\ast}$ relation. Therefore, the  difference between the {\mbh} versus {\snxv} plots for the rev and VD samples when the light curve segments are short is expected to be related to limitations in accurately measuring the X-ray variability of the sources in the VD sample. In the following section, we quantify this limitation and present a prescription for the use of $\rm \sigma_{NXV}^2$ in the BH mass determination.  

%%%%%%% Figure 4.
\begin{figure*}
    \begin{center}
    \includegraphics[height=0.62\columnwidth]{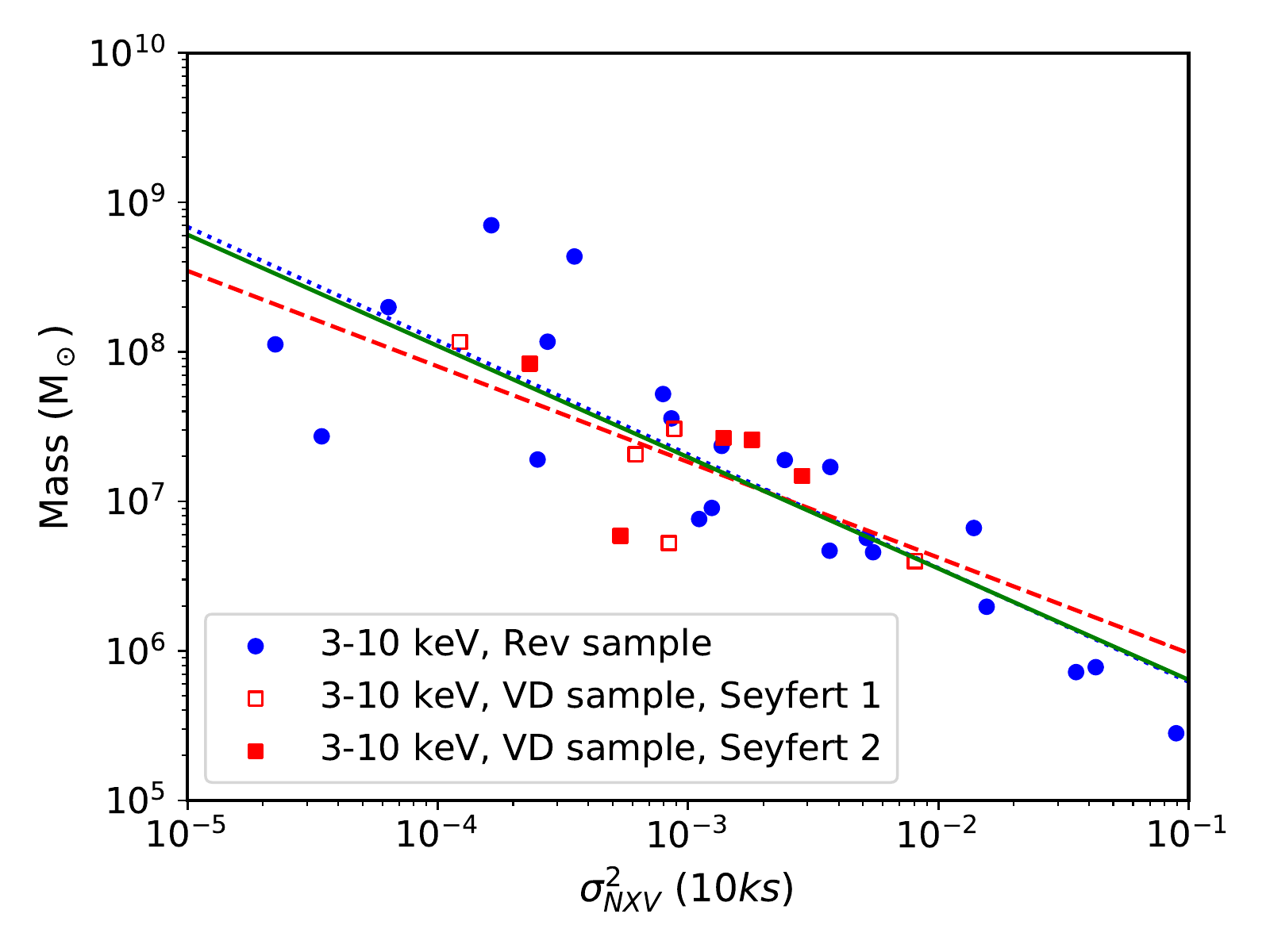}
    \includegraphics[height=0.62\columnwidth]{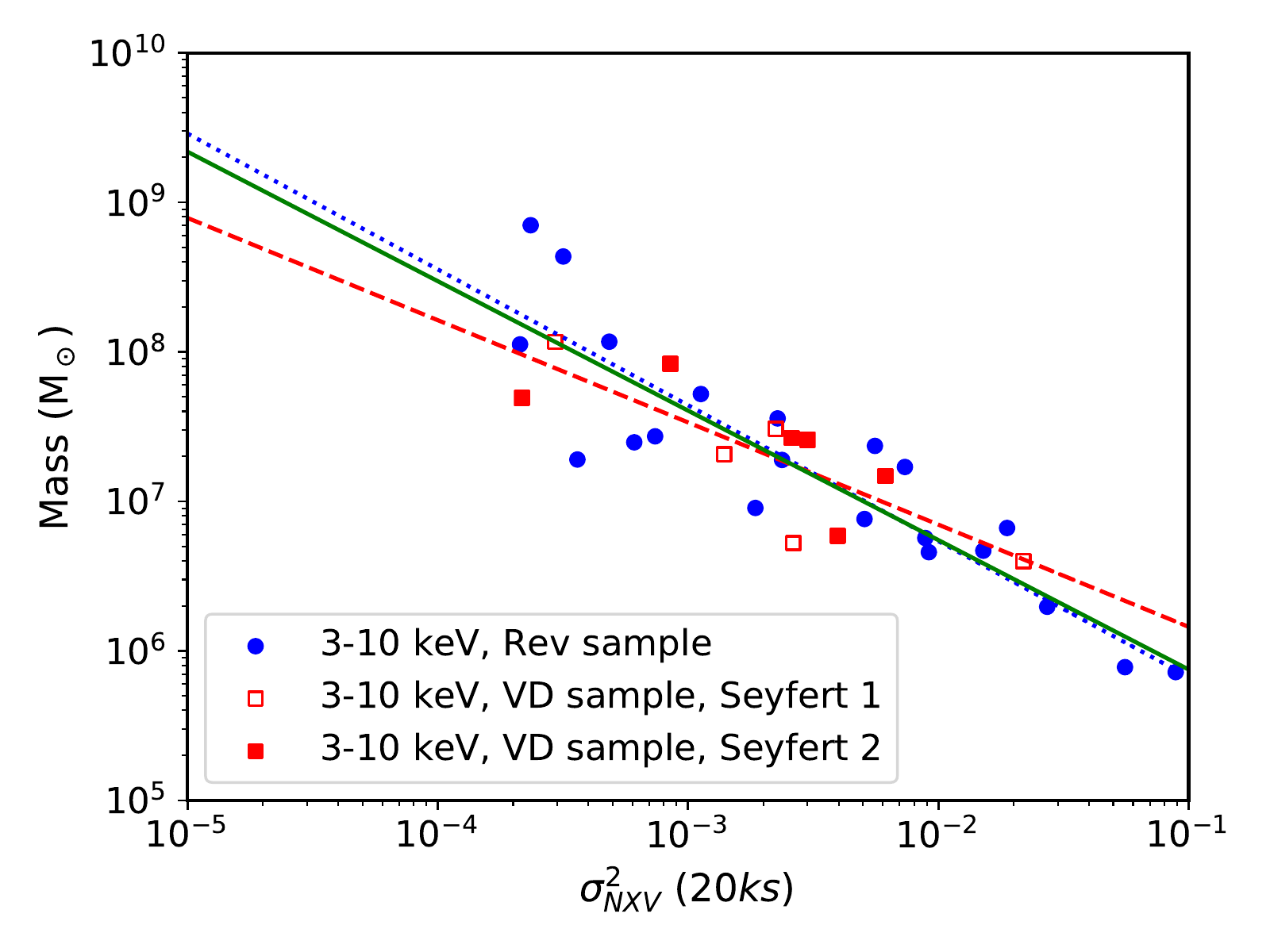}
    \includegraphics[height=0.62\columnwidth]{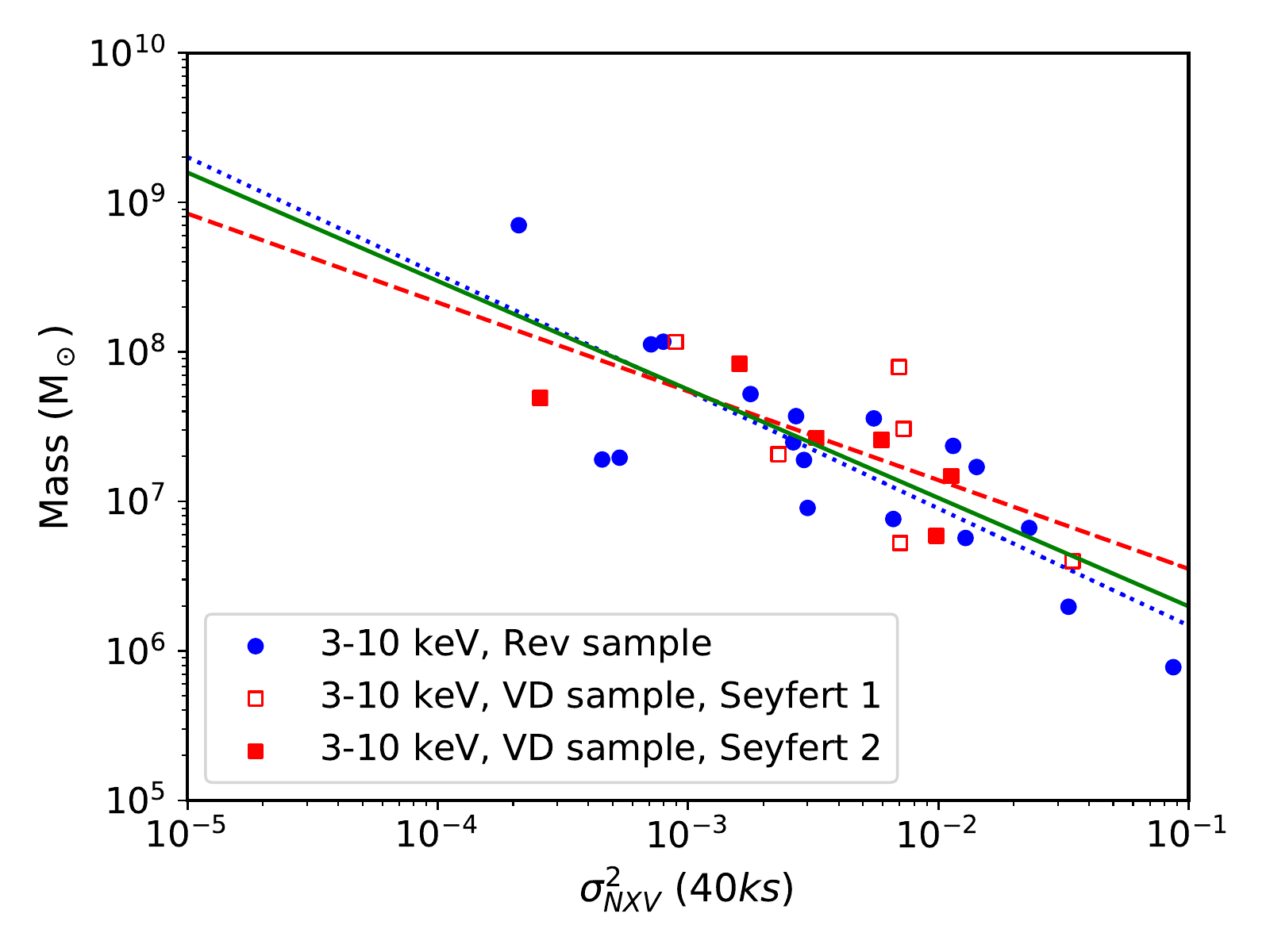}
    \includegraphics[height=0.62\columnwidth]{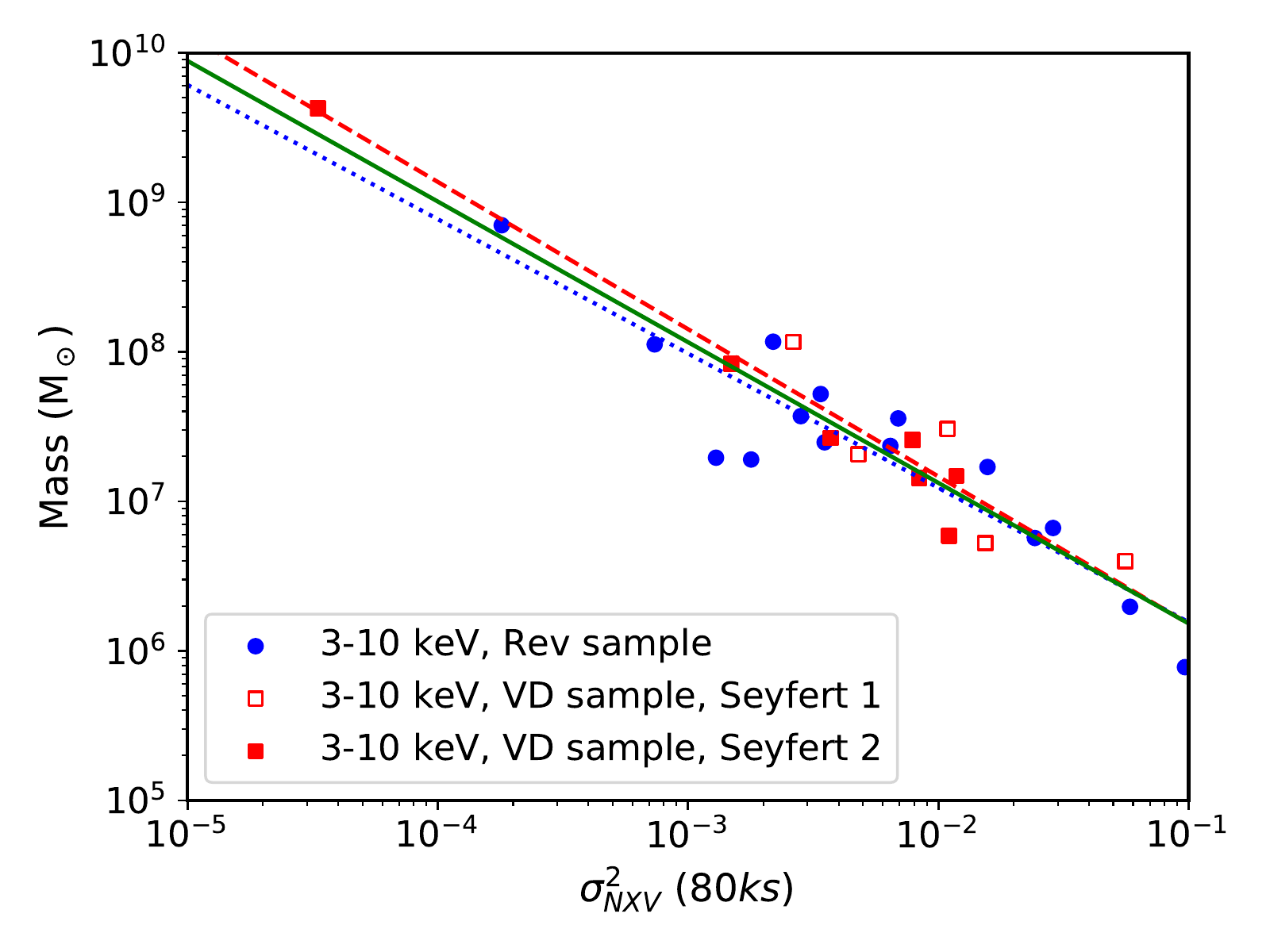}
    \end{center}
\caption{$\rm M_{BH}$ versus $\rm \sigma_{NXV}^2$, in the 3--10 keV band, for the 
final rev and VD sub-samples (chosen as explained in \S  \ref{choosesamples}).
{Filled-blue circles indicate the rev sample data (Type I AGN) while open-red  and filled-red squares indicate Type I and Type II AGN from the VD sample, respectively. The blue-dotted and the red-dashed lines show the best-fit lines to the rev and VD (Type I \& Type II) data respectively. The solid green line shows the best fit relation using all the data in each panel.}}
    \label{merged1}
    \end{figure*}

    \begin{figure*}
    \begin{center}
    \includegraphics[height=0.62\columnwidth]{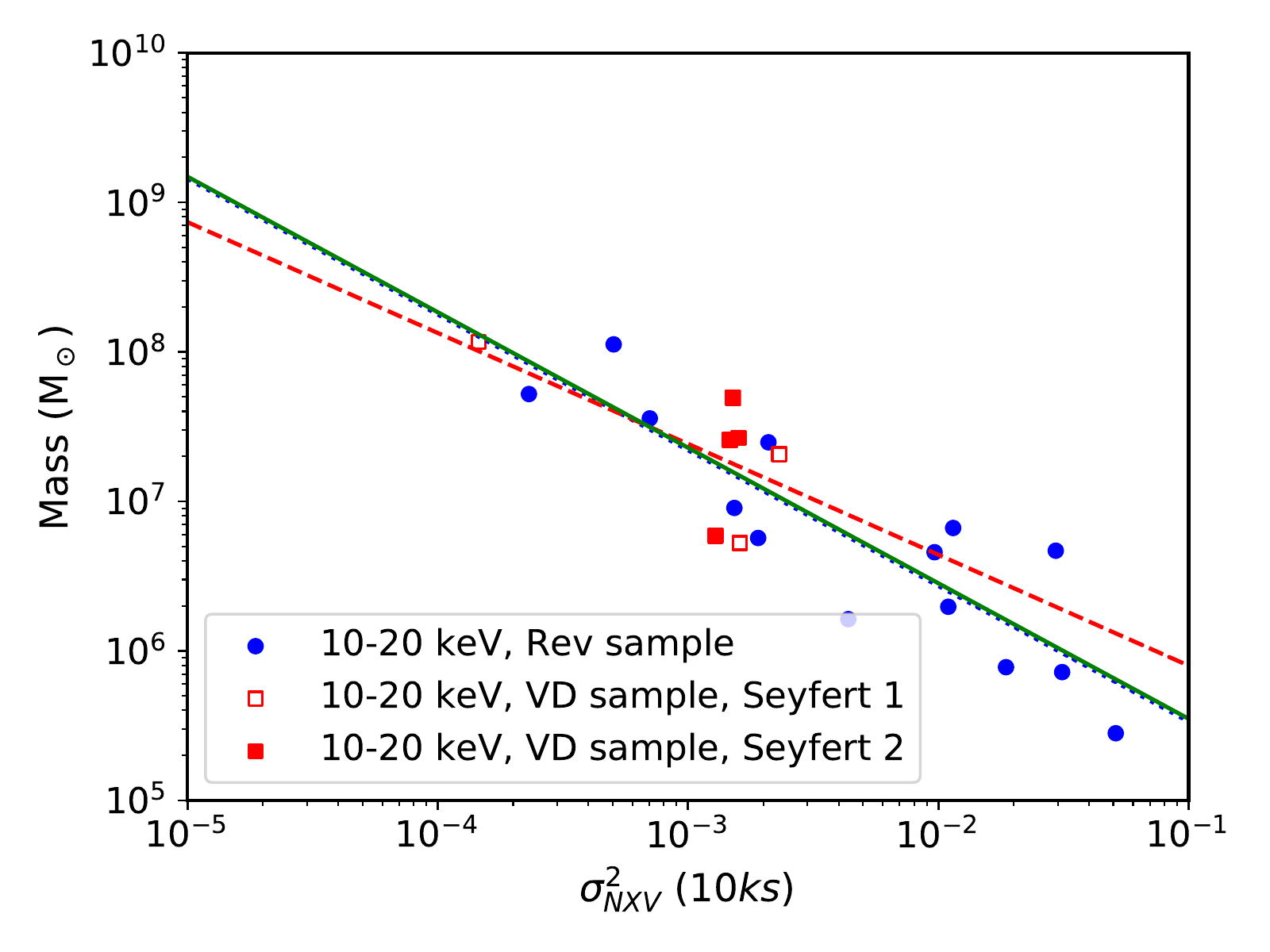}
    \includegraphics[height=0.62\columnwidth]{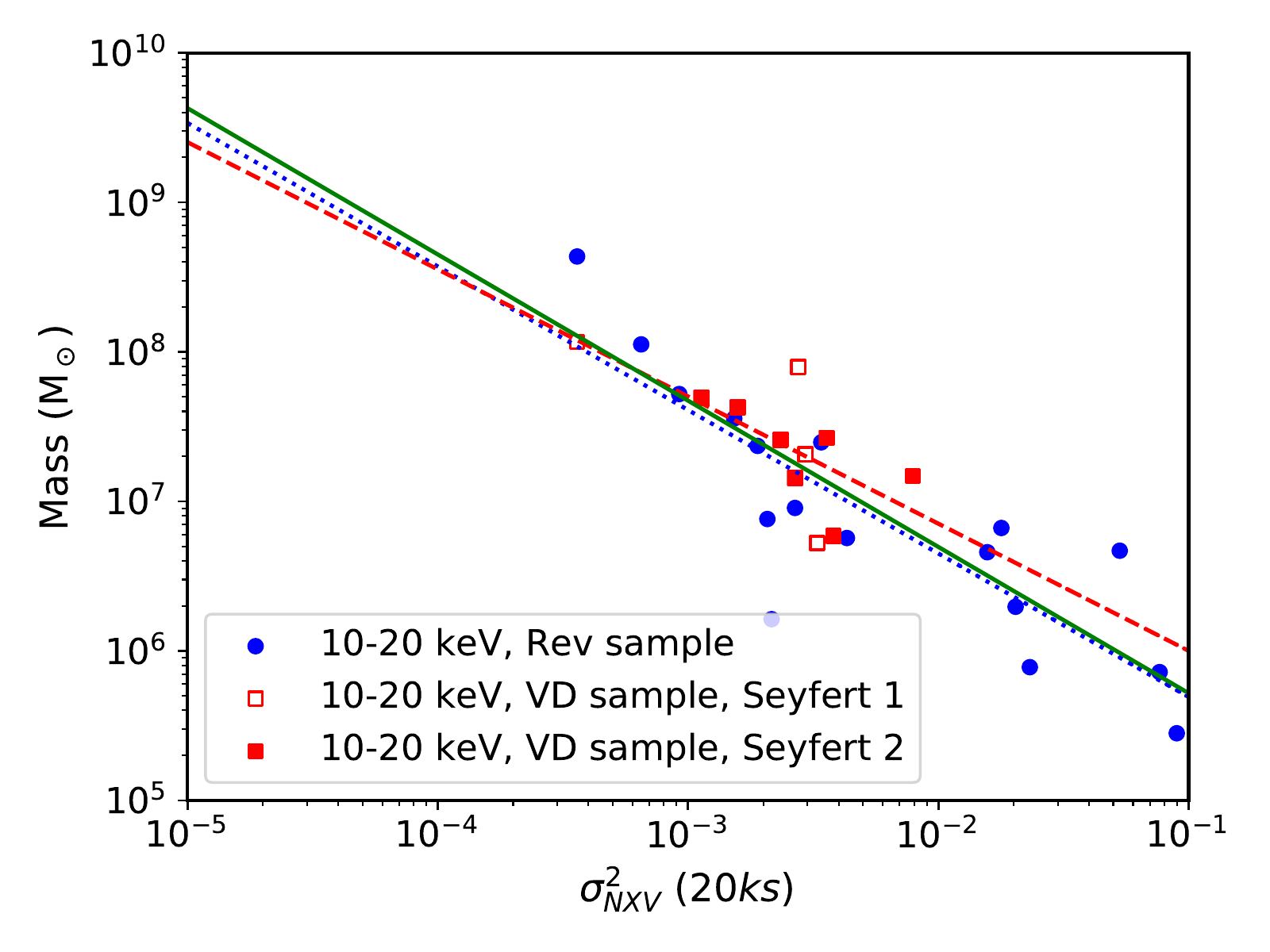}
    \includegraphics[height=0.62\columnwidth]{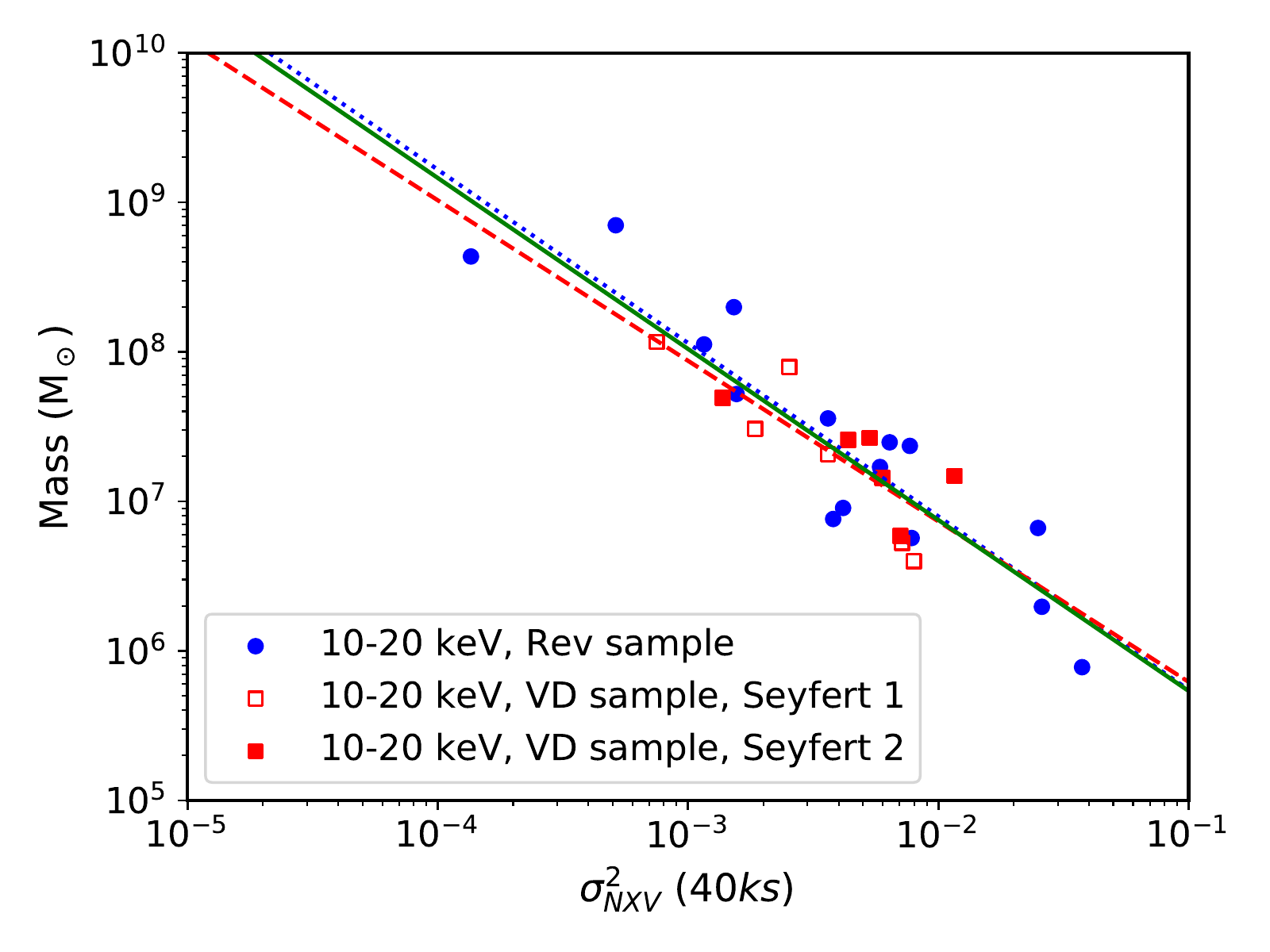}
    \includegraphics[height=0.62\columnwidth]{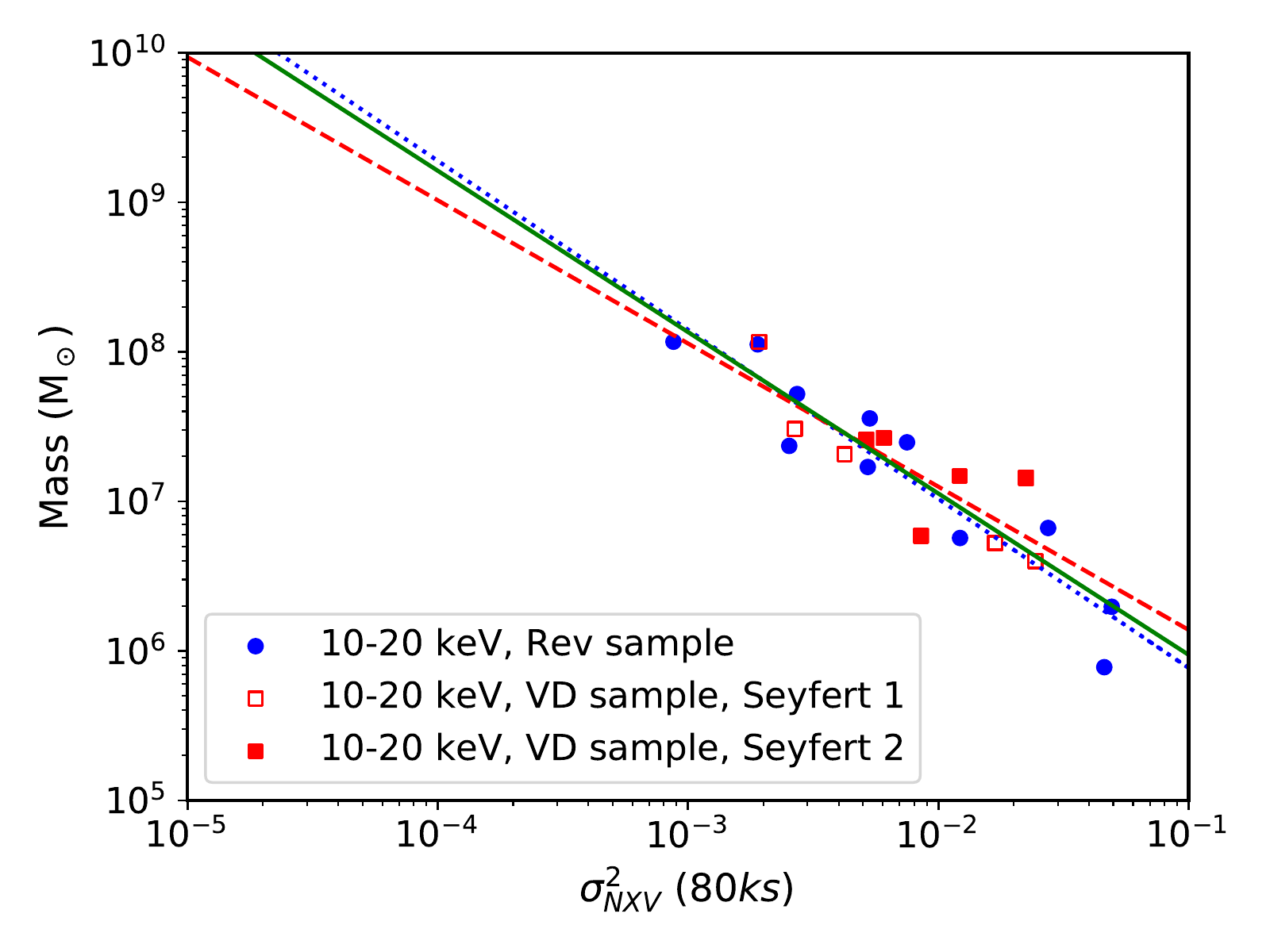}
    \end{center}
\caption{$\rm M_{BH}$ versus $\rm \sigma_{NXV}^2$, in the 10--20 keV band, for the final rev and VD sub-samples (chosen as explained in \S\ref{choosesamples}).
{Filled-blue circles indicate the rev sample data (Type I AGN) while open-red  and filled-red squares indicate Type I and Type II AGN from the VD sample, respectively. The blue-dotted and the red-dashed lines show the best-fit lines to the rev and VD (Type I \& Type II) data respectively. The solid green line shows the best fit relation using all the data in each panel.}}
    \label{merged2}
    \end{figure*}
    
%----------------------------------------------
%%%%%%% S 5.2

\subsection{Significance of the observation duration}\label{criteria}\label{choosesamples}

The main difference between the rev and VD samples is the average number of observations per target and their duration. There are, on average, 3.3 observations per source in the rev sample, with a mean duration of 80 ks each. On the contrary, there are only 1.3 observations per target in the VD sample, with a mean duration of 60 ks each. In order to quantify the impact of the available exposure time of a source in the estimation of the $\sigma_{NXV}^2$, we performed the following experiment.

For each timescale, we progressively increased  the minimum number of light curve segments required in order to keep a target in the sample. Effectively, this leads to smaller and smaller sub-samples, with an increased average duration for the observations. 
For example, when we request a minimum of at least two time segments of 20 ks for the estimation of \snxv, then only sources with one observation of at least $\rm \Delta t=40$ ks or two observations of at least 20 ks each are considered. When, for instance, we increase the minimum number of 20 ks segments to five, then it is only those sources with at least a $\rm \Delta t=100$ ks observation or five observations of 
20 ks each are considered. We repeated this exercise, namely, increasing the minimum number of segments for both the rev and the VD samples, until there were less than six sources left in the sample and we stopped. 

For each new sub-sample, we again fit  a straight line to obtain the best fit slope and intercept of the $\rm M_{BH}$ versus $\sigma_{NXV}^2$ relation. 
Figures \ref{duration1} and \ref{duration2} show the resulted best-fit parameters plotted as a function of the average duration of the observations for each sub-sample. 
The numbers next to each point indicate the number of sources left in each sub-sample after increasing the minimum number of segments.

The best-fit parameters remain roughly constant for the rev sample, irrespective of the number of segments required to compute \snxv. An increase in that number does not affect the fitting results. However, this is not the case with the VD sample. The VD sample includes many sources with observations with a significantly shorter average duration than the average duration of the rev sample.  The best-fit line parameters change 
as we increase the number of segments we use to compute \snxv. When the average duration of the observations in the VD sub-sample is similar to that of the rev sample, then the best-fit slopes and intercepts become consistent. 

We also note, as explained in \citet{isobe1990}, that the best-fit parameter errors may be underestimated for the smaller-sized samples (i.e.\ for the best-fit parameters determined by fitting the objects with the longest observations). This actually makes the agreement between the rev and VD best-fit parameters even more impressive. It also shows that the systematic bias present in the best-fit parameters when we consider all the VD sample sources is much greater than the statistical uncertainty of the best-fit parameters. 

The black circles in Figs.\, \ref{duration1} and \ref{duration2} indicate the first VD and rev sub-samples (i.e. those with a shorter average duration and, therefore, a larger size) for which the best-fit line slope and intercept are consistent (within errors). For all subsequent sub-samples, the best-fit results remain roughly constant. 

Our results indicate that an average of $\sim250$ ks in the duration of the observations per source is necessary for the best-fit results, from both samples, to agree when using 10 ks segments. This number decreases to $\sim180-200$ ks when the estimation of the $\sigma_{NXV}^2$ is based on 40 ks and 80 ks segments. This is probably due to the fact that as the segment duration increases, the intrinsic variance increases as well; therefore, it is easier (for a given S/N) to measure it accurately using slightly shorter average duration of observations. The results are similar for both the 3-10 and the 10-20 keV bands.

%%%%%%% SECTION 6
\section{A prescription for measuring \mbh\ in Seyfert galaxies.}

Figures \ref{merged1} and \ref{merged2} show the {\mbh} versus {\snxv} data for the largest rev and VD samples, where the best-fit line slope and intercept are consistent with each other (see the encircled points in Figs. \ref{duration1} and \ref{duration2}). 
The blue-dotted and red-dashed lines indicate these best-fit models to the rev and VD samples, respectively (for the same $\rm \Delta t$ value and energy band). 

Our rev sample is comprised of Type I AGN only, since the mass estimation is obtained by the reverberation method. On the other hand, the VD sample, where the mass is estimated using stellar velocity dispersion measurements from high-quality spectra,  contains both Type I and Type II AGNs. We note, however, that we do not expect any systematic difference in the {\mbh} -- {\snxv} relation for Type I and Type II AGNs. For example, \citet{rani2017} presented their results on the flux variability on hourly timescales for a large sample of AGN using {\nustar} and suggested no difference in the variability behaviour between Seyfert 1 and 2 galaxies in all X-ray bands. 

To further explore this notion, in Figures \ref{merged1} and \ref{merged2}, we plot  Type I AGN separately from the rev sample (filled-blue circles), Type I AGN from the VD sample (open-red squares), and Type II AGN from the VD sample (filled-red squares). The small number statistics in the individual  VD sub-samples  does not allow for a separate line fitting analysis. However, the plots clearly show that there are no systematic differences or  inconsistencies between the different populations. We therefore chose to combine the samples and fit all the data simultaneously.

The solid green  lines in Figs.\, \ref{merged1} and \ref{merged2} show the best-fit lines obtained from the combined (rev+VD) sample. The best-fit results are listed in Table \ref{bestsample} (columns 5 and 6 for the 3--10 keV band, columns 11 and 12  for the 10--20 keV band). 
We propose for these best fit-line models to be used to measure the BH mass in Seyfert galaxies. 

Columns 2 and 8 in the same table present a list of the total number of sources in the combined (rev+VD) sample for each band. The samples are smaller in the 10--20 keV band, mainly because the S/N of the light curves in this band is significantly smaller; hence, there are fewer objects with positive excess variance measurements. Columns 3 and 9 list the average duration of the observations ($\rm \Delta t_a$) as well as the average ($N_A$) and the minimum ($N_m$) number of segments for the the objects in each sample, rounded to the lower integer. Here, 
$N_{m}$ corresponds to the number of segments of the target with the shortest duration for the observations in the combined sample. Columns 4 and 10  list 
the average and the minimum S/N of the light curves that we used to compute the excess variance in the combined sample. Lastly, columns 7 and 13 list the average scatter of the points around the best-fit line. 

%%%%%%%   TABLE 2
\begin{table*}
\setlength{\tabcolsep}{4pt}
\caption{Best-fit results of the $\rm M_{BH}$ vs $\rm \sigma_{NXV}^2$ relation for the combined rev and VD samples.} 
\label{bestsample}      
\centering                         
\begin{tabular}{c c c c c c c c c c c c c}        
\hline                
& \multicolumn{6}{c}{Rev+VD (3--10 keV)} & \multicolumn{6}{c}{Rev+VD(10--20 keV)}  \\
\hline
$\rm \Delta t^1$ & 
N$_{\rm s}^2$ & 
$\rm \Delta t$$_{\rm a}$/N$_{\rm a}$/N$_{\rm m}^3$ & 
$\rm S/N_{a/m}^4$ & 
${\rm \alpha}^5$ & $\beta^6$ & 
$\sigma_{sc}^7$ & 
N$_{\rm src}$ & 
$\rm \Delta t$$_{\rm a}$/N$_{\rm a}$/N$_{\rm m}$ & 
$\rm S/N_{a/m}$ & ${\rm \alpha}$ & ${\rm \beta}$ & $\sigma_{sc}$  \\   
\hline \hline                     

10 & 32 & 240/24/10 & 13.6/4.1 & -0.74$\pm$0.10  &  6.78$\pm$0.06 & 0.42 & 21 & 250/25/11 & 8.5/2.5 & -0.91$\pm$0.11 & 6.73$\pm$0.09 & 0.35 \\
20 & 33 & 230/11/5  & 13.3/4.2 & -0.87$\pm$0.08  &  7.00$\pm$0.05 & 0.32 & 28 & 210/10/4  & 8.5/2.5 & -0.98$\pm$0.11 & 6.99$\pm$0.07 & 0.38 \\
40 & 30 & 230/5/2   & 13.6/4.7 & -0.73$\pm$0.13  &  7.24$\pm$0.06 & 0.36 & 27 & 230/5/2   & 8.3/2.9 & -1.14$\pm$0.12 & 7.22$\pm$0.06 & 0.29 \\
80 & 27 & 200/2/1   & 13.7/3.3 & -0.94$\pm$0.06  &  7.41$\pm$0.05 & 0.26 & 19 & 200/2/1   & 8.9/2.9 & -1.08$\pm$0.12 & 7.38$\pm$0.05 & 0.23 \\

\hline                                   %inserts single line

\multicolumn{13}{l}{\small $^1$ Segment duration in ks.} \\
\multicolumn{13}{l}{\small $^2$ Number of sources in the combined rev+VD samples.} \\
\multicolumn{13}{l}{\small $^3$ Average duration in ks/average number of segments/minimum number of segments} \\
\multicolumn{13}{l}{\small $^4$ Average signal-to-noise ratio /minimum signal-to-noise ratio} \\
\multicolumn{13}{l}{\small  $^5$ Best-fit slope, $^6$ best-fit intercept.$^7$ Average scatter of the points around the best fit.} \\

\end{tabular}
\end{table*}

%%% SECTION 6.1
\subsection{ The prescription}
\label{prescription}

Based on the results presented in the previous section, we propose the following prescription for the measurement of the BH mass in AGN excess variance measurements.

First, the light curve(s) should be divided into a number of segments, the average \snxv\ should be computed, and then Equation \ref{linefit} can be used to derive \mbh, with $\alpha$ and $\beta$ listed in Table \ref{bestsample}, depending on the duration of the segment that was chosen and the energy band. 

Second, the number of light curve segments that will be used to compute the excess variance should ideally be comparable to the average number of segments listed in Table \ref{bestsample} and (at least) larger than the minimum number of segments listed in the same table. Since the scatter of the points around the best-fit lines decreases with increasing  segment duration, we propose the use of the longest possible segment to estimate the BH mass. 

Finally, the average S/N of the light curves should be comparable to the average S/N and (at least) greater than the minimum S/N listed in Table \ref{bestsample} (see columns 4 and 10 in Table \ref{bestsample}). The smallest S/N values suggest that S/N ratios at least equal to $\sim$3 are necessary to achieve a reliable estimate of the BH mass. 

%%%%%%% The CT sources
\subsection{Compton-thick sources}

We used the prescription mentioned above to measure the BH mass of the CT sources. Therefore, we  apply the same minimum duration and S/N criteria for the measurement of $\sigma_{NXV}^2$ presented above for the  sources in the CT sample. Due to the small size of the CT sample and the reduced photon statistics due to the absorption, we were able to measure $\sigma_{NXV}^2$ in only four sources when using the 80 ks segments. The number of sources with positive \snxv\ measurements on shorter timescales is very small (only one or two). 

\begin{figure}
    \begin{center}
    \includegraphics[height=0.62\columnwidth]{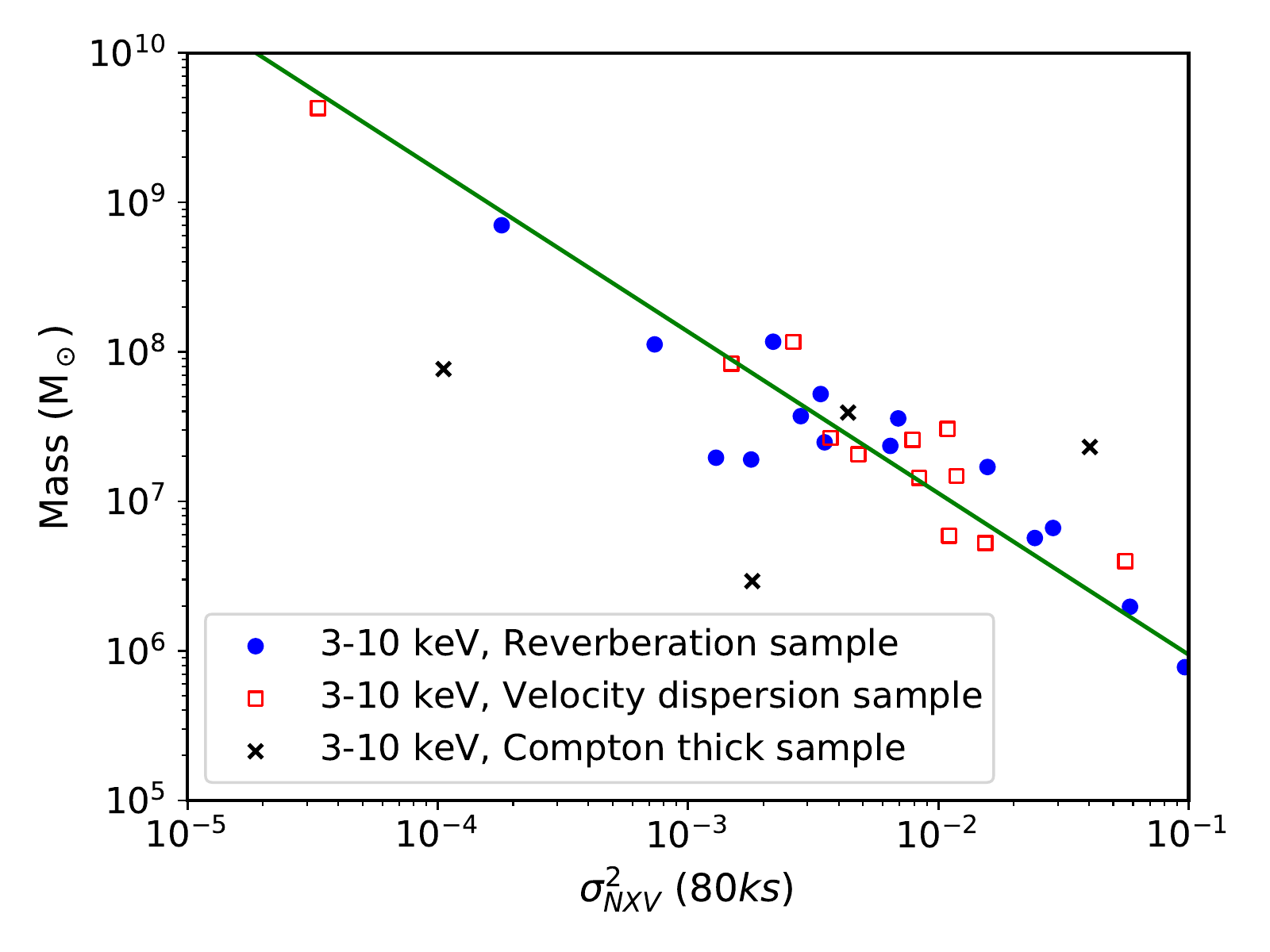}
    \includegraphics[height=0.62\columnwidth]{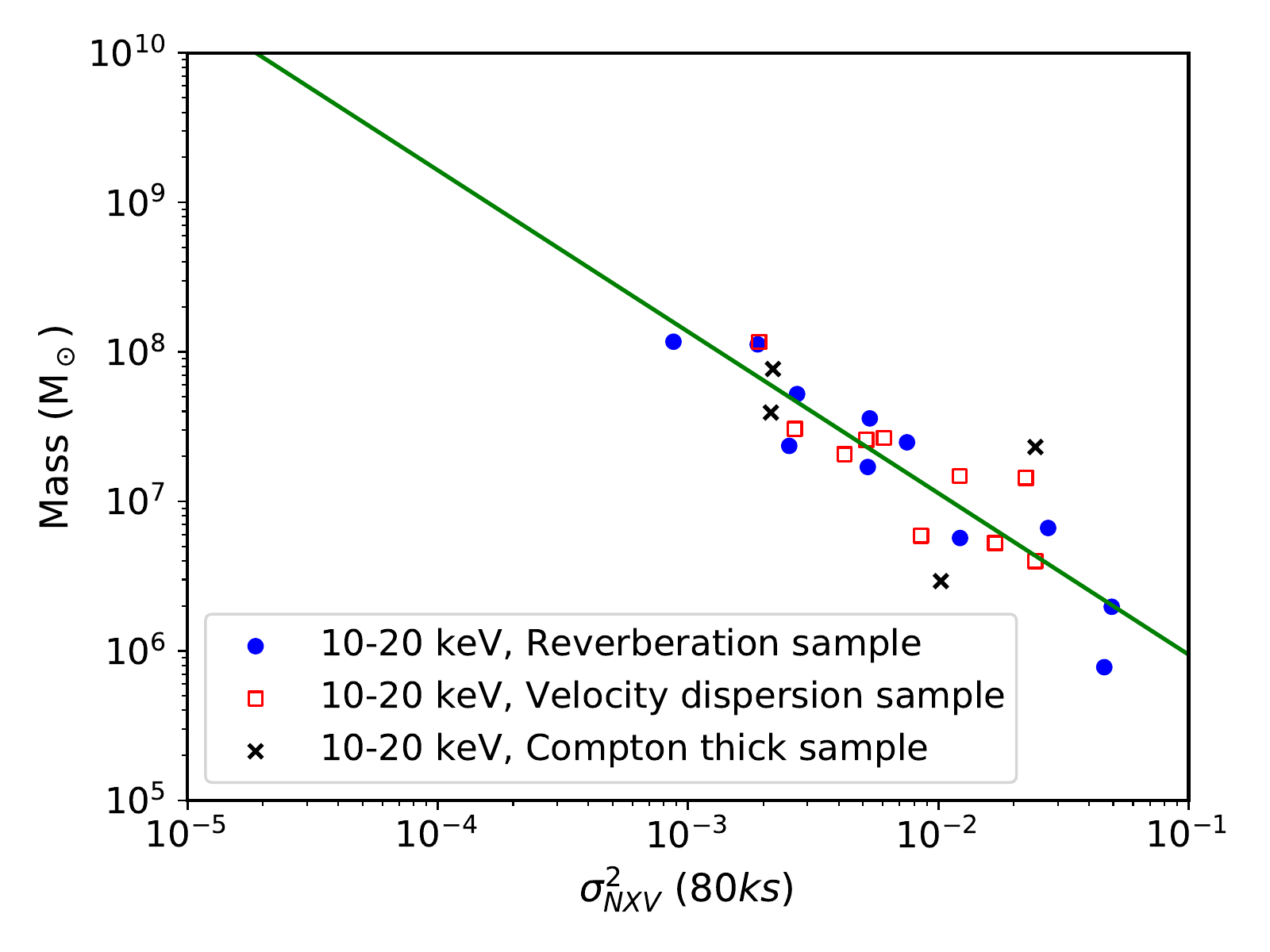}
    \end{center}
    \caption{$\rm M_{BH}$ vs $\rm \sigma_{NXV}^2$ data for the CT sources in the 3-10 and 10--20 keV bands (upper and lower panel, respectively). 
%For comparison we also present results for the Rev sample (filled-blue circles) and the VD sample (open-red squares).  
$\rm \Delta t=80$ ks, and the solid green  line shows the best fit $\rm M_{BH}$-$\rm \sigma_{NXV}^2$ relation for the combined rev+VD sample. }
    \label{ct}
    \end{figure}

Black cross-marks in Fig. \ref{ct} show the \mbh versus $\rm \sigma_{NXV}^2$ for the CT sources, in the 3-10 keV (upper panel) and the 10-20 keV band (lower panel) for the 80 ks segments. 
%For comparison we also present results for the Rev sample (filled-blue circles) and the VD sample (open-red squares).  
The solid green  line shows the best-fit relation to the combined rev+VD sources. An inspection of Fig. \ref{ct}  suggests that the 
$\sigma_{NXV}^2$ measurements in the 10-20 keV band are in good agreement with the model. Although the number of the CT sources is small, this result suggests that we could use the variability method to measure the BH mass in CT sources, when we use data in the 10--20 keV band. 

On the other hand, the scatter of the BH-mass versus excess variance CT points around the best-fit line in the 3--10 keV is significantly greater. The CT data in the top panel of Fig.\, \ref{ct} may even indicate a lack of correlation between BH mass and variability for the CT sources in the 3--10 keV band. This could be due to the fact that most of the observed flux in the 3--10 keV band may be due to scattered radiation rather than the primary continuum in CT sources, although the limited number of CT measurements do not allow us to reach a firm conclusion in this regard.

\section{Application of the proposed prescription}

We applied the proposed prescription to obtain BH mass estimates for AGN that are not part of either the rev, VD, or CT samples and that have \nustar archival observations which fulfill the minimum number of segments and S/N criteria (as  presented in \S \ref{prescription}). There are 22 such sources, which are listed in Table \ref{mass_estimates} and these constitute our 'prediction' sample (prd sample, hereafter). The observation log of these sources is presented in Table \ref{prediction_sample}. We followed the same steps as detailed in Sections \ref{data_reduction} and \ref{calc_nxv} to compute $\rm \sigma_{NXV}^2$ in the 3-10 and the 10-20 keV bands, using $\rm \rm \Delta t=10$, 20, 40, and 80 ks segments. Then we used the best-fit models listed in Table \ref{bestsample} to estimate the BH mass for each source. Our BH mass predictions are listed in Table \ref{mass_estimates}.

There are two sources (i.e. ESO511-30 and 2MASXJ19301380) where the excess variance is negative, at all timescales and in both energy bands. This is despite the fact that the light curves have a S/N value greater than 3 in both bands and are relatively long, specially in the case of ESO511-30 (BAT ID=719)m  where the observation is $\sim$320 ks long. Since all the sources in the rev+VD sample have a positive excess variance measurement and their BH mass is (mostly) smaller than 10$^8$M$_{\odot}$, we suspect that the mass of the BH in these objects is significantly larger than 10$^8$ M$_{\odot}$. 

Within each energy band, the BH mass estimates from the various light curve segments agree with each other within $1-2\sigma$ (where $\sigma$ is equal to the average scatter of the points around the best-fit line, as listed in Table \ref{bestsample}). This is not surprising, as the various light curve segments are all part of the same observations for each source. 

The 3--10 keV and the 10--20 keV band BH mass estimates agree within $1-2\sigma$, at all timescales, in all sources, except IRAS09149-6206. In this case, we are able to measure the BH mass using segments with durations of 10, 20, 40m and 80 ks in the 3--10 keV band and the log(\mbh) estimates ranging between 7.35 and 7.55. 
On the other hand, we cannot measure the BH mass using 10 and 20 ks long segments in the 10--20 keV band, because the resulting excess variance is negative. Moreover, the 10--20 keV excess variance using the 40 and 80 ks segments implies a much larger BH mass. The reason for this discrepancy is the fact that a (relatively) large amplitude variation appears in one of the 3--10 keV band light curves, while it is absent in the respective 10--20 keV light curves (see also \citet{walton2020}).

For two more sources (i.e. Pictor A and HE1143-1810), the excess variance is positive in the 3--10 keV band, but negative in the 10--20 keV band. There are low-amplitude variations in the 3--10 keV 
band light curves, which do not appear in the 10--20 keV light curves. In both cases, log(\mbh) is greater than $\sim 7.6-7.7$. It is possible that the variability detection in the 3--10 keV band only is a statistical effect: due to the high BH mass, it just might happen that the variability is detected only in the 3--10 keV band, where the S/N is greater anyway. Another possibility may be that these sources are dominated by the reflection component at energies above 10 keV; hence, the continuum variations are better sampled at lower energies. 

There are also two cases (i.e 3C111 and Mrk 926) where the $\rm \sigma_{NXV}^2$ was measured only in the 10-20 keV band. Mrk 926 is the source with the largest BH mass estimate in  the prd sample. We believe that the detection of variability in the 10-20 keV and not in the 3--10 keV band is purely statistical in nature. 
It just happens to measure a positive excess variance in this source at energies above 10 keV; and this is probably the case for 3C111 as well. 

Finally, we note that BH mass estimates should always be treated with care. For example, NGC6240, is a galaxy that is known to harbour two active nuclei, each with a mass in excess of $9\times 10^7$ M$_{\odot}$ \citep[e.g.][]{kollatschny2020b}. Such galaxies are excluded from both the rev and the VD samples. We estimate  a log(\mbh) of 7.5 - 8.2 (for $\rm \Delta t=40$ ks, in the 10--20 and 3--10 keV bands, respectively), in agreement with the previous estimates. However, we have not explored how the presence of two active nuclei may affect the accuracy of our method.

%%%%% TABLE 3
\begin{table*}
\caption{Black hole mass estimates for the prediction sample} 
\label{mass_estimates}      
\centering                         
\begin{tabular}{c c c c c c c c c c }        
\hline    
BAT ID & Name & \multicolumn{8}{c}{Log(M/M$_{\odot}$)} \\
\hline
& & \multicolumn{2}{c}{80 ks} & \multicolumn{2}{c}{40 ks} & \multicolumn{2}{c}{20 ks} & \multicolumn{2}{c}{10 ks} \\

& & 3-10keV & 10-20keV & 3-10keV & 10-20keV & 3-10keV & 10-20keV & 3-10keV & 10-20keV \\
\hline
\hline
163 & NGC1194 & 7.2 & 6.7 & 7.07 & 6.5 & -- & 6.6 & -- & 6.44\\
184 & NGC1365 & 6.01 & 6.16 & 6.39 & 6.22 & 6.19 & 6.27 & 6.35 & 6.33\\
214 & 3C111 & -- & 7.7 & -- & 7.56 & -- & 7.39 & -- & 7.25\\
229 & HE0436-4717 & 6.67 & 6.9 & 6.72 & -- & 6.94 & -- & 6.87 & --\\
270 & PictorA & 7.78 & -- & 7.68 & -- & 7.7 & -- & 7.59 & --\\
447 & IRAS09149-6206 & 7.54 & 8.54 & 7.35 & 8.37 & 7.42 & -- & 7.53 & --\\
472 & MCG-5-23-16 & 7.3 & 7.34 & 7.16 & 7.24 & 7.03 & 7.19 & 6.95 & 7.05\\
567 & HE1143-1810 & -- & -- & 7.63 & -- & 7.83 & -- & -- & --\\
657 & ESO323-77 & 7.34 & 7.42 & 7.55 & 7.65 & 7.12 & 7.17 & 7.24 & 7.18\\ 
670 & MCG-3-34-64 & 6.56 & 6.24 & 6.84 & 6.39 & 6.62 & 6.5 & 7.06 & 6.66\\
692 & 4U1344-60 & 7.07 & 7.05 & 7.28 & 7.32 & 7.19 & 7.19 & 7.05 & 6.98\\
719 & ESO511-30 & -- & -- & -- & -- & -- & -- & -- & --\\
750 & LEDA3076910 & 6.61 & 6.3 & 6.73 & 6.31 & 6.7 & 6.52 & 6.69 & 6.45\\
837 & ESO138-1 & -- & -- & -- & -- & -- & -- & -- & --\\
841 & NGC6240 & -- & -- & 7.99 & 7.7 & 8.22 & 7.52 & -- & 7.96\\
995 & Fairall51 & 6.7 & 6.83 & 6.83 & 6.77 & 6.88 & 6.92 & 6.8 & 6.82\\
1032 & ESO141-55 & 7.14 & 7.39 & 7.47 & -- & 7.56 & -- & 7.51 & --\\
1040 & 2MASXJ19301380 & -- & -- & -- & -- & -- & -- & -- & --\\
1111 & IGRJ21277+5656 & 6.85 & 6.82 & 6.95 & 6.83 & 6.74 & 6.74 & 6.62 & 6.56\\
1172 & MR2251-178 & -- & -- & 8.22 & 8.36 & 8.1 & 8.41 & 7.94 & --\\
1183 & Mrk926 & -- & 8.32 & -- & 8.59 & -- & 8.5 & -- & --\\
1194 & IRAS23226-3843 & 7.79 & 8.52 & 7.61 & -- & 7.83 & -- & -- & --\\

\hline
\end{tabular}
\end{table*}

\section{Summary \& guidelines}

Here, we present a prescription for measuring the mass of the central BH in AGN using excess variance measurements. We emphasise that we did not study the variability versus BH mass relation in AGN, as this has been widely considered in previous studies. Our objective is to select a well-defined sample of sources with available BH mass measurements and then to identify the minimum duration and S/N of the light curves that are necessary for the \mbh versus \snxv\, relation to be well defined, that is, for the scatter of the points around the best-fit lines to be similar to the average uncertainty of BH mass estimates computed with other methods. If the X-ray variability mechanism is the same for all AGN, we can measure BH mass in any AGN following the prescription we outlined in \S \ref{prescription}, as long as the duration and S/N of the available light curves satisfy the 'quality' criteria discussed there. 

We emphasise that the prescription to measure AGN that we present in this work is only applicable to the case of light curves with a bin size of 250 sec that can be divided in segments of 80, 40, 20, and 10 ks. Such light curves can be obtained from observations taken by {\rm NuSTAR} and  {\rm XMM-Newton}, as well as from past X--ray observatories, such as {\rm Suzaku} and {\rm ASCA}. There are many AGN observations in the {\rm RXTE} and {\it Swift}/XRT archive, however, their mean bin size is significantly larger than 250 sec and, hence, they cannot be used to estimate BH mass using the prescription in Section \ref{prescription}. Nevertheless, there are already quite a few AGN whose \nustar\, and {\rm XMM-Newton} archival light curves can meet the criteria described in Section \ref{prescription}. We already applied our prescription to archival light curves of 22 AGN with \nustar\ light curves that fulfill the necessary criteria and we computed BH mass estimates for them. In the near future, we plan to search the {\rm XMM-Newton} archive for AGN with light curves that can satisfy the necessary criteria and estimate their BH mass. 

The average scatter of the rev+VD sources around the best-fit line, $\bar{\sigma}_{SC}$,  is 0.25, 0.35, and 0.4, for the segments with $\rm \Delta t$=80 ks, 40 and 20 ks, and 10 ks, respectively (these numbers are equal to the mean of $\sigma_{SC,3-10 {\rm keV}}$ and $\sigma_{SC,10-20 {\rm keV}}$ values, listed in Table \ref{bestsample}). On average, the error of log(\mbh) should be comparable to $\bar{\sigma}_{SC}$. We stress that $\bar{\sigma}_{SC}$ is not the formal error on the BH mass estimate of an individual source. Instead, it should be considered as representative of the expected standard deviation of these estimates. It is possible to compute the error of log(\mbh) for an individual source, but only if the available light curves are long enough (see below). 

Our results demonstrate that it is possible to measure the BH mass in AGN using excess variance measurements if the available light curves are long enough and have the necessary S/N values. The resulting BH mass estimates will be as reliable as the estimates which are based on other techniques (e.g. the continuum and emission line reverberation and velocity dispersion measurements). 
The 'quality' requirements on the light-curve duration and S/N are quite strong. They are mainly imposed by the statistical properties of \snxv. The normalised excess variance is a statistic which follows a highly asymmetric distribution function, with a broad variance \cite[see][]{allevato2013}. We strongly advise against the use of light curves  that do not follow the duration and S/N criteria that we list in Section \ref{prescription}. If such light curves are used to compute \snxv\, and then log(\mbh) using the best-fit results listed in Table \ref{bestsample}, the resulting estimate may be heavily biased. 
%As we said above, our aim was not to establish the \mbh\, vs \snxv\, relation for the AGN population as a  whole. 
Due to the \snxv\, statistical properties, the \mbh\, versus \snxv\ relation may actually depend on the characteristics of the  light curves we use (for the same segment duration). For example, the difference between the slope of the best-fit line to the VD data plotted in the top left and bottom right panels in Fig.\,\ref{comparison1} is 1.09$\pm 0.19$, which is significant at the 5.7$\sigma$ level. This result clearly shows that it will be inappropriate to use the best-fit results listed in Table \ref{bestsample} to compute the BH mass for sources with low-quality light curves. 

It may be worth investigating the relation between BH mass and excess variance for shorter light curves or light curves with lower S/N and to establish a prescription for estimating their \mbh\, values as well. In this way, BH masses could be measured for many more sources. This aspect, however, is beyond the scope of this work and, in any case, we suspect that the resulting estimates will have significantly larger uncertainties, hence, they may be of limited applicability.

Table \ref{bestsample} lists the best-fit parameters of the lines we fitted to the data of the combined (rev+VD) sample, when the excess variance is computed using light curve segments which are $\rm \Delta t=10$, 20, 40 and 80 ks long (Figs. \ref{merged1} and \ref{merged2}). We expect that the BH mass estimates should be independent of the segment duration. However, whenever possible, it would be advisable to compute \mbh\, using all four different light-curve segment durations to verify that there are no unexpected complications with the observations. Since the average error of the BH estimates should decrease with increasing segment duration, we propose to adopt BH mass estimates based on the longest $\rm \Delta t$.  

The accuracy of the BH mass estimate will not increase even if the duration of the light curves is longer than the longest light curves among the sources in the rev+VD sample. This is because the uncertainty on \mbh\ mainly depends on the uncertainty of the best-fit line parameters, which should be representative of the typical scatter of the points plotted in Figs \ref{merged1} and \ref{merged2}. On the other hand, if there are more than 50 segments, with an average S/N ratio greater than 3, then it is possible to compute the error on the mean \snxv\ \citep[see][]{allevato2013}. In this case, an error on the BH mass measurement of the individual source can be measured, using the error of the best-fit line parameters (listed in Table \ref{bestsample}) and the commonly used rule of error propagation. 

The best-fit results and the scatter of the rev and VD points around the best-fit lines are comparable for the 3--10 and the 10--20 keV bands (see Figs.\, \ref{merged1} and \ref{merged2}). Black hole mass estimates should be similar when using light curves in both bands to compute the excess variance. Nevertheless, we suggest computing the BH mass in both bands when there are light curves in both bands (i.e. from NuSTAR observations). The 10--20 keV band should be preferred for sources which are heavily absorbed (as our results, from a limited number of CT sources, indicate). On the other hand, the 3--10 keV band should be preferred for sources which may be dominated by the X--ray reflection component at energies above 10 keV. 

The \mbh\--\snxv\ relations studied here are well calibrated for sources with BH mass between 10$^6-10^8$M$_{\odot}$. We caution against the use of the proposed method for sources with \mbh$\le 10^6$M$_{\odot}$. In this regime, the relation between \mbh\ and \snxv\ could significantly deviate from the linear assumption adopted in this paper. On the other hand, the method we present should be valid for AGN with larger BH masses, although the available data sets for such sources may not be sufficient. The best-fit line in Figs. \ref{merged1} and \ref{merged2} show that the normalised excess variance in sources with BH mass larger than 10$^8$ M$_{\odot}$ should be smaller than $\sim 10^{-3}$, even when $\rm \Delta t=80$ ks  (the limit is smaller for shorter segments). 
This is a small excess variance limit and light curves with the minimum S/N and duration requirements will probably result in a negative \snxv. It will be necessary to use light curves with a much higher S/N or many light curve segments in  order to measure such a low intrinsic excess variance. 

The proposed prescription should be valid both for Type I and Type II objects. Although all AGN in the rev sample are Type I, the VD sample contains a mixture of Type I and Type II sources. Our plots show no difference
in the  {\mbh} -- {\snxv} relation for these two classes and, moreover, 
the best-fit lines to the rev and VD log(\mbh) -- \snxv\, plots are almost identical. Therefore we expect the method to perform equally well for objects in either class.

In this work, we study mainly radio-quiet AGN, although a limited number of objects from the 3C (and one for the 4C) catalogue are included in our sample. Emission from the jet is not supposed to be dominant in the bands we study for any of these sources and we therefore expect the proposed prescription to be valid for these sources as well. In any case, we strongly advise against the use of the proposed prescription to blazars. In this class of sources, in constrast to their radio-quiet counterparts, the X-ray emission could be dominated  by radiative processes in the relativistic jet. In agreement to this scenario, \citet{rani2017} suggested that blazars  show much greater X-ray flux variation amplitudes as compared to Seyfert galaxies.

In order to  compare or combine BH mass measurements using the prescription presented here with those derived using different methods,  it is necessary to take into account (when relevant) that our estimators are normalised to the \mbh\ -- $\sigma_{\ast}$ relation of \citet{woo2013}. We note that our results are applicable for AGN in the nearby Universe (i.e. sources with z$\le 0.07$). Our prescription should also be applicable to AGN further away, as long as rest-frame light curves satisfy the necessary quality-criteria. Obviously, the reliability of the BH mass estimates in this case will also depend on the assumption that the AGN variability properties do not change with redshift. 

It would be interesting to see if a similar method can be established for the AGN light curves resulting from e-ROSITA sky survey data. The method will have to be calibrated appropriately, as the light curve sampling properties will be entirely different to the light curves that are currently delivered by the pointed observations of XMM-Newton and \nustar. It is not certain that it will be possible to measure AGN using variability measurements with the e-ROSITA light curves, but given the large amount of sources that will be identified, it may be worth investigating this possibility. 

\begin{acknowledgements}
We have made use of data from the $\rm NuSTAR$ mission, a project led by the California Institute of Technology, managed by the Jet Propulsion Laboratory, and funded by the National Aeronautics and Space Administration. We thank the \nustar Operations, Software and Calibration teams for support with the execution and analysis of these observations. This research has made use of the \nustar Data Analysis Software (NuSTARDAS) jointly developed by the Space Science Data Center (SSDC; ASI, Italy) and the California Institute of Technology (USA). This work is based on archival data, software or online services provided by the SSDC. This research has made use of the High Energy Astrophysics Science Archive Research Centre Online Service, provided by the NASA/Goddard Space Flight Centre and NASA’s Astrophysics Data System.
  \end{acknowledgements}

\bibliography{ref}
\bibliographystyle{aa}

\clearpage
\onecolumn

\begin{appendix}

\section{Plots of the best-fit results as function of  observation durations}

\begin{figure*}
\begin{center}
    \includegraphics[height=0.30\columnwidth]{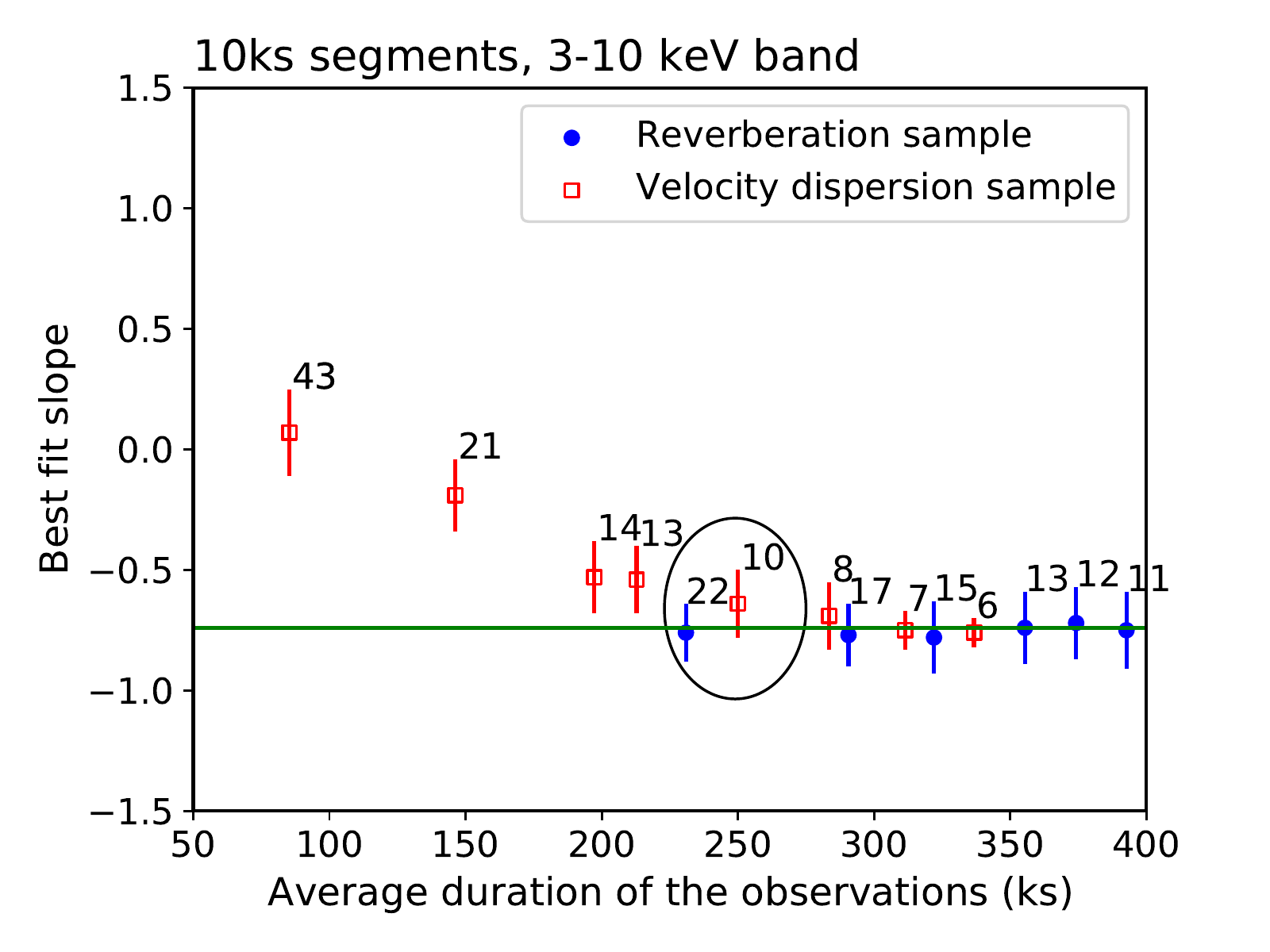}
    \includegraphics[height=0.30\columnwidth]{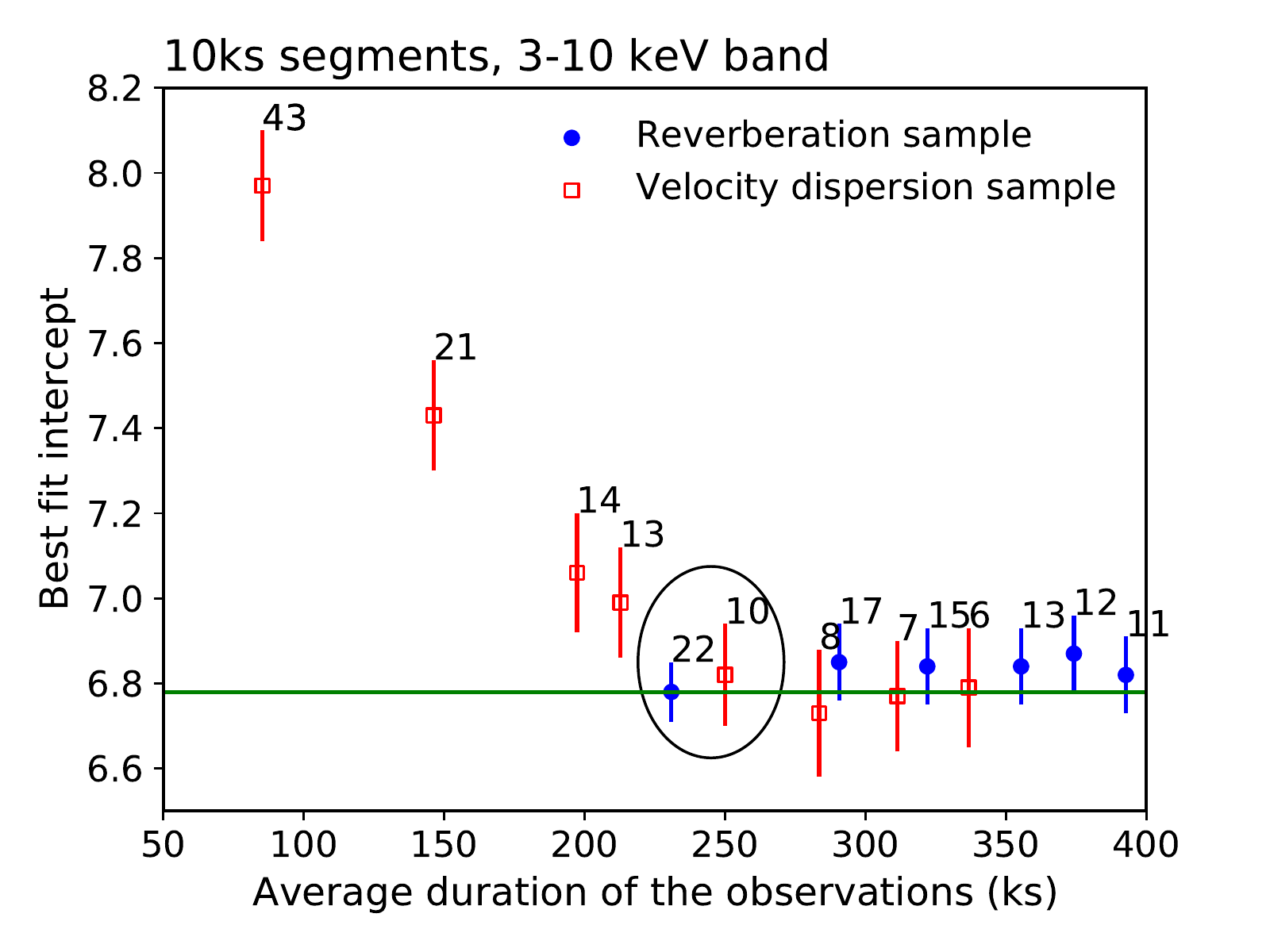}
    \includegraphics[height=0.30\columnwidth]{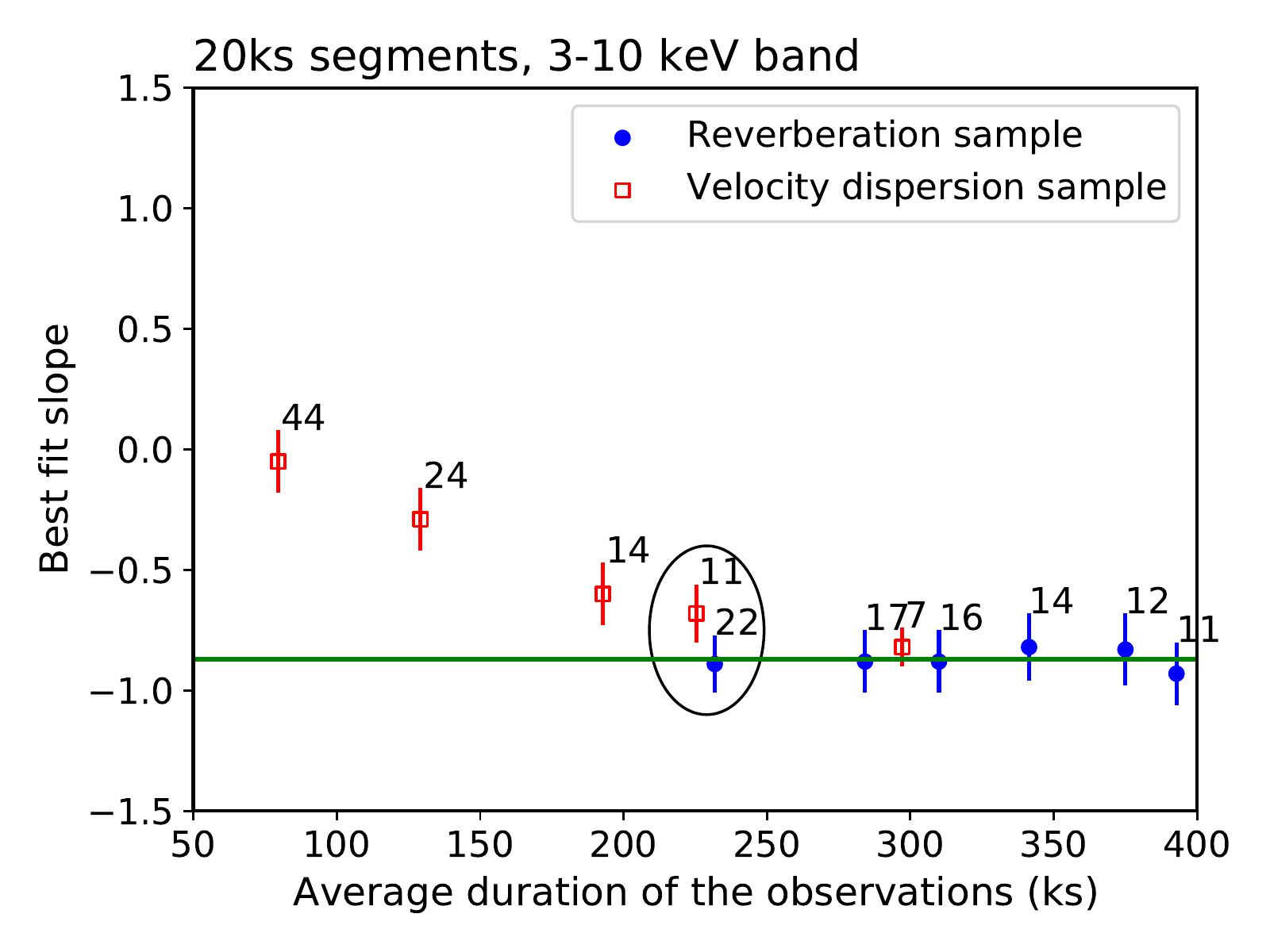}
    \includegraphics[height=0.30\columnwidth]{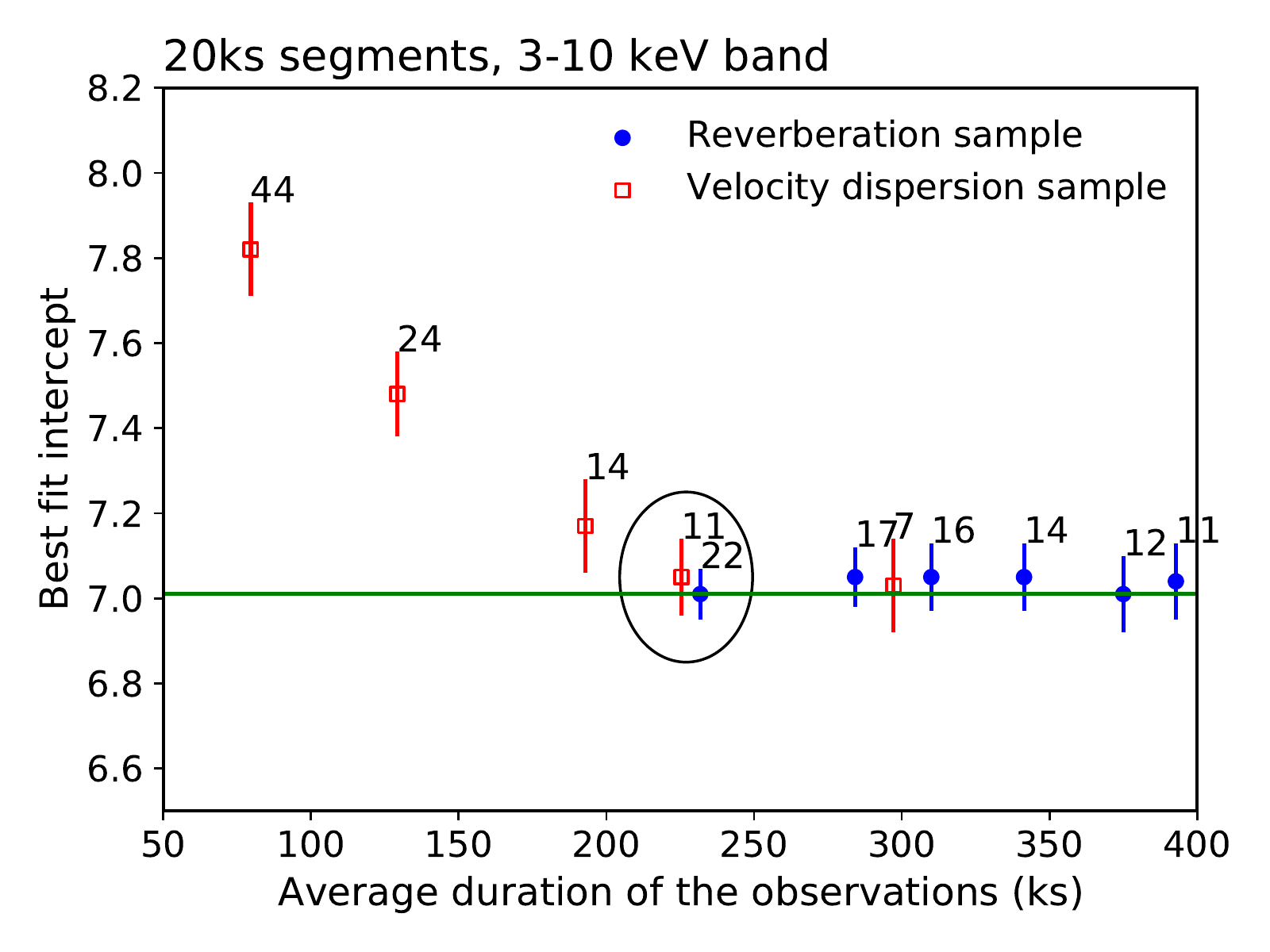}
    \includegraphics[height=0.30\columnwidth]{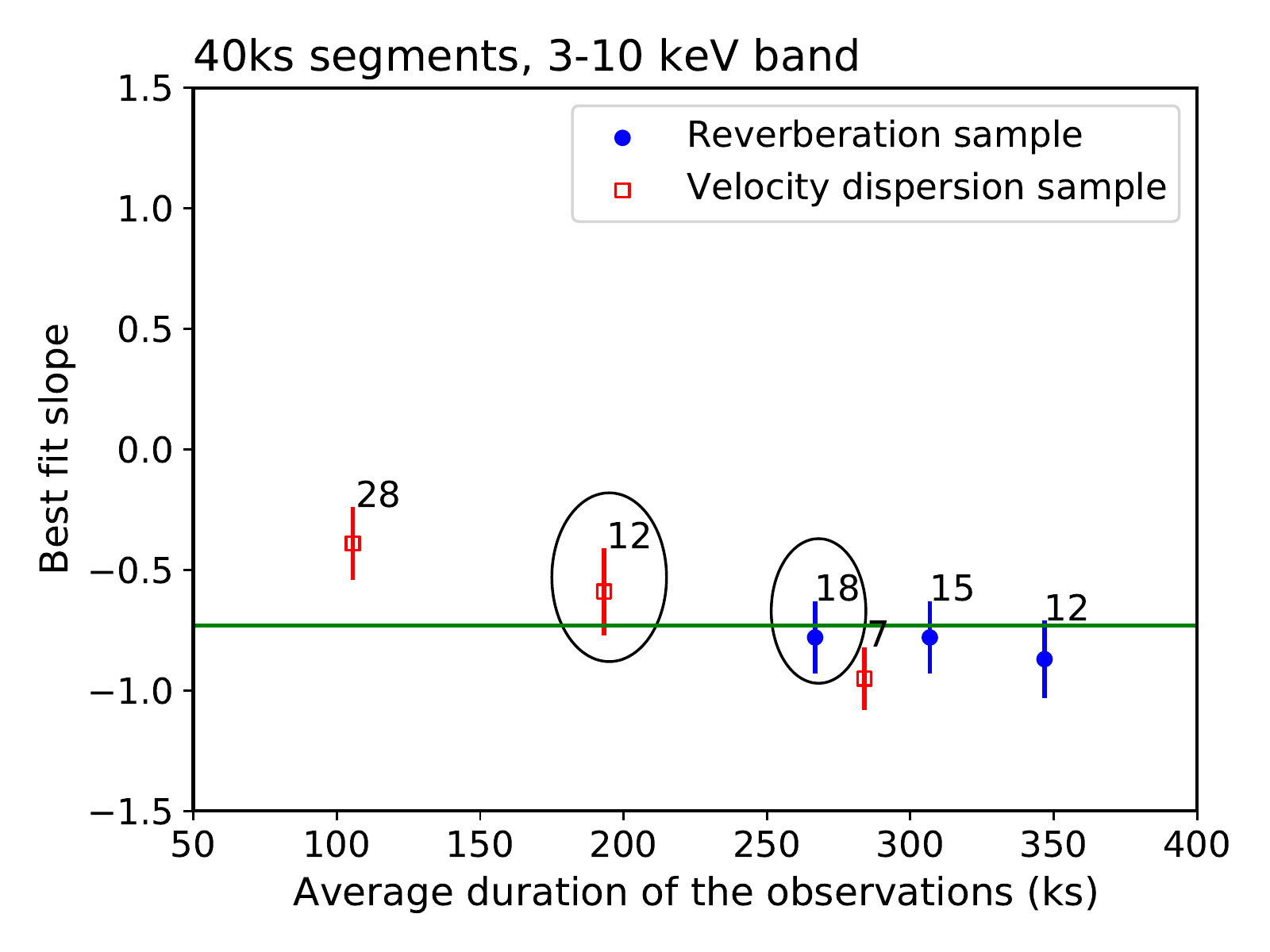}
    \includegraphics[height=0.30\columnwidth]{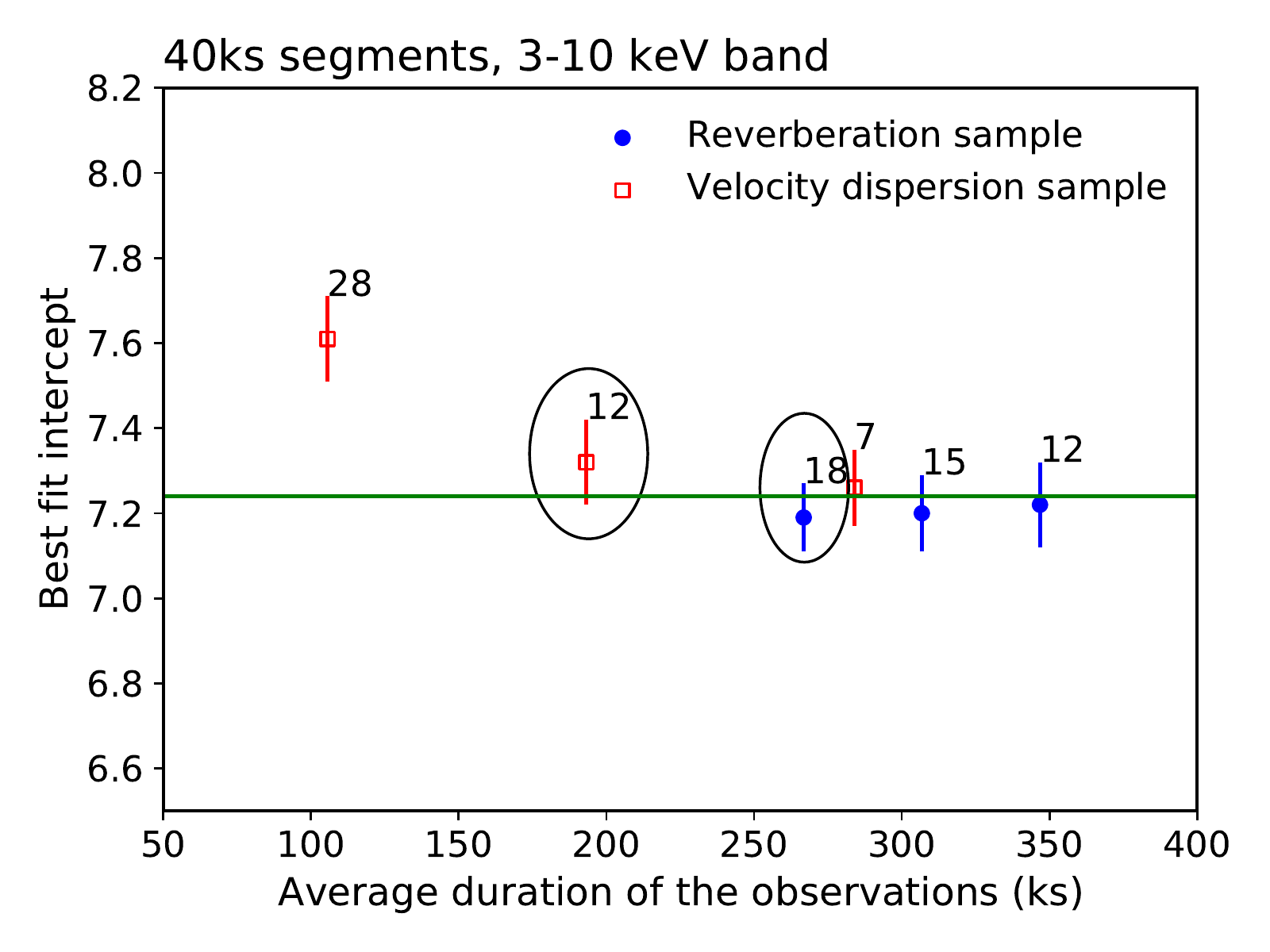}
    \includegraphics[height=0.30\columnwidth]{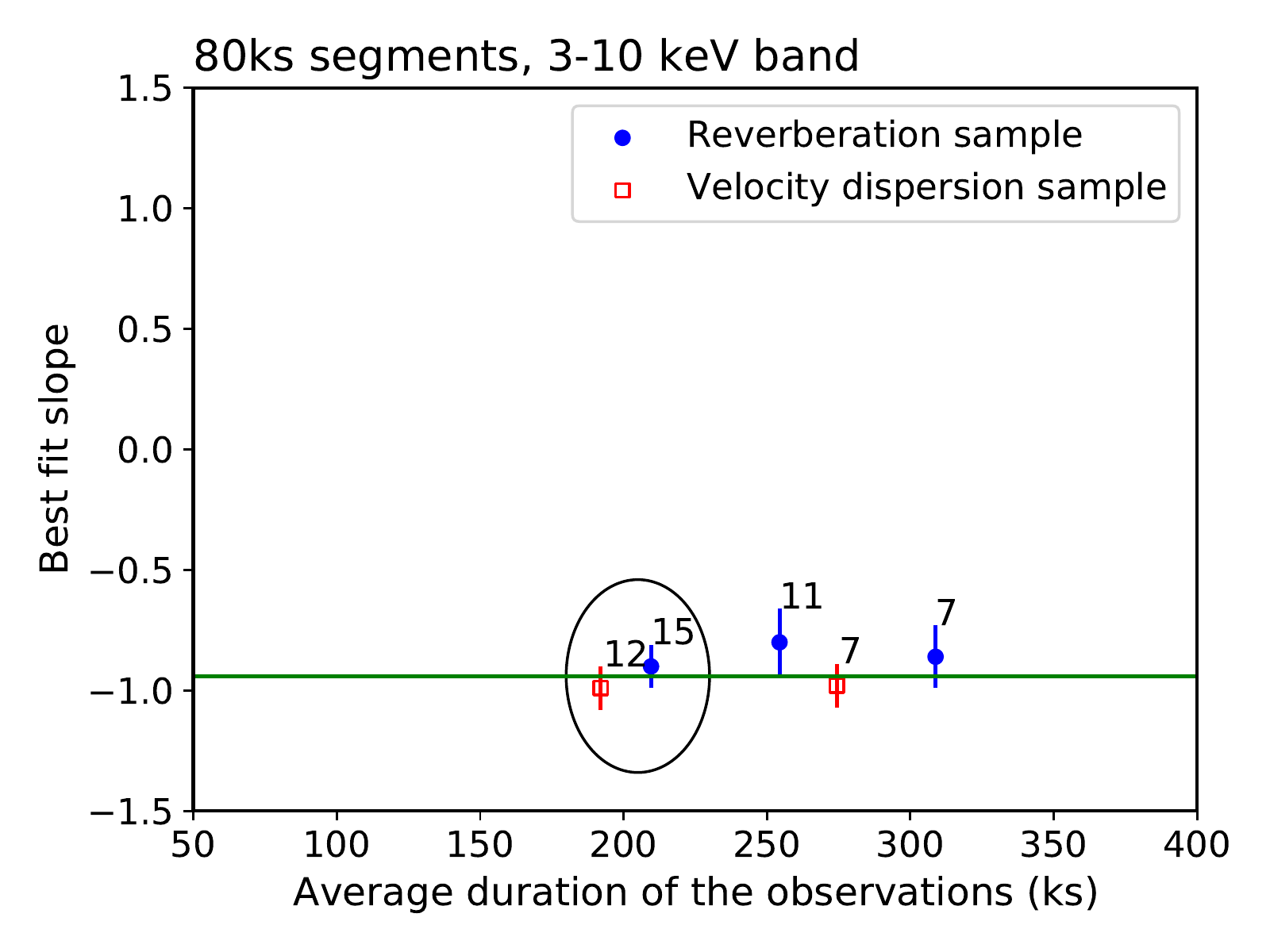}
    \includegraphics[height=0.30\columnwidth]{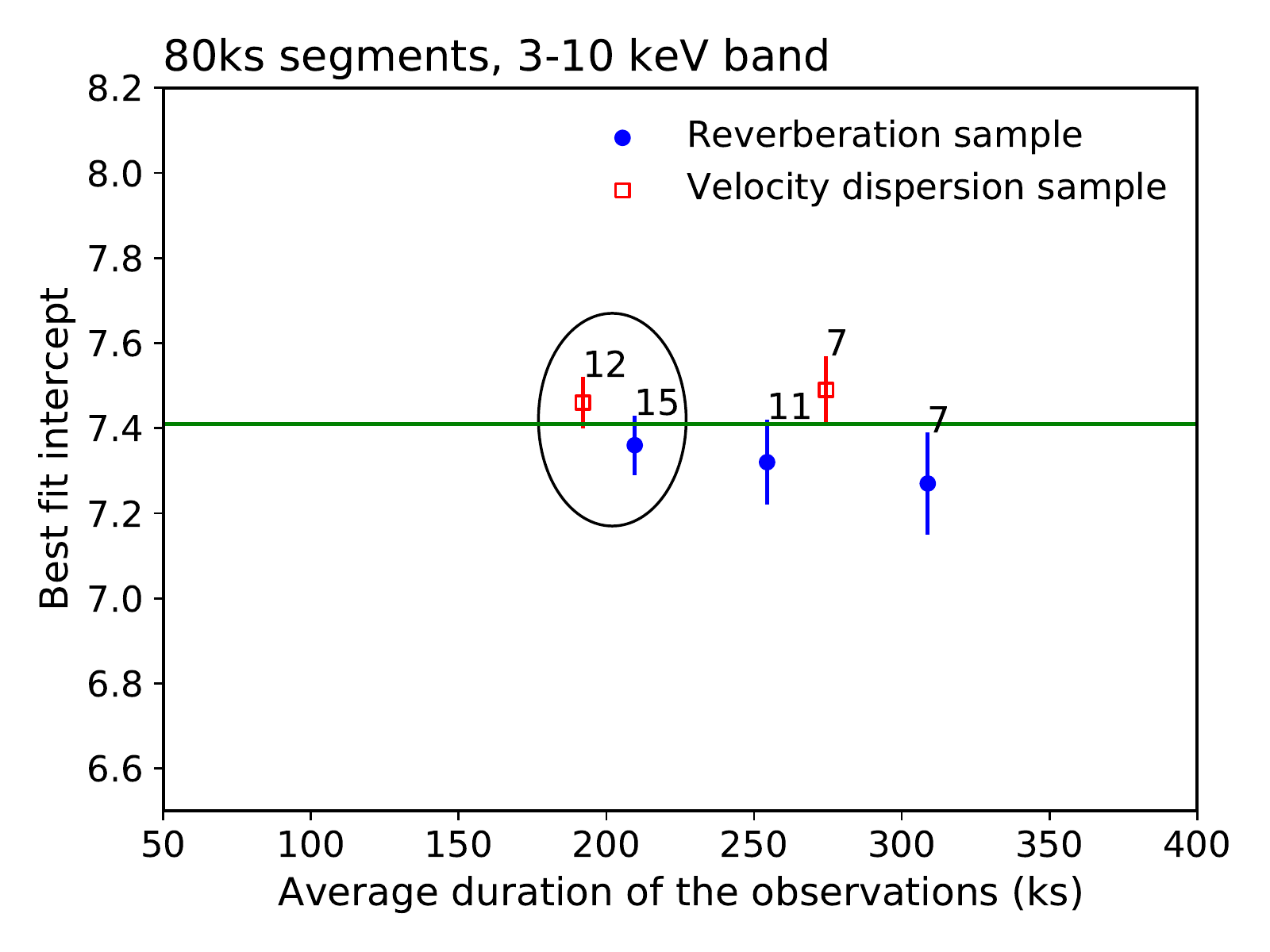}
\caption{Best-fit slope and intercept in the 3-10 keV band for the rev sub-samples (filled-blue circles) and the VD sub-samples (open-red squares) as a function of the average duration of their sources. The circles indicate the first  rev and VD sub-samples (i.e those with a shorter duration) for which the best-fit results are in agreement (within errors). The horizontal lines indicates the best fit value when we fit the combined data from the two sub-samples in agreement (rev and VD). }
\label{duration1}
\end{center}
\end{figure*}

\begin{figure*}
\begin{center}
    \includegraphics[height=0.30\columnwidth]{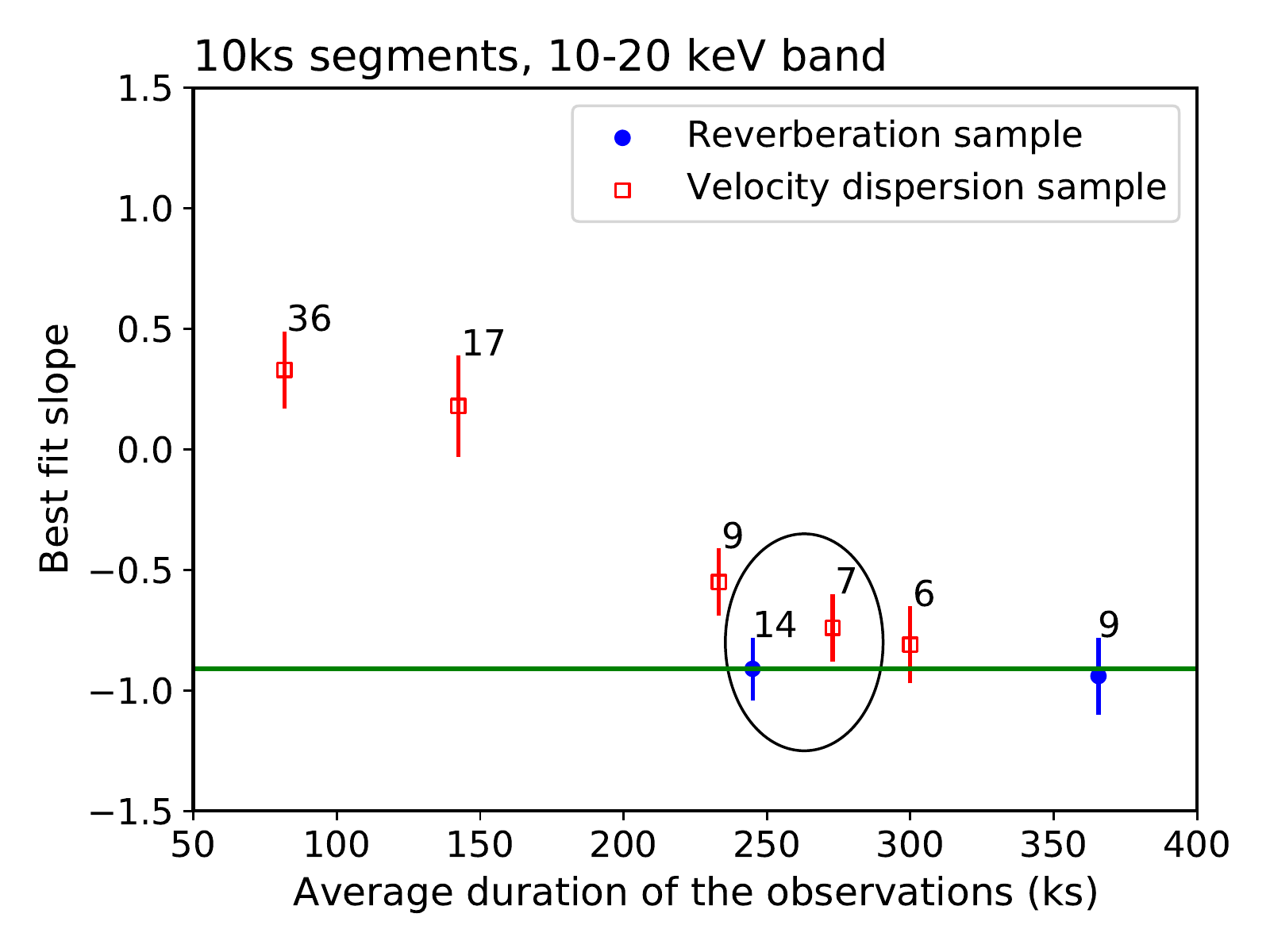}
    \includegraphics[height=0.30\columnwidth]{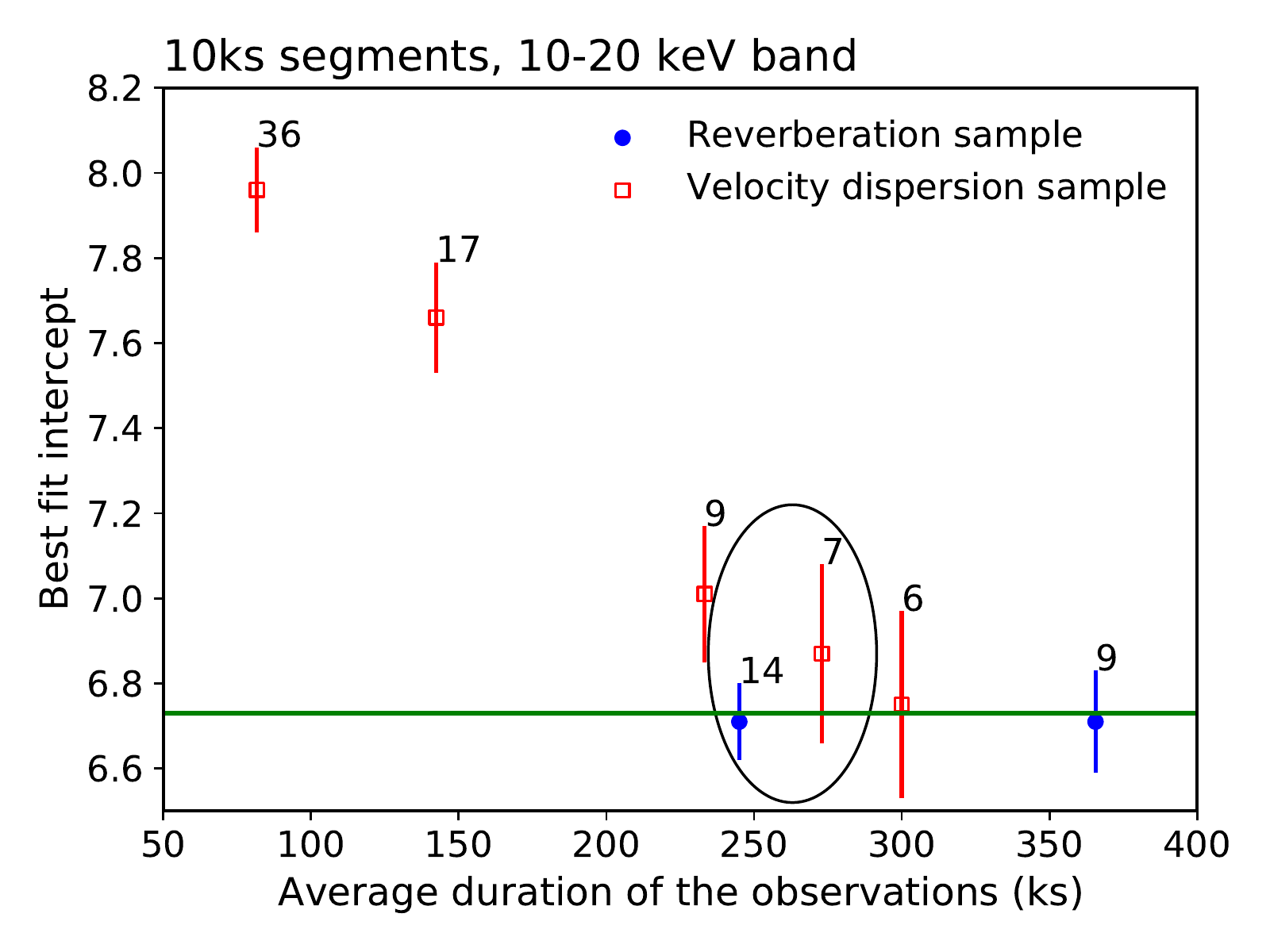}
    \includegraphics[height=0.30\columnwidth]{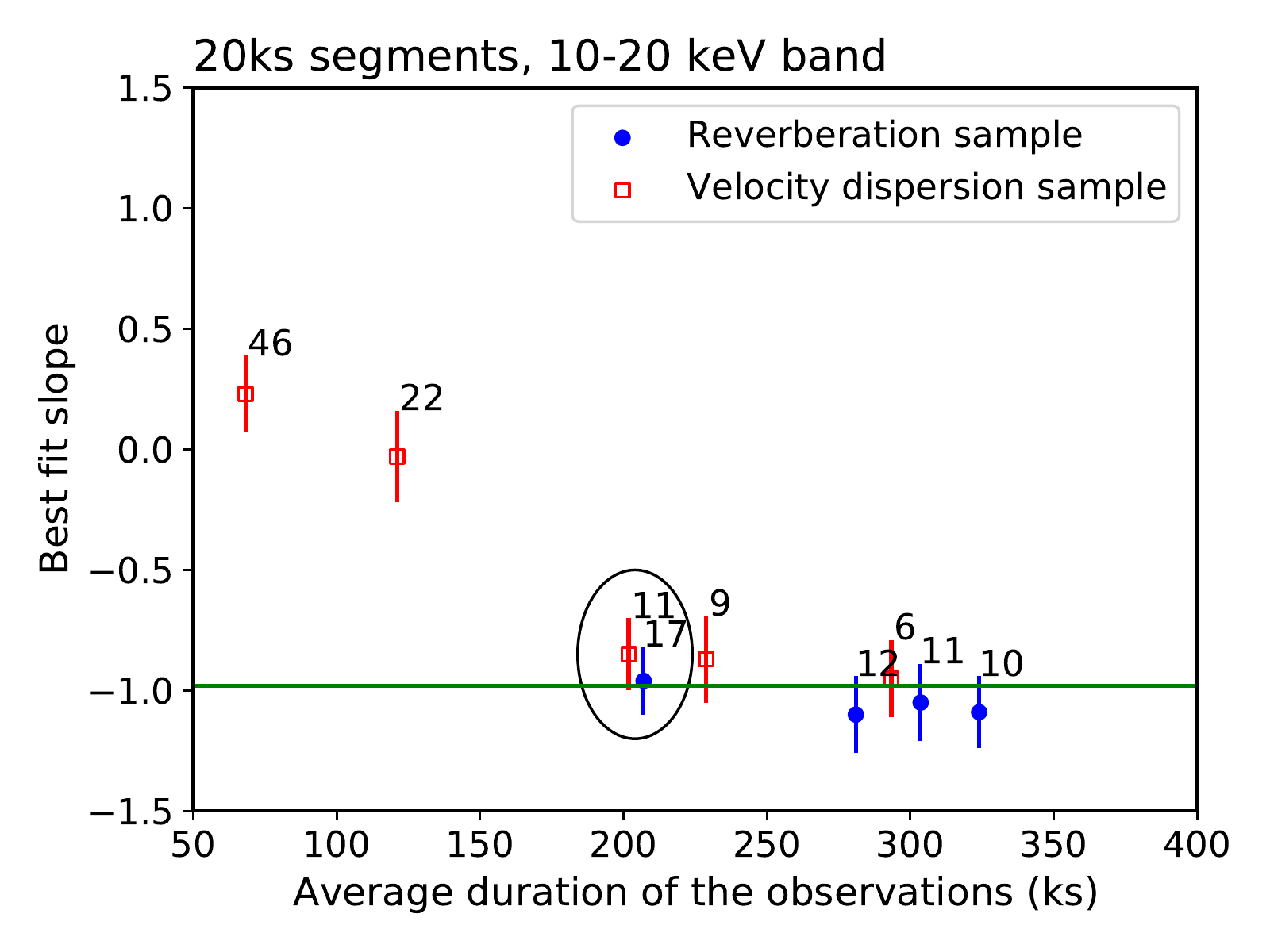}
    \includegraphics[height=0.30\columnwidth]{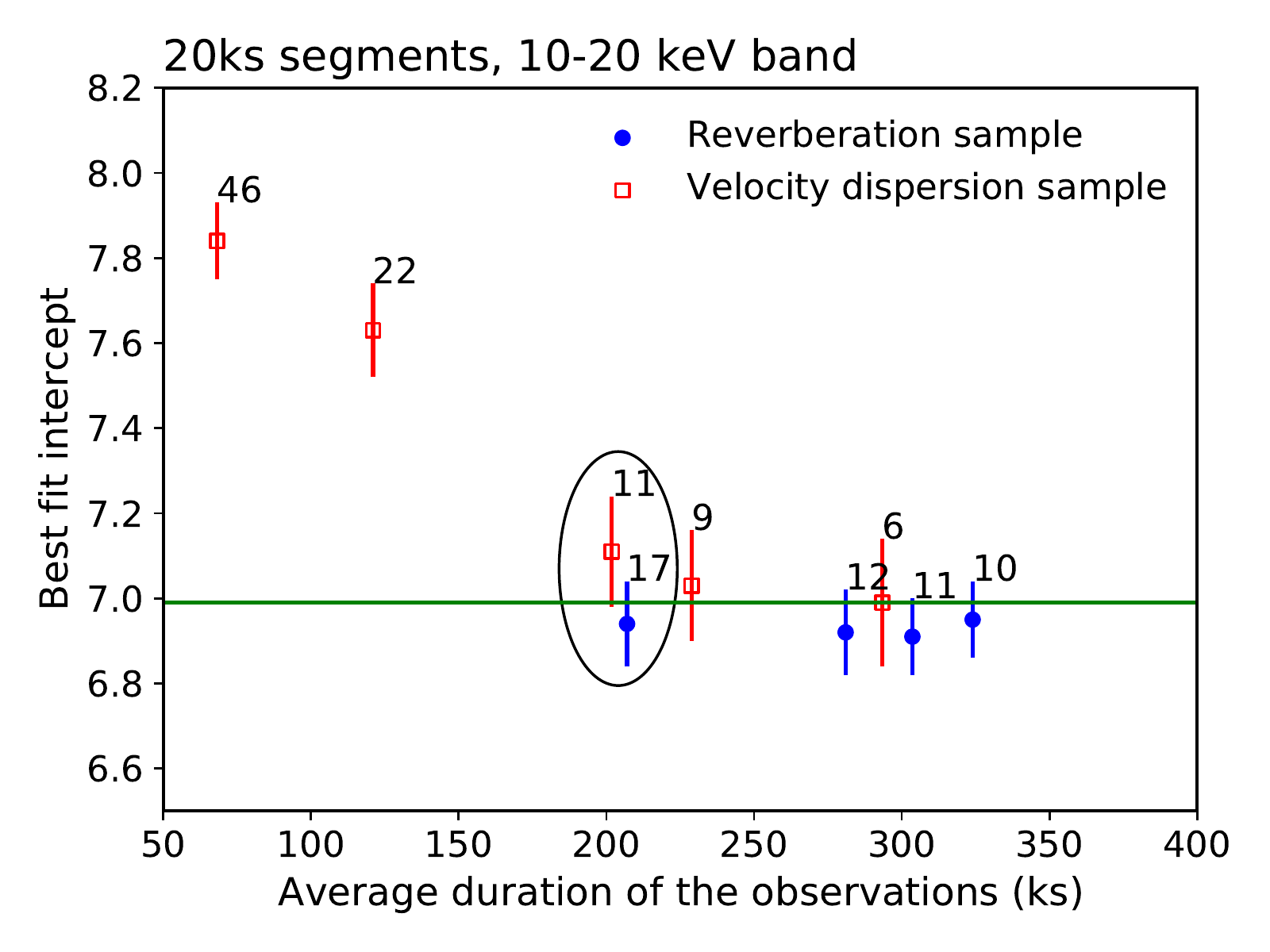}
    \includegraphics[height=0.30\columnwidth]{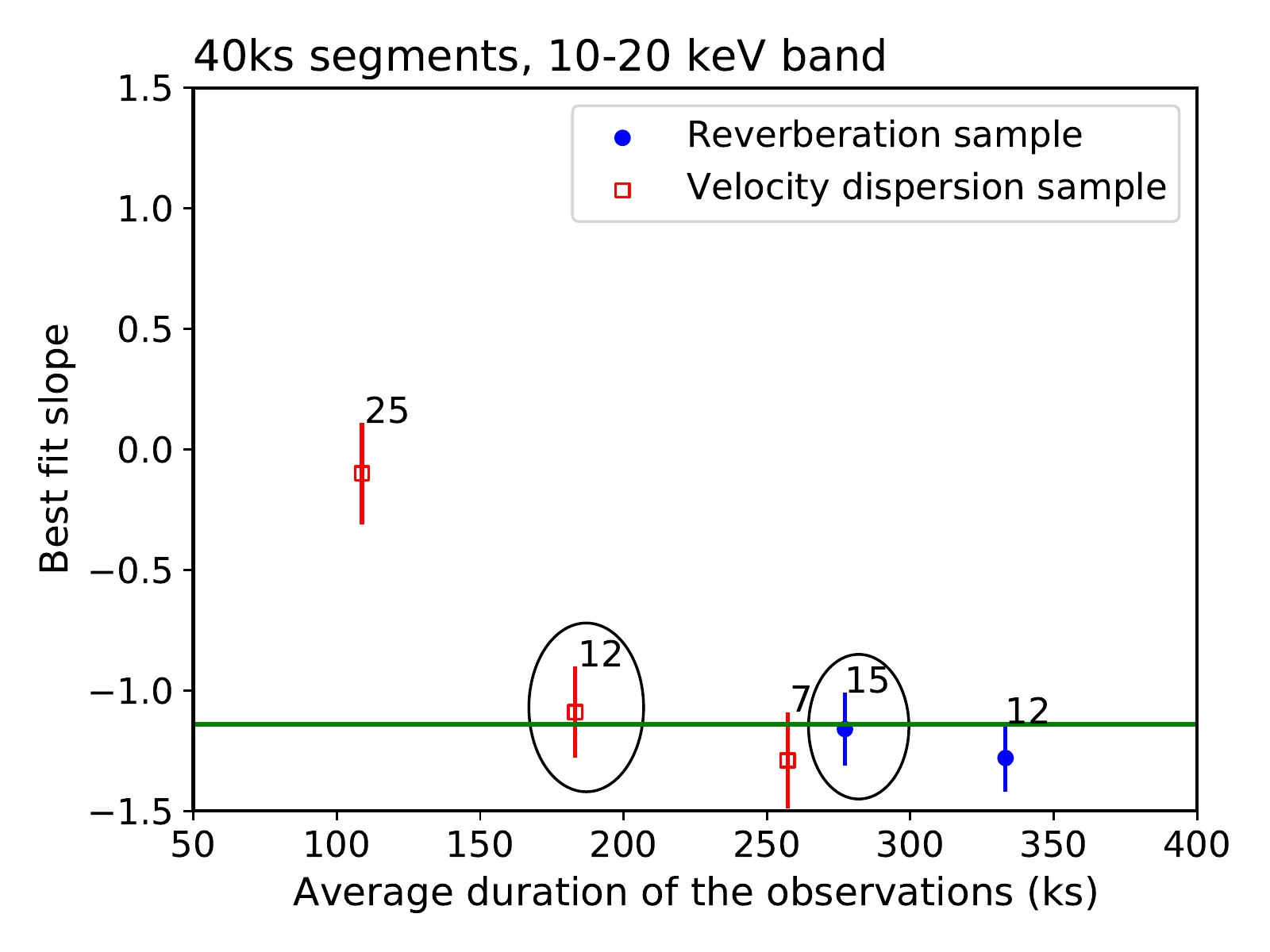}
    \includegraphics[height=0.30\columnwidth]{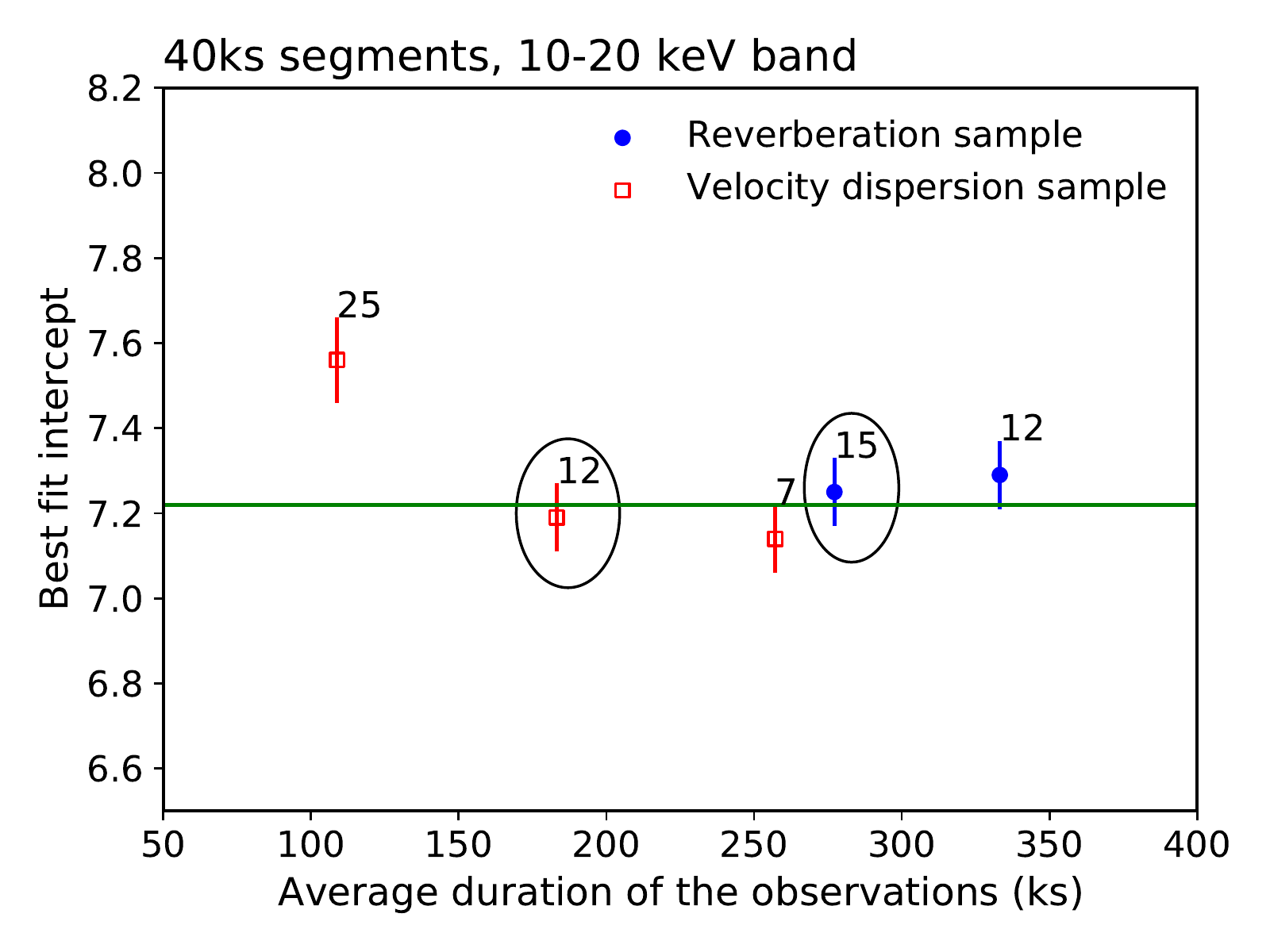}
    \includegraphics[height=0.30\columnwidth]{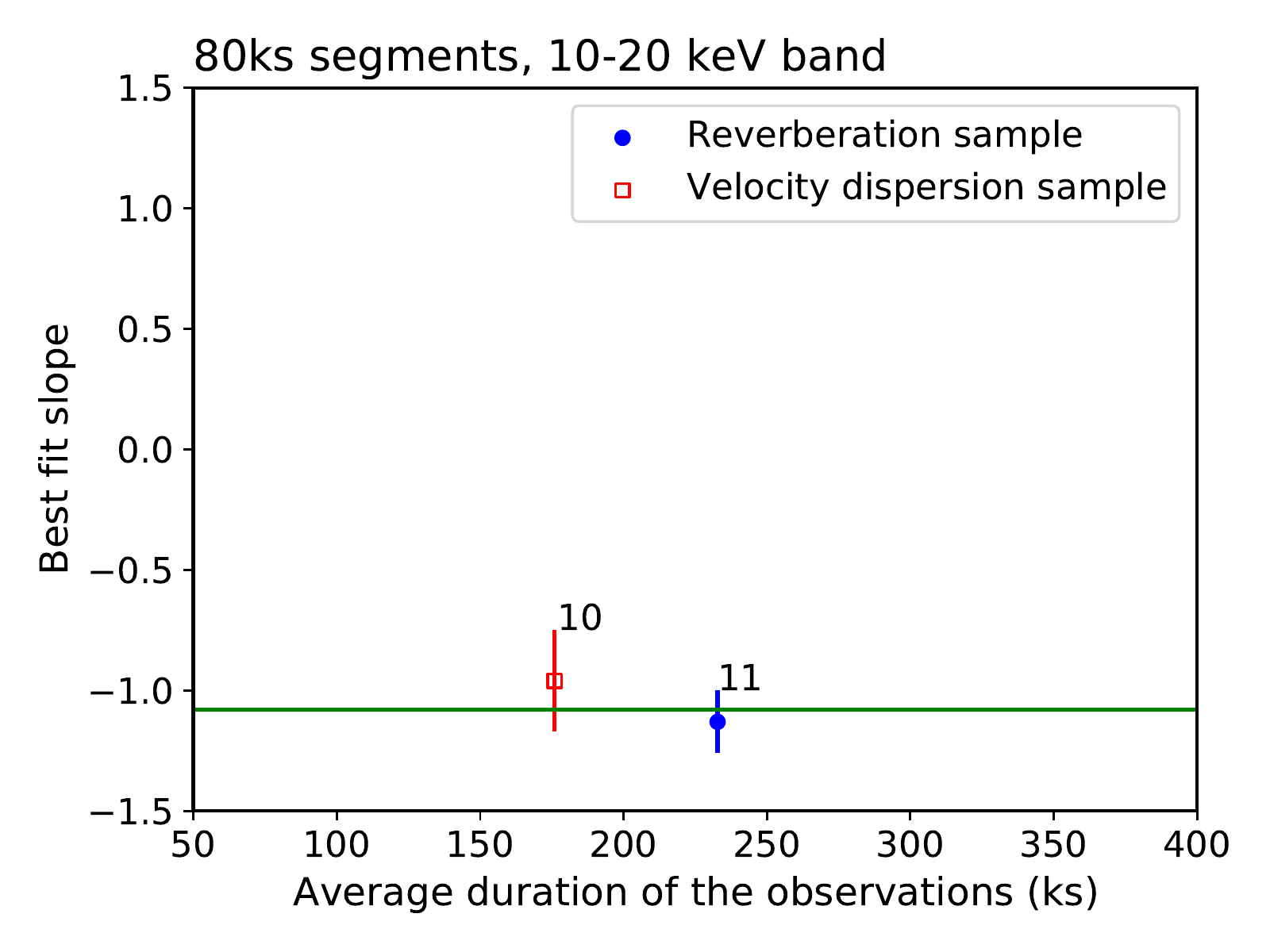}
    \includegraphics[height=0.30\columnwidth]{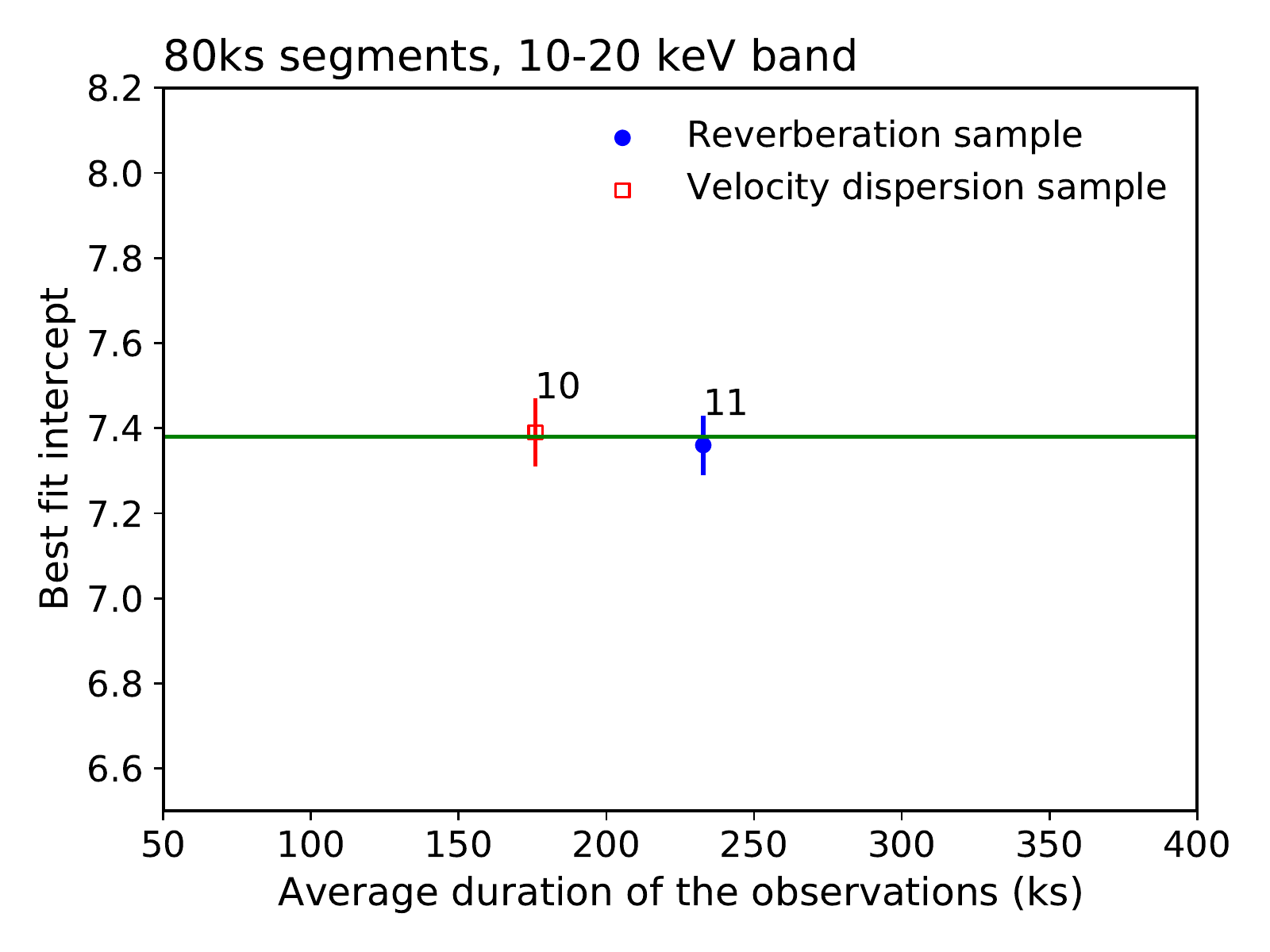}
\caption{Best-fit slope and intercept in the 10-20 keV band for the rev sub-samples (filled-blue circles) and the VD sub-samples (open-red squares) as a function of the average duration of their sources. The circles indicate the first  rev and VD sub-samples (i.e those with a shorter duration) for which the best-fit results are in agreement (within errors). The horizontal lines indicates the best fit value when we fit the combined data from the two sub-samples in agreement (rev and VD).}
\label{duration2}
\end{center}
\end{figure*}

\section{Observation logs}

\begin{longtable}{rrcccccp{2cm}}
\caption{\label{rev_sample} Log of the rev sample}\\
\hline\hline
BAT ID  &       Name    &       NuSTAR ObsID    &       z &     Type    &       Duration         &       $\rm log(M/M_\odot)$    &       Ref.     \\
& & & & & (sec) & & \\
(1) & (2) & (3) & (4) & (5) &  (6) & (7) & (8) \\
\hline
\endfirsthead
\caption{continued.}\\
\hline\hline
BAT ID  &       Name    &       NuSTAR ObsID    &       z &     Type    &       Duration        &       $\rm log(M/M_\odot)$         &       Ref.    \\
& & & & & (sec) & & \\
(1) & (2) & (3) & (4) & (5) &  (6) & (7) & (8) \\

\hline
\endhead
\hline
\endfoot
6       &       Mrk335  &       60001041005     &       0.025 & Sy1.2   &       191700  &       7.23    &       1, 2, 3, 5, 6, 7, 8, 9     \\
        &               &       80201001002     &        &              &       167400  &               &               \\
        &               &       80001020002     &        &              &       138000  &               &               \\
        &               &       80502636006     &        &              &       96000   &               &               \\
        &               &       80502636002     &        &              &       94500   &               &               \\
        &               &       90602619006     &        &              &       64800   &               &               \\
        &               &       90602619004     &        &              &       63900   &               &               \\
        &               &       80502636004     &        &              &       48900   &               &               \\
        &               &       90602619008     &        &              &       48000   &               &               \\
        &               &       60001041002     &        &              &       45900   &               &               \\
        &               &       60001041003     &        &              &       42900   &               &               \\
        &               &       90602619002     &        &              &       24600   &               &               \\
73      &       Fairall9&       60001130002     &       0.047 & Sy1.2   &       74700   &       8.29    &       2, 7, 9, 10, 11    \\
116     &       Mrk590  &       90201043002     &       0.026 & Sy1.5   &       96900   &       7.57    &       2, 6, 7, 9 \\
        &               &       60160095002     &       &               &       39900   &       &               \\
        &           &   80402610002     &       &               &       39300   &       &               \\
266     &       Ark120  &       60001044004     &       0.032 & Sy1.0   &       132600  &       8.06    &       2, 6, 7, 9 \\
310     &       UGC3374 &       60201027002     &       0.020 & Sy1.5   &       184200  &       7.28    &       12, 13      \\
458     &       Mrk110  &       60201025002     &       0.035 & Sy1.5   &       387000  &       7.29    &       2, 6, 9 14 \\
497     &       NGC3227 &       60202002014     &       0.003 & Sy1.5   &       97800   &       6.75    &       2, 9, 13, 15, 16   \\
        &               &       60202002002     &       &               &       96000   &               &               \\
        &               &       60202002008     &       &               &       86400   &               &               \\
        &               &       60202002010     &       &               &       86100   &               &               \\
        &               &       60202002004     &       &               &       84000   &               &               \\
        &           &   80502609004     &       &               &       63600   &               &               \\
        &               &       80502609002     &       &               &       57000   &               &               \\
530     &       NGC3516 &       60002042004     &       0.008 & Sy1.2   &       110100  &       7.39    &       2, 9, 13, 15, 16   \\ 
        &               &       60002042002     &        &              &       71400   &               &               \\
        &               &       60160001002     &        &          &   70500   &               &               \\
        &               &       60302016012     &        &          &   57900   &               &               \\
        &               &       60302016010     &        &              &       55500   &               &               \\
        &               &       60302016004     &        &              &       53400   &               &               \\
        &               &       60302016006     &        &              &       53100   &           &            \\
        &               &       60302016002     &        &              &       45600   &               &               \\
        &               &       60302016008     &        &              &       37800   &               &               \\
542     &       Arp151  &       60160430002     &       0.021 & Sy1.2   &       39000   &       6.67    &       9, 17, 18, 19, 20  \\
558     &       NGC3783 &       60101110004     &       0.009 & Sy1.2   &       88800   &       7.37    &       2, 9, 21, 22, 23, 24       \\
        &               &       60101110002     &        &              &       86100   &               &               \\
        &               &       80202006002     &        &              &       56400   &               &               \\
566     &       UGC6728 &       60160450002      &      0.006 & Sy1.2   &       27600   &       5.85    &       25, 26      \\
583     &       Mrk1310 &       60160465002     &       0.019 & Sy1.5   &       39300   &       6.21    &       9, 17, 18  \\
585     &       NGC4051 &       60001050008     &       0.002 & Sy1.5   &       102900  &       5.89    &       2, 9, 12, 13, 27   \\ 
        &               &       60001050006     &        &              &       95400   &               &               \\
        &               &       60001050003     &        &              &       86100   &               &               \\
        &               &       60001050002     &        &              &       16800   &               &               \\
595     &       NGC4151 &       60001111005     &       0.003 & Sy1.5   &       119700  &       7.55    &       2, 7, 9, 13, 15, 28, 29, 30, 31, 32        \\
        &               &       60502017004     &        &              &       87000   &               &               \\
        &               &       60502017006     &        &          &   62700   &               &               \\
        &               &       60502017002     &        &              &       62100   &               &               \\
        &               &       60502017008     &        &              &       61800   &               &               \\
        &               &       60502017012     &        &              &       57000   &               &               \\
        &               &       60502017010     &        &              &       57000   &               &               \\
        &               &       60001111002     &        &              &       46200   &               &               \\
608     &       NGC4253 &       60001048002     &       0.012 & Sy1.5   &       172800  &       6.82    &       9, 17, 18, 33      \\
616     &       NGC4395 &       60061322002     &       0.001 & Sy1.0   &       39000   &       5.44    &       34      \\
631     &       NGC4593 &       60001149006     &       0.009 & Sy1.0   &       47700   &       6.88    &       2, 9, 35, 36       \\
        &               &       60001149004     &        &      &       47700   &               &               \\
        &               &       60001149010     & &             &       39900   &               &               \\
680     &       MCG-6-30-15     &       60001047003     &       0.007 & Sy1.0   &       261000  &       6.29    &       37      \\
        &               &       60001047005     &        &              &       57000   &               &               \\
        &               &       60001047002     &        &              &       46500   &               &               \\
686     &       NGC5273 &       60061350002     &       0.003 & Sy1.5   &       39900   &       6.66    &       38      \\
697     &       Mrk279  &       60160562002     &       0.030 & Sy1.5   &       37800   &       7.43    &       2, 7, 9, 47, 48\\
717     &       NGC5548 &       60002044008     &       0.017 & Sy1.5   &       98100   &       7.71    & 2, 7, 13, 15, 17, 39, 40, 41, 42, 43, 44, 45, 46        \\
        &               &       60002044006     &        &              &       97800   &               &               \\
        &               &       60002044005     &        &              &       97200   &               &               \\
        &               &       90701601002     &        &              &       73800   &               &               \\
774     &       Mrk290  &       60061266004     &       0.029 & Sy1.5   &       41700   &       7.27    &       2, 9, 16   \\
984     &       3C382   &       60001084002     &       0.057 & Sy1.2   &       155400  &       8.84    &       12, 13      \\
        &               &       60202015002     &        &              &       47400   &               &               \\
        &               &       60202015004     &        &              &       44400   &               &               \\
        &               &       60202015010     &        &              &       40200   &               &               \\
        &               &       60202015008     &        &              &       38700   &               &               \\
        &               &       60202015006     &        &              &       38700   &           &            \\
        &               &       60061286002     &        &          &   32700   &               &               \\
994     &       3C390.3 &       60001082003     &       0.056 & Sy1.5   &       78000   &       8.63         &       2, 7, 9, 41, 49, 50, 51, 52, 53, 54     \\ 
    &           &       60001082002     &        &              &       38100   &               &               \\
1090    &       Mrk509  &       60101043002     &       0.034 & Sy1.2   &       320400  &       8.05    &       2, 6, 7, 9 \\
        &               &       60101043004     &        &              &       75000   &               &               \\
1182    &       NGC7469 &       60101001004     &       0.016 & Sy1.5   &       45600   &       6.95    &       2, 4, 7, 13, 55    \\
        &               &       60101001008     &        &              &       45300   &               &               \\
        &               &       60101001006     &        &              &       45300   &               &               \\
        &               &       60101001014     &        &              &       45000   &               &               \\
        &               &       60101001002     &        &              &       41400   &               &               \\
        &               &       60101001012     &        &              &       39600   &               &               \\
        &               &       60101001010     &        &              &       39600   &               &               \\ 
\hline 
\end{longtable}
{Notes: (1) {\it  Gehrels SWIFT}/BAT catalogue identification number (2) Optical counterpart name (3) \nustar observation ID (4) Spectroscopic redshift (5) Optical classification (6) Duration of the \nustar observation (7) Logarithm of the source black hole estimate in solar mass units (8) Reference for the stellar velocity dispersion measurement: (1) \citet{grier2012a}, (2) \citet{zu2011}, (3) \citet{grier2012b}, (4) \citet{collier1998}, (5) \citet{hu2015}, (6) \citet{peterson1998}, (7) \citet{peterson2004}, (8) \citet{du2014}, (9) \citet{bentz2013}, (10) \citet{rodriguez1997}, (11) \citet{santos1997}, (12) \citet{fausnaugh2017}, (13) \citet{dalla2020}, (14) \citet{kollatschny2001}, (15) \citet{derosa2018}, (16) \citet{denney2010}, (17) \citet{grier2013}, (18) \citet{bentz2010}, (19) \citet{valenti2015}, (20) \citet{bentz2008}, (21) \citet{bentz2021a}, (22) \citet{reichert1994}, (23) \citet{onken2002}, (24) \citet{stirpe1994}, (25) \citet{bentz2016b}, (26) \citet{bentz2021b}, (27) \citet{denney2009}, (28) \citet{clavel1990}, (29) \citet{metzroth2006}, (30) \citet{ulrich1996}, (31) \citet{maoz1991}, (32) \citet{bentz2006}, (33) \citet{bentz2009}, (34) \citet{peterson2005}, (35) \citet{barth2013}, (36) \citet{denney2006}, (37) \citet{bentz2016a}, (38) \citet{bentz2014}, (39) \citet{derosa2015}, (40) \citet{pei2017}, (41) \citet{Kovacevik2014}, (42) \citet{clavel1991}, (43) \citet{korista1995}, (44) \citet{dietrich1993}, (45) \citet{peterson2013}, (46) \citet{peterson2002}, (47) \citet{maoz1990}, (48) \citet{santos2001}, (49) \citet{dietrich2012}, (50) \citet{obrien1998}, (51) \citet{sergeev2017}, (52) \citet{dietrich1998}, (53) \citet{sergeev2011}, (54) \citet{shapovalova2010}, (55) \citet{peterson2014}}

\begin{longtable}{rrrrrccrc}
\caption{\label{vd_sample} Log of VD sample}\\
\hline\hline
BAT ID  &       Name    &       NuSTAR ObsID    &       z &      Type   &       Duration  &      $\rm Log(M/M_\odot)$            &       Ref.    & $\rm N_H$ \\
& & & & & (sec) & & & $(\rm cm^{-2})$  \\
(1) & (2) & (3) & (4) & (5) &  (6) & (7) & (8) & (9)\\
\hline
\endfirsthead
\caption{continued.}\\
\hline\hline
BAT ID  &       Name    &       NuSTAR ObsID    &       z &     Type    &       Duration        &       $\rm Log(M/M_\odot)$ &       Ref.  & $\rm N_H$       \\
& & & & & (sec) & & &  $\rm (cm^{-2})$  \\
(1) & (2) & (3) & (4) & (5) &  (6) & (7) & (8) & (9)\\
\hline
\endhead
\hline
\endfoot
62      &       IC1657  &       60261007002     &       0.012   &       Sy2     &       86100   &       6.93    &       1       &       23.40   \\
        &               &       60061008003     &       &       &       23400   &       &               &               \\
63      &       NGC454E &       60061009002     &       0.012   &       Sy2     &       41100   &       8.41    &       1       &       23.30   \\
74      &       NGC513  &       60061012002     &       0.019   &       Sy2     &       27600   &       7.64    &       1       &       22.78   \\
77      &       Mrk359  &       60402021010     &       0.017   &       Sy1.5   &       100200  &       6.60    &       1       &       20.61   \\
        &               &       60402021006     &               &               &       98700   &               &               &               \\
    &           &       60402021004     &               &               &       98400   &               &               &               \\
84      &       NGC612  &       60061014002     &       0.030   &       Sy2     &       31500   &       8.98    &       1       &       23.97   \\
102     &       NGC788  &       60061018002     &       0.013   &       Sy2     &       28800   &       8.03    &       1       &       23.82   \\
157     &       NGC1142 &       60368001002     &       0.028   &       Sy2     &       41400   &       9.29    &       1       &       23.76   \\
171     &       UGC2638 &       60160148002     &       0.023   &       Sy2     &       44400   &       7.56    &       1       &       22.86   \\
216     &       NGC1566 &       80401601002     &       0.005   &       Sy1.5   &       151800  &       6.72    &       2       &       20.00   \\
        &                   &   60501031006     &               &       &       149400  &               &               &           \\
        &                   &   60501031004     &               &       &       134400  &               &               &               \\
        &                   &   80502606002     &               &       &       116700  &               &               &               \\
        &                   &   80301601002     &               &       &       111000  &               &               &               \\
        &                   &   60501031002     &               &       &       104700  &               &               &               \\
260     &       2MASXJ05081967+1721483  &       60006011002     &       0.017   &       Sy1.9   &       27600   &       8.49    &       1       &       22.33   \\
279     &       ESO553-43       &       60160236002     &       0.027   &       Sy2     &       38700   &       7.84    &       1       &       23.30   \\
308     &       NGC2110 &       60061061002     &       0.007   &       Sy2     &       33300   &       9.29    &       1       &       22.94   \\
337     &       VIIZw73 &       60061067002     &       0.041   &       Sy2     &       30900   &       7.30    &       1       &       23.74   \\
345     &       2MASXJ06411806+3249313  &       60061071002     &       0.047   &       Sy2     &       44400   &       7.92    &       1       &       23.12   \\
349     &       UGC3601 &       60160278002     &       0.017   &       Sy1     &       42300   &       8.50    &       1       &       21.40   \\
385     &       UGC3995B        &       60061352002     &       0.016   &       Sy2     &       43200   &       7.55    &       1       &       23.92   \\
405     &       UGC4211 &       60260001002     &       0.034   &       Sy2     &       39300   &       8.03    &       1       &       23.02   \\
416     &       Fairall272      &       60061080002     &       0.021   &       Sy2     &       45300   &       7.95    &       1       &       23.53   \\
426     &       4C+29.30        &       60061083002     &       0.064   &       Sy2     &       40200   &       8.71    &       1       &       23.80   \\
430     &       CASG218 &       60260002002     &       0.054   &       Sy2     &       39001   &       8.32    &       1       &       23.61   \\
439     &       Mrk18   &       60061088002     &       0.011   &       Sy1.9   &       36600   &       7.59    &       1       &       23.08   \\
451     &       IC2461  &       60061353002     &       0.007   &       Sy2     &       62700   &       6.68    &       1       &       22.78   \\
453     &       MCG-1-24-12     &       60061091010     &       0.019   &       Sy2     &       28800   &       7.63    &       1       &       22.81   \\
        &               &       60061091006     &               &       &       28500   &               &               &               \\
        &               &       60061091012     &           &   &       22200   &               &               &               \\
        &               &       60061091002     &               &       &       22200   &           &            &               \\
        &               &       60061091004     &               &       &       20100   &               &               &               \\
471     &       NGC2992 &       60160371002     &       0.007   &       Sy1.9   &       37800   &       7.44    &       1       &       21.72   \\
474     &       4C+73.08        &       60160374002     &       0.058   &       Sy2     &       22500   &       9.17    &       1       &       23.79   \\
480     &       NGC3081 &       60561044002     &       0.008   &       Sy2     &       108900  &       7.17    &       2       &       23.91   \\
509     &       LEDA93974       &       60061202002     &       0.020   &       Sy2     &       39900   &       8.41    &       1       &       22.60   \\
515     &       UGC5856 &       60061359002     &       0.025   &       Sy2     &       44100   &       6.77    &       1       &       22.60   \\
517     &       UGC5881 &       60160409002     &       0.020   &       Sy2     &       39000   &       7.96    &       1       &       22.90   \\
519     &       Mrk417  &       60061206002     &       0.032   &       Sy2     &       39600   &       7.96    &       1       &       23.90   \\
528     &       Z291-28 &       60160420002     &       0.047   &       Sy2     &       28500   &       8.21    &       1       &       23.15   \\
532     &       Mrk732  &       60061208002     &       0.029   &       Sy1.5   &       51300   &       8.41    &       1       &       20.00   \\
533     &       2MASXJ11140245+2023140  &       60061324002     &       0.026   &       Sy2     &       43200   &       8.07    &       1       &       23.28   \\
548     &       NGC3718 &       60301031004     &       0.003   &       Sy1.9   &       167400  &       9.63    &       1       &       21.96   \\
        &               &       60301031008     &               &               &       110400  &               &               &               \\
        &               &       60301031006     &               &       &       106500  &       &               &               \\
557     &       HE1136-2304     &       80002031003     &       0.027   &       Sy1.5   &       127800  &       7.32    &       1       &       21.30   \\
        &               &       80002031002     &               &               &       51300   &       &               &               \\
560     &       NGC3786 &       60061349002     &       0.008   &       Sy1.9   &       43200   &       7.46    &       2       &       22.36   \\
568     &       UGC6732 &       60161452002     &       0.009   &       Sy2     &       40200   &       8.52    &       1       &       22.28   \\
580     &       IC751   &       60061217002     &       0.031   &       Sy2     &       21600   &       8.50    &       1       &       23.88   \\
584     &       LEDA38038       &       60061219002     &       0.028   &       Sy2     &       39600   &       7.75    &       1       &       22.46   \\
600     &       Z187-22 &       60160481002     &       0.024   &       Sy2     &       39000   &       7.47    &       1       &       23.74   \\
605     &       Was49b  &       60061335002     &       0.063   &       Sy1     &       38100   &       8.18    &       1       &       23.41   \\
609     &       NGC4258 &       60101046004     &       0.001   &       Sy1.9   &       195600  &       7.43    &       2       &       23.00   \\
        &               &       60101046002     &               &               &       103500  &               &               &               \\
615     &       NGC4388 &       60501018002     &       0.008   &       Sy2     &       97200   &       6.74    &       2       &       23.52   \\
        &               &       60061228002     &               &               &       38400   &               &               &               \\
626     &       NGC4507 &       60102051004     &       0.011   &       Sy1.9   &       68700   &       7.69    &       2       &       23.95   \\
        &               &       60102051008     &               &       &       57300   &       &               &               \\
629     &       ESO506-27       &       60469006002     &       0.025   &       Sy2     &       39300   &       8.98    &       1       &       23.95   \\
630     &       LEDA170194      &       60061232002     &       0.036   &       Sy2     &       39900   &       7.97    &       1       &       22.76   \\
653     &       NGC4941 &       60061236002     &       0.004   &       Sy2     &       39600   &       6.60    &       1       &       23.91   \\
654     &       NGC4939 &       60002036002     &       0.010   &       Sy2     &       42000   &       7.19    &       1       &       23.29   \\
659     &       NGC4992 &       60061239002     &       0.025   &       Sy2     &       43500   &       8.43    &       1       &       23.69   \\
669     &       LEDA46599       &       60160536002     &       0.031   &       Sy2     &       38700   &       8.39    &       1       &       22.88   \\
674     &       ESO509-38       &       60260010002     &       0.026   &       Sy1.9   &       45600   &       8.45    &       1       &       20.00   \\
678     &       ESO509-IG066    &       60061244002     &       0.034   &       Sy1.9   &       39000   &       7.98    &       1       &       22.84   \\
684     &       NGC5283 &       60465006002     &       0.010   &       Sy2     &       51600   &       7.56    &       2       &       23.15   \\
685     &       Mrk268  &       60061246002     &       0.039   &       Sy1.9   &       39600   &       8.43    &       1       &       23.53   \\
694     &       IC4329A &       60001045002     &       0.016   &       Sy1.5   &       310200  &       8.66    &       2       &       21.52   \\
710     &       2MASXJ14104482-4228325  &       60160571002     &       0.033   &       Sy2     &       39900   &       8.10    &       1       &       22.78   \\
712     &       NGC5506 &       60501015002     &       0.006   &       Sy1.9   &       121200  &       8.06    &       2       &       22.44   \\
        &               &       60501015004     &               &               &       93900   &               &               &               \\
723     &       NGC5610 &       60160581002     &       0.016   &       Sy2     &       45300   &       7.55    &       1       &       22.56   \\
733     &       NGC5674 &       60061337002     &       0.024   &       Sy2     &       39300   &       7.35    &       1       &       22.84   \\
734     &       NGC5683 &       60160589002     &       0.036   &       Sy1.2   &       40200   &       7.54    &       2       &       20.00   \\
754     &       Mrk1392 &       60160605002     &       0.036   &       Sy1.5   &       39300   &       8.39    &       2       &       20.00   \\
755     &       2MASXJ15064412+0351444  &       60301023002     &       0.037   &       Sy2     &       125400  &       7.16    &       1       &       22.18   \\
        &               &       60061261002     &               &       &       39000   &       &               &               \\
757     &       Mrk1393 &       60376005002     &       0.054   &       Sy1     &       60300   &       7.90    &       1       &       20.28   \\
        &               &       60160607002     &               &               &       44100   &               &               &               \\
766     &       NGC5899 &       60061348002     &       0.008   &       Sy2     &       43500   &       8.58    &       1       &       23.03   \\
783     &       NGC5995 &       60061267002     &       0.025   &       Sy1.9   &       38400   &       8.55    &       1       &       21.97   \\
804     &       Z367-9  &       60061270002     &       0.027   &       Sy2     &       38700   &       9.98    &       1       &       23.02   \\
817     &       SDSSJ163115.52+235257.4 &       60260011002     &       0.059   &       Sy1     &       42000   &       8.67    &       1       &       21.70   \\
836     &       LEDA214543      &       60061273002     &       0.032   &       Sy2     &       39600   &       9.99    &       1       &       22.58   \\
875     &       NGC6300 &       60061277002     &       0.003   &       Sy2     &       28800   &       6.80    &       2       &       23.31   \\
960     &       MCG+7-37-31     &       60061283002     &       0.041   &       Sy2     &       21900   &       7.91    &       1       &       22.48   \\
        &               &       60061283004     &               &       &       15600   &               &               &               \\
971     &       CGMW5-03333     &       60160686002     &       0.067   &       Sy1.9   &       44400   &       7.75    &       1       &       22.43   \\
978     &       LEDA3097193     &       60061354002     &       0.022   &       Sy2     &       28800   &       7.44    &       1       &       22.98   \\
        &               &       60061354004     &               &       &       16500   &               &               &               \\
981     &       CGMW5-04382     &       60061285002     &       0.019   &       Sy2     &       42900   &       8.25    &       1       &       23.18   \\
1009    &       Fairall189      &       60160700002     &       0.028   &       Sy1     &       33300   &       7.96    &       1       &       20.00   \\
1027    &       ESO231-26       &       60160706002     &       0.062   &       Sy2     &       42000   &       8.76    &       1       &       23.38   \\
1049    &       2MASXJ19471938+4449425  &       60061292002     &       0.053   &       Sy2     &       33900   &       9.00    &       1       &       22.84   \\
1051    &       3C403   &       60061293002     &       0.059   &       Sy2     &       39300   &       9.17    &       1       &       23.69   \\
1135    &       NGC7172 &       60061308002     &       0.008   &       Sy2     &       60000   &       8.32    &       1       &       22.91   \\
1156    &       ESO533-50       &       60061312002     &       0.026   &       Sy2     &       42300   &       7.58    &       1       &       23.49   \\
1161    &       Mrk915  &       60002060002     &       0.024   &       Sy1     &       103800  &       7.49    &       1       &       20.00   \\
1162    &       UGC12138        &       60061343002     &       0.025   &       Sy1.5   &       42000   &       7.35    &       2       &       20.00   \\
1177    &       UGC12282        &       60160812002     &       0.017   &       Sy1     &       50400   &       9.96    &       1       &       23.76   \\
1202    &       UGC12741        &       60061321002     &       0.017   &       Sy2     &       39600   &       8.61    &       1       &       23.82   \\
1409    &       NGC4579 &       60201051002     &       0.005   &       Sy1.9   &       253800  &       7.94    &       2       &       20.50   \\
\hline 
\end{longtable}

\begin{longtable}{rrrrrccrc}
\caption{\label{ct_sample} Log of CT sample}\\
\hline\hline
BAT ID  &       Name    &       NuSTAR ObsID    &       z &     Type    &       Duration        &       $\rm log(M/M_\odot)$         &       Ref.    & $\rm N_H$ \\
& & & & & (sec) & & & $\rm (cm^{-2})$  \\

\hline
\endfirsthead
\caption{continued.}\\
\hline\hline
BAT ID  &       Name    &       NuSTAR ObsID    &       z &     Type    &       Duration        &       $\rm log(M/M_\odot)$         &       Ref. & $\rm N_H$        \\
& & & & & (sec) & & & $\rm (cm^{-2})$  \\
(1) & (2) & (3) & (4) & (5) &  (6) & (7) & (8) & (9)\\
\hline
\endhead
\hline
\endfoot
70      &       MCG+8-3-18      &       60061010002     &       0.02    &       Sy2     &       58200   &       8.27    &       1       &       24.12   \\
81      &       ESO244-30       &       60468001002     &       0.025   &       Sy2     &       59700   &       7.04    &       1       &       24.36   \\
144     &       NGC1068 &       60302003004     &       0.003   &       Sy1.9   &       110700  &       7.88    &       2       &       25.00   \\
        &               &       60002030002     &               &               &       109200  &               &               &               \\
        &               &       60302003008     &               &               &       108600  &               &               &               \\
        &               &       60002033004     &               &               &       108300  &               &               &               \\
        &               &       60002033002     &               &               &       103800  &               &               &               \\
        &               &       60302003002     &               &               &       97800   &               &               &               \\
        &               &       60302003006     &               &               &       97200   &               &               &               \\
        &               &       60002030004     &               &               &       90900   &               &               &               \\
153     &       NGC1125 &       60510001002     &       0.011   &       Sy2     &       64500   &       6.98    &       1       &       24.21   \\
165     &       NGC1229 &       60061325002     &       0.036   &       Sy2     &       46500   &       8.10    &       1       &       24.00   \\
245     &       Z420-15 &       60061053004     &       0.029   &       Sy2     &       34800   &       8.15    &       1       &       24.08   \\
    &           &       60061053002     &               &               &       27300   &       &               &               \\
325     &       Mrk3    &       60002048004     &       0.013   &       Sy1.9   &       52200   &       8.56    &       1       &       24.06   \\
        &               &       60002048002     &               &               &       51900   &               &               &               \\
        &               &       60002048006     &               &                   &    51000   &               &               &               \\
        &               &       60002048010     &               &                   &    46800   &               &               &               \\
        &               &       60002048008     &               &                   &    46200   &               &               &               \\
        &               &       60002049004     &               &                &       42000   &               &               &               \\
        &               &       60002049002     &               &                    &   40500   &               &               &               \\
        &               &       60002049006     &               &                &       40200   &               &               &               \\
        &               &       60002049010     &               &                &       39900   &       &               &               \\
        &               &       60002049008     &               &               &       37800   &       &               &               \\
362     &       UGC3752 &       60061072002     &       0.015   &       Sy1.9   &       45300   &       6.98    &       1       &       24.78   \\
440     &       NGC2788A        &       60160344002     &       0.013   &       Sy2     &       40500   &       8.64    &       1       &       24.26   \\
        &               &       60469001002     &               &               &       37500   &               &               &               \\
467     &       UGC5101 &       60001068004     &       0.039   &       Sy1.9   &       44400   &       8.20    &       1       &       24.28   \\
484     &       NGC3079 &       60061097002     &       0.003   &       Sy2     &       40200   &       8.06    &       2       &       24.56   \\
518     &       NGC3393 &       60061205002     &       0.012   &       Sy2     &       27900   &       8.34    &       2       &       24.40   \\
590     &       NGC4102 &       60160472002     &       0.002   &       Sy2     &       39600   &       8.69    &       1       &       24.14   \\
711     &       Circinus &      60002039002     &       0.001   &       Sy2     &       94200   &       7.67    &       2       &       24.36   \\
        &               &       30002038004     &       &               &       74400   &               &               &       24.36   \\
        &               &       30002038006     &               &               &       66300   &               &               &               \\
        &           &   30002038002     &               &               &       34200   &               &               &               \\
739     &       NGC5728 &       60061256002     &       0.009   &       Sy1.9   &       49800   &       8.33    &       2       &       24.14   \\
740     &       Z164-19 &       60061327006     &       0.029   &       Sy1.9   &       37200   &       6.70    &       1       &       24.64   \\
828     &       NGC6232 &       60061328004     &       0.014   &       Sy2     &       63000   &       7.08    &       1       &       24.35   \\
        &               &       60061328002     &               &               &       31200   &               &               &               \\
1127    &       NGC7130 &       60261006002     &       0.016   &       Sy1.9   &       81000   &       7.30    &       1       &       24.22   \\
        &               &       60061347002     &               &                &       39600   &               &               &               \\
1184    &       NGC7479 &       60061316002     &       0.007   &       Sy1.9   &       43500   &       7.21    &       1       &       24.16   \\
1188    &       NGC7582 &       60201003002     &       0.005   &       Sy2     &       100500  &       7.15    &       2       &       24.15   \\
        &               &       60061318002     &               &               &       38100   &               &               &               \\
        &               &       60061318004     &           &           &       27900   &               &               &               \\
1198    &       NGC7682 &       60368002002     &       0.017   &       Sy2     &       47100   &       7.17    &       2       &       24.27   \\
        &               &       60061319002     &               &               &       43200   &               &               &               \\
1262    &       NGC1320 &       60061036004     &       0.008   &       Sy2     &       52200   &       6.99    &       2       &       24.10   \\
        &               &       60061036002     &               &               &       26400   &               &               &               \\
1302    &       NGC2273 &       60001064002     &       0.006   &       Sy2     &       39600   &       7.56    &       2       &       24.10   \\
1425    &       NGC5194 &       60201062003     &       0.002 & Sy2     &       324000  &       6.47    &       2       &       24.70   \\
        &               &       60002038002     &               &               &       28500   &               &               &               \\
\hline 
\end{longtable}
{Notes: (1) {\it  Gehrels SWIFT}/BAT catalogue identification number. (2) Optical counterpart name. (3) \nustar observation ID. (4) Spectroscopic redshift. (5) Optical Classification. (6) Duration of the \nustar observation. (7)Logarithm of the source black hole estimate in solar mass units. (8) Reference for the stellar velocity dispersion measurement: (1) \citet{koss2017}, (2) \citet{paturel2003}-Hyperleda database. (9) X-ray column density }

\begin{longtable}{rrrrrr}
\caption{\label{prediction_sample} Log of the prediction sample}\\
\hline\hline
BAT ID & Name & NuSTAR ObsID    & z &   Type    & Duration  \\
& & & & & (sec)  \\
(1) & (2) & (3) & (4) & (5) &  (6)  \\

\hline
\endfirsthead
\caption{continued.}\\
\hline\hline
BAT ID & Name & NuSTAR ObsID    & z &   Type    & Duration \\
& & & & & (sec)  \\
(1) & (2) & (3) & (4) & (5) &  (6)  \\
\hline
\endhead
\hline
\endfoot
%49  & MCG-7-3-7 & 60561039002 & 0.03 & Sy2 & 97500\\
%106     &  Mrk1018 & 60160087002 & 0.042 & Sy1.2 & 43200\\
%  &     Mrk1018 & 60301022005  &   &   & 80400\\
%  &     Mrk1018 & 60301022002  &   &   & 86400\\
%  &     Mrk1018 & 60301022003  &   &   & 90900\\
163     &  NGC1194 & 60501011002 & 0.013 & Sy2 & 114900\\
184     &  NGC1365 & 60002046003 & 0.005 & Sy2 & 75900\\
  &      & 60002046009  &   &   & 132300\\
  &      & 60002046007  &   &   & 146100\\
214     &  3C111 & 60202061002 & 0.048 & Sy1.2 & 39900\\
  &      & 60202061004  &   &   & 91800\\
  &      & 60202061006  &   &   & 104100\\
229     &  HE0436-4717 & 60160197002 & 0.053 & Sy1 & 39000\\
  &     & 30001061004  &   &   & 116100\\
270     &  PictorA & 60101047002 & 0.035 & Sy2 & 225300\\
%330     &  ESO426-2 & 60561040002 & 0.022 & Sy2 & 105900\\
447     &  IRAS09149-6206 & 90401630002 & 0.057 & Sy1 & 162600\\
  &      & 60401020002  &   &   & 214500\\
%465     &  ESO565-19 & 60561043002 & 0.016 & Sy2 & 102300\\
472     &  MCG-5-23-16 & 60001046006 & 0.008 & Sy1.9 & 207900\\
567     &  HE1143-1810 & 60302002010 & 0.032 & Sy1.2 & 42600\\
  &      & 60302002006  &   &   & 45000\\
%599     &  NGC4180 & 60160480002 & 0.007 & Sy2 & 44400\\
%  &      & 60201038002  &   &   & 68400\\
%621     &  NGC4500 & 60161497002 & 0.01 & Sy2 & 39300\\
%  &      & 60510002002  &   &   & 98100\\
657     &  ESO323-77 & 60202021004 & 0.015 & Sy1.5 & 80100\\
  &      & 60202021002  &   &   & 80400\\
  &      & 60202021006  &   &   & 80700\\
  &      & 60202021008  &   &   & 86400\\
670     &  MCG-3-34-64 & 60101020002 & 0.016 & Sy1.9 & 151200\\
692     &  4U1344-60 & 60201041002 & 0.012 & Sy1.9 & 163800\\
719     &  ESO511-30 & 60502035006 & 0.022 & Sy1 & 60300\\
  &      & 60502035002  &   &   & 62100\\
  &      & 60502035004  &   &   & 63000\\
  &      & 60502035010  &   &   & 63300\\
  &      & 60502035008  &   &   & 80400\\
%731     &  NGC5643 & 60061362004 & 0.004 & Sy2 & 38700\\
%  &      & 60061362002  &   &   & 39900\\
%  &      & 60061362006  &   &   & 89400\\
750     &  LEDA3076910 & 60061259002 & 0.016 & Sy1.5 & 39900\\
  &      & 60401022002  &   &   & 188700\\
837     &  ESO138-1 & 60201040002 & 0.009 & Sy2 & 91800\\
  &      & 60061274002  &   &   & 91800\\
841     &  NGC6240 & 60102042006 & 0.024 & Sy1.9 & 45000\\
  &      & 60102042004  &   &   & 51300\\
  &      & 60002040002  &   &   & 59700\\
%942     &  NGC6552 & 60561046002 & 0.026 & Sy2 & 92100\\
995     &  Fairall51 & 60402014004 & 0.014 & Sy1.5 & 59400\\
  &      & 60402014006  &   &   & 63300\\
  &      & 60402014002  &   &   & 118200\\
1032     &  ESO141-55 & 60201042002 & 0.036 & Sy1.2 & 174300\\
1040     &  2MASXJ19301380+3410495 & 60160713002 & 0.062 & Sy1.5 & 39600\\
  &      & 60376001002  &   &   & 95100\\
1111     &  IGRJ21277+5656 & 60001110003 & 0.014 & Sy1 & 50100\\
  &      & 60001110007  &   &   & 75000\\
  &      & 60001110002  &   &   & 89100\\
  &      & 60402008004  &   &   & 132300\\
  &      & 60402008010  &   &   & 133800\\
  &      & 60402008008  &   &   & 137400\\
  &      & 60402008006  &   &   & 142800\\
1172     &  MR2251-178 & 60102025004 & 0.064 & Sy1.2 & 43500\\
  &      & 60102025008  &   &   & 45600\\
  &      & 60102025002  &   &   & 51600\\
  &      & 90601637002  &   &   & 52800\\
1183     &  Mrk926 & 60201029002 & 0.046 & Sy1.5 & 213600\\
%1186     &  UGC12383 & 60561047002 & 0.035 & Sy2 & 107700\\
1194     &  IRAS23226-3843 & 80502607002 & 0.035 & Sy2 & 118800 \\

\hline 
\end{longtable}
{Notes: (1) {\it  Gehrels SWIFT}/BAT catalogue identification number. (2) Optical counterpart name. (3) \nustar observation ID. (4) Spectroscopic redshift. (5) Optical Classification. (6) Duration of the \nustar observation.}

\end{appendix}
\end{document}